\pdfoutput=1
\documentclass[%
 reprint,
 amsmath,amssymb,
 aps,
 prb,
]{revtex4-2}

\usepackage{graphicx}%
\usepackage{dcolumn}%
\usepackage{bm}%
\usepackage{color}

\begin{document}

\title{Origin of the tetragonal-to-hexagonal phase transitions in Fe-doped BaTiO$_3$}

\author{Zhiyuan Li}
\author{Ruiwen Xie}
\author{Hongbin Zhang}

\email{hzhang@tmm.tu-darmstadt.de} 

\affiliation{Institute of Materials Science, Technische Universität Darmstadt, 64287 Darmstadt, Germany}

\date{\today}

\newcommand{\TODO}[1]{\textcolor{red}{[\textbf{TODO} #1]}}
\newcommand{\VO}{$V_{\mathrm{O}}$}

\begin{abstract}
Based on detailed first-principles calculations, we investigate the
tetragonal-to-hexagonal phase transition in Fe-doped BaTiO$_3$. Free energy calculations confirm a crossover from the tetragonal to hexagonal phases at 2.7--6\% Fe on cooling from the sintering temperature, in agreement with experimental observations, where comparative calculations show that neither CaTiO$_3$ nor SrTiO$_3$ exhibits similar behavior under equivalent substitution.
Furthermore, three possible mechanisms are quantified: oxygen vacancies shift the crossover concentration from 2.7--6\% to 1.1--4.7\%, Jahn-Teller distortions are confined to the tetragonal phase, both favoring tetragonal-to-hexagonal phase transitions; whereas the tolerance factor is reduced in comparison with that of pristine BaTiO$_3$ for reasonable Fe valence states, disfavoring the occurrence of the hexagonal phases. Detailed analysis on the electronic structure reveals that the charge redistribution induced by oxygen vacancy is strongly orbital dependent, with the added charge occupying the $d_{z^2}$ orbital.
\end{abstract}

\maketitle

\section{Introduction}\label{sec:intro}
Structural phase transitions in ABO$_3$ perovskite oxides give rise to diverse functional properties, enabling tunable ferroelectricity~\cite{catalan2009physics,spaldin2019advances,lines2001principles,rabe2007modern}, piezoelectricity~\cite{lines2001principles,setter2006ferroelectric}, and electro-optic responses~\cite{lines2001principles,setter2006ferroelectric} through symmetry-breaking distortions.
This underscores the critical potential of tailoring phase transitions in perovskites for enabling enhanced functionalities.
For instance, BaTiO$_3$ (BTO) stands as an archetypal system for investigating such structure-property relationships, distinguished by its rich polymorphism and exceptional sensitivity to external perturbations. \label{ln:seq}Upon heating, BTO undergoes a definitive structural evolution: rhombohedral ($R3m$) $\to$ orthorhombic ($Amm2$) $\to$ tetragonal ($P4mm$) $\to$ cubic ($Pm\bar{3}m$) at approximately 183 K, 278 K, and 403 K, respectively, while the hexagonal structure ($P6_3/mmc$) stabilizes only above 1733 K~\cite{kirby1991phase}. Strategically engineering these phase transitions yields exceptional performance metrics. For example, constructing a morphotropic phase boundary (MPB) in (Ba,Ca)(Zr,Ti)O$_3$ minimizes the polarization anisotropy energy, achieving a giant piezoelectric coefficient ($d_{33} \sim 620$ pC/N) comparable to lead-based ceramics~\cite{liu2009large}. Similarly, stabilizing the tetragonal symmetry via epitaxial strain in BTO thin films has enabled effective Pockels coefficients as high as 923 pm/V for integrated photonics~\cite{abel2019large}.

To tailor the desired phases with enhanced properties, three most common strategies are usually leveraged to manipulate the energetic landscape: epitaxial strain, dynamic field excitation, and chemical substitution. Epitaxial strain engineering utilizes substrate-induced lattice mismatch to stabilize non-equilibrium phases~\cite{lee2010strong,bousquet2008improper}, e.g. imparting a compressive strain of $\sim 1.7\%$ in BTO thin films elevates the Curie temperature to nearly $500^\circ$C and enhances the remanent polarization to $\sim 70$ $\mu$C/cm$^2$, far exceeding bulk values~\cite{choi2004enhancement}. Dynamic excitation offers transient access to metastable structures, where optical pulses or intense terahertz electric fields can coherently drive soft phonon modes to induce ferroelectricity in quantum paraelectrics on picosecond timescales~\cite{nova2017effective,li2019terahertz}. Complementing these physical stimuli that target metastable or transient states, chemical substitution, distinguished by the concentration regime into doping versus alloying (low vs. high concentration of extrinsic contents), provides a distinct pathway for intrinsic bulk phase control. This approach enables the tailoring of phase stability through precise modifications of local structure, electronic configuration, and defect chemistry~\cite{morrison1999electrical,choi2012wide,reaney2006microwave,smyth2000defect}. For instance, isovalent alloying (e.g., Sr$^{2+}$ for Ba$^{2+}$) is widely utilized to shift transition temperatures for tunable dielectrics~\cite{alexandru2004oxides}, while aliovalent doping/alloying causes more fundamental changes depending on the possible charge compensation mechanisms~\cite{klein2023fermi}. 

Among chemical substitution strategies, transition metal (TM) doping in BTO has revealed complex and concentration-dependent tailoring behaviors that remain elusive.
\label{ln:keith}Keith et al.~\cite{keith2004synthesis} stabilized the hexagonal 6H polymorph at room temperature with a range of $B$-site substituents, among them Cr, Mn, Fe, Co, Ni and Zn. Fe substitution in particular retains the high-temperature hexagonal phase (space group P6$_3$/mmc) at room temperature~\cite{keith2004synthesis,xu2009room,mangalam2009multiferroic}. 
The hexagonal polymorph exhibits fundamentally different coordination environments characterized by the coexistence of corner-sharing and face-sharing octahedral sites, the latter forming distinct M$_2$O$_9$ dimers. This connectivity contrasts sharply with the exclusively corner-sharing framework of the tetragonal phase, leading to distinct electronic and magnetic properties~\cite{ray2008high,chakraborty2011defect}. 
Systematic experimental investigations have identified critical concentration thresholds for hexagonal phase emergence in Fe-substituted BTO. X-ray diffraction studies reveal tetragonal-hexagonal phase coexistence beginning at Fe concentrations of 2–4~at.\%, with hexagonal fraction progressively increasing to dominance ($>$90\%) at 10~at.\%~\cite{nguyen2011tetragonal,tho2024crystal}. The origin of this phase transition has been attributed to oxygen vacancy formation~\cite{nguyen2011tetragonal,langhammer2000crystal} and Jahn-Teller (JT) distortions~\cite{langhammer2000crystal,zorko2015strain}, as well as tolerance factor, though the contributions of these mechanisms need to be quantified. In contrast, Fe substitution in CaTiO$_3$(CTO) and SrTiO$_3$(STO) produces no hexagonal phase even at concentrations exceeding 20~at.\%, while affecting only the temperature-dependent sequence of orthorhombic-tetragonal-cubic transitions in CaTiO$_3$~\cite{becerro2002displacive,shafique2021magnetic}.

Recent experimental investigations by Pal \textit{et al.} demonstrated that strategic A-site co-doping can dramatically alter the tetragonal-hexagonal phase equilibrium in Fe-substituted BTO: modest Bi$^{3+}$ substitution (5~at.\%) reduces the hexagonal phase fraction from 98.8\% to 18.6\%, while even isovalent substitutions (Sr$^{2+}$, Ca$^{2+}$) produce similar phase transitions despite maintaining charge neutrality\cite{pal2020origin}. These findings imply Goldschmidt's tolerance factor as a critical control parameter, with small variations ($\sim$0.3\%) producing disproportionately large structural responses. However, a comprehensive theoretical understanding of the microscopic mechanisms, particularly the interplay between tolerance factor variations, local structural distortions, and oxygen vacancy formation, remains elusive\cite{rabe2007modern}. While previous density functional theory (DFT) studies have examined individual dopant effects in transition metal-doped BTO\cite{sambrano2005theoretical,yin2014ab,yang2017first,adeagbo2019theoretical,li2021coexistence}, they provide limited insight into the origin of tetragonal to hexagonal phase transitions. The quantitative evaluation of defect-induced phase stability and establishment of design principles for phase transition control remain largely unexplored from a first-principles perspective\cite{zunger2018inverse,curtarolo2013high,bartel2019new,goldschmidt1926gesetze}.

In this work, we employ comprehensive DFT calculations to investigate phase stability in Fe-substituted BTO to elucidate the underlying mechanisms. Systematic DFT calculations have been performed to evaluate the energetics of tetragonal-hexagonal phase transitions in Fe-substituted BTO, in comparison with CTO and STO. Furthermore, three fundamental mechanisms, {\it i.e.}, oxygen vacancies, Jahn-Teller distortions, and tolerance factor, are scrutinized based on detailed DFT-informed analysis. Unfolded band structure analysis, Fe $3d$ occupation matrix decomposition, and charge-density difference calculations are applied to quantify the orbital-resolved charge compensation at the Fe site induced by oxygen vacancies and reveal a symmetry-lowering orbital reconstruction of the Fe$_\text{Ti}$--$V_\text{O}$ defect complex.

\section{Computational Details}\label{sec:methods}

\begin{figure}[htbp]
\centering
\includegraphics[width=\columnwidth]{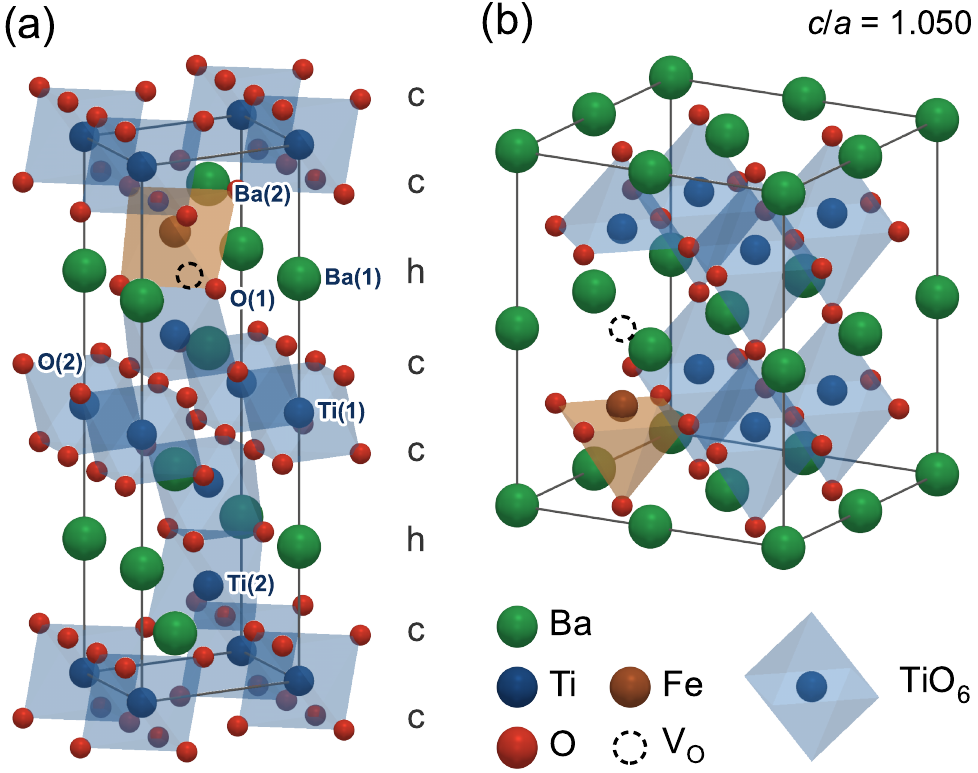}
\caption{Relaxed structures of the two BaTiO$_3$ phases, drawn to the same scale. (a) Hexagonal 6H BaTiO$_3$ ($P6_3/mmc$), one
conventional cell. The corner-sharing Ti(1) and face-sharing Ti(2) sites, the
two Ba sites, the O(1) site on the shared face and an O(2) site are labelled;
c and h mark the BaO$_3$ layers through which the octahedra share corners and
faces. (b) Tetragonal BaTiO$_3$ ($P4mm$), $2\times2\times2$ primitive cells.
TiO$_6$ octahedra are drawn translucent, with Ti at their centres. In (b)
$c/a = 1.050$, drawn with $c/a - 1$ four times larger for illustration, and
Ti lies 0.22~\AA{} off the centre of its octahedron along $c$. The lowest-energy Fe--$V_{\mathrm{O}}$ pair of each phase at 1.85\% Fe is marked on the pristine structure, with Fe and its polyhedron in brown and the removed O as a dashed circle. In (a) Fe is on Ti(2) and the vacancy on O(1), and in (b) the vacancy is on the apical O nearest Fe.}
\label{fig:structures}
\end{figure}
\begin{table*}[htbp]
\caption{Supercells and configurations. The supercell is given in multiples of the
perovskite, six-layer hexagonal, 20-atom or 10-atom cell; $a$, $b$, $c$ are its lengths, taken from the relaxed pristine host and held
fixed in every substituted cell. $n$ Fe atoms on the 54 B sites give
$A_{54}$Ti$_{54-n}$Fe$_n$O$_{162}$ ($A$ = Ba, Sr), and on the 48 B sites
Ca$_{48}$Ti$_{48-n}$Fe$_n$O$_{144}$; a vacancy removes one O. The last column gives
the number of symmetry-inequivalent configurations at each Fe content, in the same
order; f.u.\ is the number of formula units. The tetragonal CaTiO$_3$ supercell is built on the body-centred 10-atom cell
and is not orthogonal ($\alpha = \beta = 119.6^{\circ}$).}
\label{tab:cells}
\begin{ruledtabular}
\begin{tabular}{lllcccc}
Host & Phase & Supercell & $a$, $b$, $c$ (\AA) & \begin{tabular}[b]{@{}c@{}}f.u./\\atoms\end{tabular} & Fe (at.\%) & Configurations \\
\hline
BaTiO$_3$ & tetragonal, $P4mm$ & $3\sqrt{2}\times3\sqrt{2}\times3$ perovskite & 16.99, 16.99, 12.61 & 54/270 & 1.85, 3.70, 5.56 & 1, 11, 96 \\
          & \quad with $V_{\mathrm{O}}$ & & & 54/269 & 1.85, 3.70 & 3, 68 \\
          & hexagonal 6H, $P6_3/mmc$ & $3\times3\times1$ six-layer & 17.36, 17.36, 14.10 & 54/270 & 1.85, 3.70, 5.56 & 2, 22, 176 \\
          & \quad with $V_{\mathrm{O}}$ & & & 54/269 & 1.85, 3.70 & 3, 104 \\
SrTiO$_3$ & cubic, $Pm\bar{3}m$ & $3\sqrt{2}\times3\sqrt{2}\times3$ perovskite & 16.74, 16.74, 11.83 & 54/270 & 1.85, 3.70 & 1, 11 \\
          & hexagonal 6H (hypothetical) & $3\times3\times1$ six-layer & 16.90, 16.90, 13.88 & 54/270 & 1.85, 3.70 & 2, 22 \\
CaTiO$_3$ & orthorhombic, $Pbnm$ & $3\times2\times2$ 20-atom & 16.23, 11.01, 15.39 & 48/240 & 2.08, 4.17 & 1, 18 \\
          & tetragonal, $I4/mcm$ & $4\times3\times2$ 10-atom & 21.67, 16.25, 10.97 & 48/240 & 2.08, 4.17 & 2, 38 \\
\end{tabular}
\end{ruledtabular}
\end{table*}

The first-principles calculations were performed using the Vienna Ab initio Simulation Package (VASP)\cite{PhysRevB.54.11169}. The electron-ion interactions were described using the projector-augmented wave (PAW) method\cite{blochl1994projector}, with the Perdew-Burke-Ernzerhof (PBE) generalized gradient approximation (GGA) functional\cite{perdew1996generalized} for the exchange-correlation potential.
\label{ln:cells}The BaTiO$_3$ phases are shown in Fig.~\ref{fig:structures} and the control hosts in Fig.~S6 of the Supplemental Material.
To simulate the dilute substitution, supercells of 270 atoms containing 54 formula units were used for BTO and STO: a $3\sqrt{2} \times 3\sqrt{2} \times 3$ supercell of the perovskite unit cell for tetragonal BTO and cubic STO, and a $3 \times 3 \times 1$ supercell of the six-layer hexagonal cell for hexagonal BTO and STO. For CTO, 240-atom cells were used: a $3 \times 2 \times 2$ supercell of the 20-atom orthorhombic cell, and a $4 \times 3 \times 2$ supercell of the 10-atom tetragonal cell.
\label{ln:hsto}Hexagonal SrTiO$_3$ is not observed experimentally. Its initial atomic configuration was constructed by substituting Sr for Ba in the hexagonal BTO lattice, and was then relaxed independently, including cell shape and volume. The CTO phases were taken from their reported structures.
\label{ln:conc}For such BTO supercells, one, two, and three substitutional Fe atoms on the Ti sites are considered, corresponding to doping concentrations of 1.85, 3.70, and 5.56 at.\%, respectively. These theoretical concentrations approximate the experimental molar percentages of 2, 4, and 6 mol.\% as in a previous theoretical study~\cite{adeagbo2019theoretical}. For CTO, one and two Fe atoms correspond to 2.08 and 4.17 at.\%.
\label{ln:table}Table~\ref{tab:cells} lists the supercells, their compositions and the number of configurations.
For each Fe concentration, multiple symmetry-inequivalent substitution sites were systematically generated
using the bsym Python package \cite{Morgan_JOSS2017b}, for all considered phases at each substitution concentration.
\label{ln:crfe}To investigate the effects of oxygen vacancies, we created oxygen-deficient structures by systematically removing one oxygen atom from the first coordination shell of Fe. Although calculations~\cite{nayak2015chromium} and EPR~\cite{bottcher2005incorporation} place the vacancy away from Cr in hexagonal BaTiO$_3$, it is found in the first shell of Fe, by experiments and calculations in tetragonal BaTiO$_3$~\cite{ren2004large,noguchi2020successive} and by EPR on single crystals in hexagonal BaTiO$_3$~\cite{bottcher2018incorporation,langhammer2020incorporation}.
Again, only symmetry-inequivalent vacancy sites were considered to avoid redundant calculations. One vacancy is introduced per supercell. \label{ln:relax}In all substituted cells, with or without a vacancy, only the atoms within 6~\AA{} of the defect were relaxed at fixed cell shape and volume, following Ref.~\cite{noguchi2020successive}. Our tests show that the lowest-energy vacancy site in the tetragonal cell can change with the supercell shape and the relaxation scheme (Sec.~S4 of the Supplemental Material). Considering the experimental evidence~\cite{ren2004large,noguchi2020successive} and for consistency, we use the scheme of Ref.~\cite{noguchi2020successive} in all cells. With this scheme the lowest-energy vacancy is on the apical O nearest Fe, as in Ref.~\cite{noguchi2020successive} and in agreement with experiment.

\label{ln:hexvo}In the hexagonal phase the position of the vacancy next to Fe has been studied extensively, by EPR, diffraction, EXAFS and calculations (Sec.~S8 and Fig.~S7 of the Supplemental Material), and remains an open question. The first shell of Fe on Ti(2) contains O(1), on the face shared with the neighbouring Ti(2), and O(2); that of Ti(1) contains only O(2) [Fig.~\ref{fig:structures}(a)]. Our lowest-energy configuration has Fe on Ti(2) and the vacancy on O(1). The DFT study of Ref.~\cite{adeagbo2019theoretical} finds the same for a vacancy next to Fe, whereas EPR on single crystals finds Fe on Ti(1) with the vacancy on O(2)~\cite{langhammer2020incorporation}.

\label{ln:avg}The total energies of the configurations with and without an oxygen vacancy were averaged with Boltzmann statistics and configurational multiplicities (degeneracies):
\begin{equation}
\langle E \rangle_{\text{weighted}} = \frac{\sum_{i} N_i \cdot e^{-E_i/(k_B T)} \cdot E_i}{\sum_{i} N_i \cdot e^{-E_i/(k_B T)}}
\label{eq:avg3}
\end{equation}
where $E_i$ is the total energy of configuration $i$, $N_i$ its degeneracy, that is, the number of symmetrically equivalent ways to arrange the Fe atoms (and the oxygen vacancy when present), $k_B$ the Boltzmann constant, and $T = 300$~K. %
With the arithmetic mean the crossover temperature also decreases with Fe content from 1.85\% Fe on, but it lies up to about 650~K above that obtained with Eq.~(\ref{eq:avg3}), because the arithmetic mean gives the same weight to configurations of high energy. The three averaging schemes are compared in Secs.~S1 and S3 of the Supplemental Material.

The averaged energies are combined into\cite{ehsan2021first}
\begin{equation}
\Delta G(x,T) = \Delta E(x) + \Delta E_{\text{ZP}} - T\,\Delta s_{\text{eff}},
\label{eq:dG}
\end{equation}
where $\Delta$ denotes hexagonal minus tetragonal and $\Delta E(x)$ is the averaged energy difference per formula unit. The zero-point term $\Delta E_{\text{ZP}} = -4.27$~meV/f.u.\ is the difference of the harmonic phonon free energies of the two pristine phases at $T = 0$, taken from the Materials Project phonon database\cite{sahasrabuddhe2026phonon,jain2013commentary}, and is used unchanged for the Fe-substituted cells. The entropy difference is approximated by a single temperature-independent value, fixed by requiring the pristine host to transform at the experimental $T_0 = 1733$~K, which gives $\Delta s_{\text{eff}} = [\Delta E(0) + \Delta E_{\text{ZP}}]/T_0 = 0.0115$~meV~K$^{-1}$~f.u.$^{-1}$; the same value is used for every concentration and for both series. The crossover temperature is then $T^{*}(x) = [\Delta E(x) + \Delta E_{\text{ZP}}]/\Delta s_{\text{eff}}$.

\label{ln:spin}All energies reported here were obtained from spin-polarized calculations on charge-neutral supercells, each structure being initialized ferromagnetically with 5~$\mu_B$ on every Fe, as ferromagnetism is reported for Fe-doped BaTiO$_3$, in particular for the hexagonal phase~\cite{xu2009room,ray2008high}. Of the 486 Fe-substituted BaTiO$_3$ configurations, 14 converged with one Fe moment antiparallel to the others and 3 with one Fe without a moment. Together they have 6.0\% of the weight in Eq.~(\ref{eq:avg3}) in the tetragonal phase at 5.56\% Fe and less than 0.01\% in every other ensemble. The dataset comprises 587 converged relaxations: 109 tetragonal and 201 hexagonal BTO, 71 and 107 of the corresponding cells with an oxygen vacancy, 13 cubic and 25 hexagonal STO, and 20 orthorhombic and 41 tetragonal CTO.
The plane-wave basis set was truncated with an energy cutoff of 500 eV.
The Brillouin zone sampling was performed on a $\Gamma$-centred grid generated from a reciprocal space sampling density of 0.25 Å$^{-1}$. Given the strongly correlated nature of Fe 3$d$ electrons in oxide environments, we employed the DFT+U approach within the Dudarev formalism\cite{PhysRevB.57.1505}. A Hubbard U correction of 4.0 eV was applied to the Fe 3d orbitals, consistent with values established for TM oxides in previous theoretical studies\cite{meng2016density,grau2006electronic}; the sensitivity of the results to this choice is examined for 2 and 6 eV in Sec.~S2 of the Supplemental Material. All structures were optimized until forces on all atoms were reduced below 0.025 eV/Å.
The projected densities of states (PDOS) were obtained by projection onto the PAW spheres with fine energy resolution to accurately resolve the Fe $3d$ and O $2p$ orbital features, and are shown broadened to a common width of 0.10 eV. Regarding the determination of the valence state required for tolerance factor calculations, we adopted the high-spin Fe$^{2+}$ and Fe$^{3+}$ configuration. This assignment is robustly supported by extensive experimental XANES and EPR characterizations\cite{padchasri2021crystal, rajan2017impact,kolodiazhnyi2003analysis}. Consequently, we utilized the corresponding Shannon ionic radii~\cite{shannon1976revised} to quantify the geometric tolerance factor. To investigate the electronic structure modifications upon Fe doping, band structure calculations were performed along high-symmetry $k$-point paths in the Brillouin zone. For direct comparison between the pristine BTO and Fe-substituted systems, band unfolding analysis, which maps the band structure of the doped supercell back to the primitive cell Brillouin zone\cite{ku2010unfolding}, was employed using the Pyprocar package \cite{herath2020pyprocar}.

\section{Results and discussion}\label{sec:results}

\subsection{Total energies for Fe-substituted BTO/CTO/STO}\label{sec:totalE}
\begin{figure*}[htbp]
    \centering
    \includegraphics[width=0.9\textwidth]{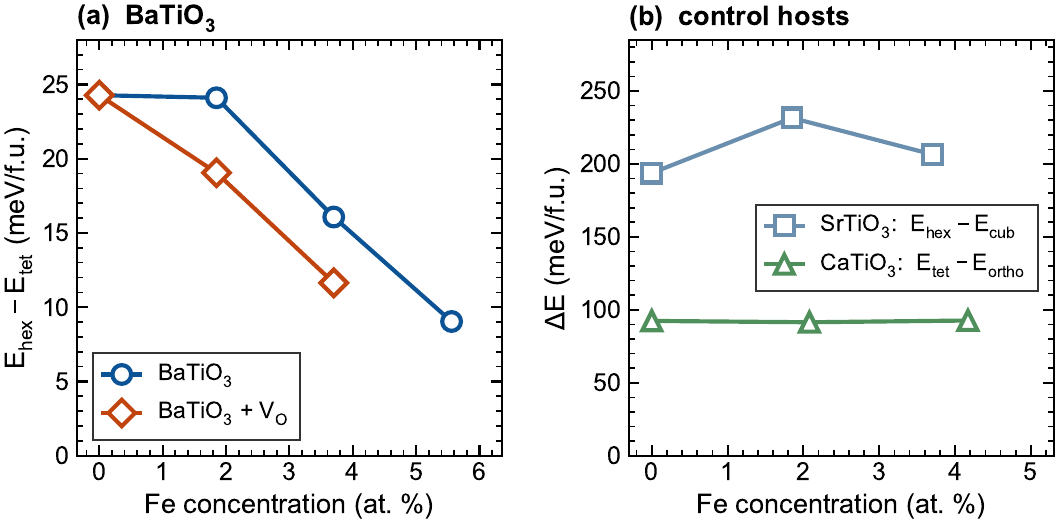}
    \caption{Concentration-dependent phase energy difference $\Delta E = E_{\text{hex}} - E_{\text{tet}}$ per formula unit, from the degeneracy-weighted average of Eq.~(\ref{eq:avg3}). (a) BaTiO$_3$ without and with an oxygen vacancy. (b) Control hosts: SrTiO$_3$ ($E_{\text{hex}} - E_{\text{cub}}$) and CaTiO$_3$ ($E_{\text{tet}} - E_{\text{ortho}}$).}
    \label{fig:total_energy}
\end{figure*}

Our total energy calculations show that Fe substitution steadily destabilizes the tetragonal phase with respect to the hexagonal one. Figure~\ref{fig:total_energy}(a) shows $\Delta E = E_{\text{hex}} - E_{\text{tet}}$ for Fe-substituted BTO, obtained with the degeneracy-weighted average of Eq.~(\ref{eq:avg3}). \label{ln:schemes}With the two other averaging schemes, compared in Secs.~S1 and S3 of the Supplemental Material, $\Delta E$ and the crossover temperature also decrease with Fe content from 1.85\% Fe on. \label{ln:dE}For the pristine host the tetragonal phase is lower by 24.3~meV/f.u. This difference falls to 24.1~meV/f.u.\ at 1.85\% Fe, 16.1~meV/f.u.\ at 3.70\% and 9.0~meV/f.u.\ at 5.56\%, so that Fe reduces it by about two thirds over the concentration range studied.

\begin{figure*}[htbp]
    \centering
    \includegraphics[width=0.9\textwidth]{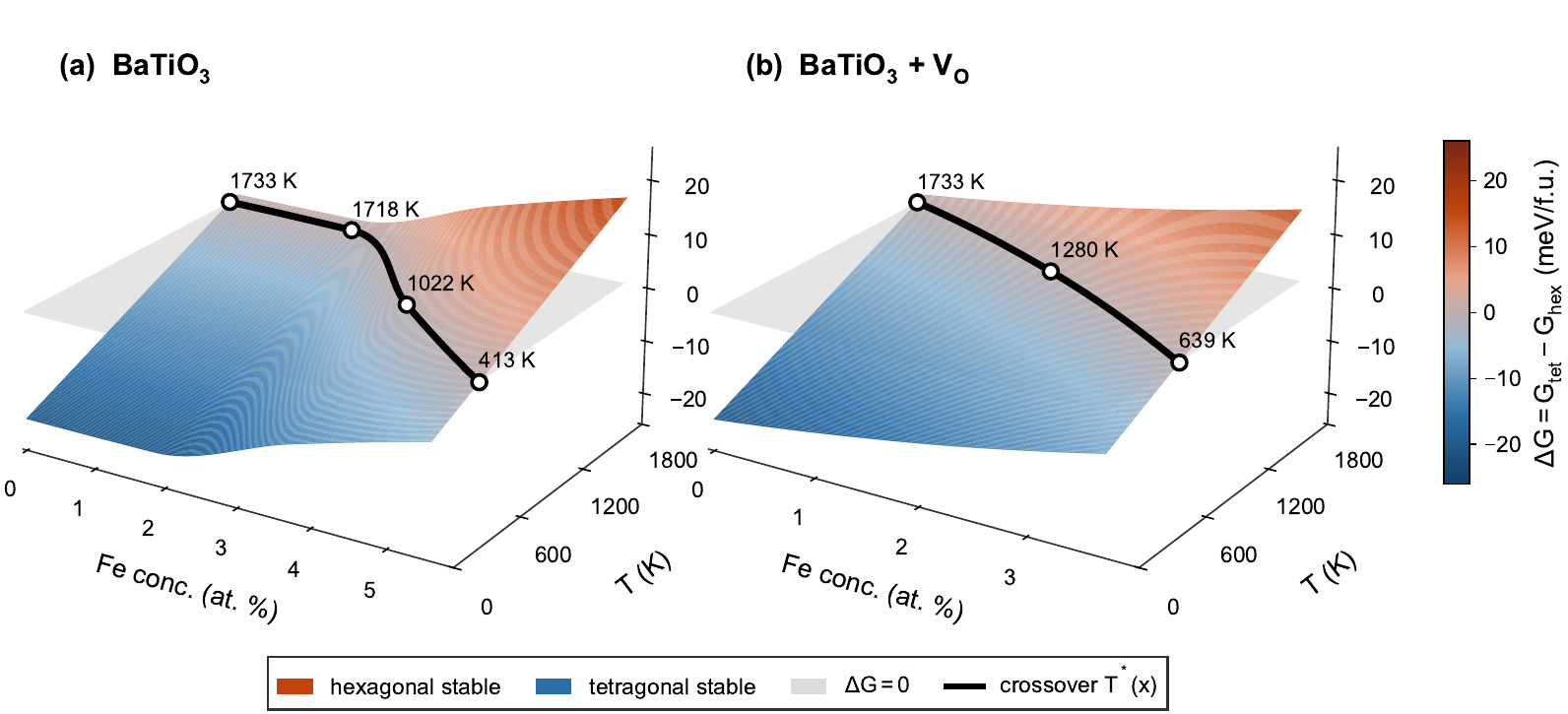}
    \caption{Free energy difference $G_{\text{tet}} - G_{\text{hex}}$, that is $-\Delta G$ of Eq.~(\ref{eq:dG}), as a function of Fe concentration and temperature, for (a) BaTiO$_3$ and (b) BaTiO$_3$ with an oxygen vacancy, which was not computed at 5.56\% Fe (Sec.~\ref{sec:ov}). The grey plane is $\Delta G = 0$; its intersection with the surface, drawn as the black line, is the crossover temperature $T^{*}(x)$, labelled at the computed compositions. Between those compositions the surface is interpolated with PCHIP.}
    \label{fig:free_energy}
\end{figure*}

The tetragonal-to-hexagonal transformation takes place at high temperature, so the two phases have to be compared through their free energies rather than through their total energies at 0~K\cite{ehsan2021first}. Figure~\ref{fig:free_energy} shows $\Delta G(x,T)$ of Eq.~(\ref{eq:dG}); the line along which $\Delta G = 0$ gives the crossover temperature $T^{*}(x)$. Starting from the experimental value of 1733~K for the pristine host, $T^{*}$ drops to 1718, 1022 and 413~K at 1.85, 3.70 and 5.56\% Fe. The same line gives the Fe content above which the hexagonal phase is stable at a given temperature: 2.7\% at the sintering temperature of about 1200~$^{\circ}$C, rising to about 6\% at room temperature. The first value is obtained by piecewise cubic Hermite interpolation (PCHIP)~\cite{fritsch1980monotone} between the computed compositions, and the second is an extrapolation beyond the highest computed composition. Ceramics are furnace-cooled from the sintering temperature and therefore pass through this whole range, which includes the onset of a hexagonal fraction at $x = 0.04$--$0.08$~\cite{tho2024crystal}. In the following section (Sec.~\ref{sec:ov}) we quantify how oxygen vacancies shift the same energy landscape.

\label{ln:control}The same Fe substitution does not alter the phase stability of CaTiO$_3$ or SrTiO$_3$ (Fig.~\ref{fig:total_energy}(b)). In CaTiO$_3$, which exhibits orthorhombic~$\to$~tetragonal~$\to$~cubic phase transitions with temperature, the tetragonal phase lies 92.4~meV/f.u.\ above the orthorhombic ground state, and this value is unchanged by Fe substitution (91.5 and 92.7~meV/f.u.\ at 2.08 and 4.17\% Fe). Fe doping in CaTiO$_3$ therefore shifts these transition temperatures but does not stabilize a different polymorph \cite{becerro2002displacive}. In SrTiO$_3$, where only the cubic phase is observed experimentally, the hypothetical hexagonal phase remains 194--232~meV/f.u.\ higher in energy over the whole concentration range \cite{shafique2021magnetic}. Therefore, only BaTiO$_3$ shows competition between tetragonal and hexagonal phases upon Fe doping.

\subsection{Origins of phase transitions}\label{sec:origins}

As suggested in \cite{pal2020origin}, there are three possible mechanisms for the tetragonal-to-hexagonal transition in Fe-substituted BaTiO$_3$, i.e.,
oxygen vacancies, Jahn--Teller distortions, and changes in the tolerance factor. Here we assess the three quantitatively. \label{ln:mech}Of the three, the vacancy lowers the crossover temperature at both concentrations studied, the JT distortion is large in the tetragonal phase and absent in the hexagonal one, and the tolerance factor does not favour the hexagonal phase.

\subsubsection{Influence of O-vacancies}
\label{sec:ov}
Figure~\ref{fig:total_energy}(a) also shows the energy difference between the tetragonal and hexagonal phases when an oxygen vacancy ($V_{\mathrm{O}}$) is introduced in the first coordination shell of Fe. All symmetry-inequivalent vacancy sites of that shell were enumerated in both phases and averaged in the same way as the substitutional configurations, giving 3 and 68 tetragonal and 3 and 104 hexagonal configurations at 1.85 and 3.70\% Fe. \label{ln:vocount}At 5.56\% Fe the same enumeration gives 1245 and 2247 configurations, too many to compute.

\label{ln:ovres}The vacancy makes the hexagonal phase more favourable at both concentrations studied: $\Delta E$ drops from 24.1 to 19.1~meV/f.u.\ at 1.85\% Fe and from 16.1 to 11.6~meV/f.u.\ at 3.70\%. The crossover temperature shifts accordingly from 1718 to 1280~K and from 1022 to 639~K [Fig.~\ref{fig:free_energy}(b)]. The Fe content above which the hexagonal phase is stable is now 1.1\% at 1200~$^{\circ}$C and about 4.7\% at room temperature (again by extrapolation), compared with 2.7 and 6\% without the vacancy. This range is consistent with the hexagonal fraction found from $x = 0.02$ on in oxygen-deficient BaTi$_{1-x}$Fe$_x$O$_{3-\delta}$~\cite{nguyen2011tetragonal}. Quenching retains the phase assemblage stable at the sintering temperature, which corresponds to the low-concentration end of this range, and quenched samples are reported to contain larger hexagonal fractions \cite{pal2020origin}.

\subsubsection{Role of Jahn-Teller Distortions}
\label{sec:JT}

\label{ln:pjt}The distortion of the FeO$_6$ octahedron depends on how the Fe $3d$ orbitals are occupied~\cite{bersuker2006jahn}. High-spin $d^{5}$ has one electron in each $3d$ orbital and no orbital degeneracy. Fe here is not an isolated $d^{5}$ ion, with $N_d = 5.81$ without and 6.04 with the vacancy (Sec.~\ref{sec:electronic}). A non-degenerate state can also distort, through the pseudo-JT effect~\cite{bersuker2020jahn}. We therefore quantify the local distortion of the FeO$_6$ octahedron in this wider sense, without assuming an orbitally degenerate ground state.

\begin{figure}[htbp]
    \centering
    \includegraphics[width=\linewidth]{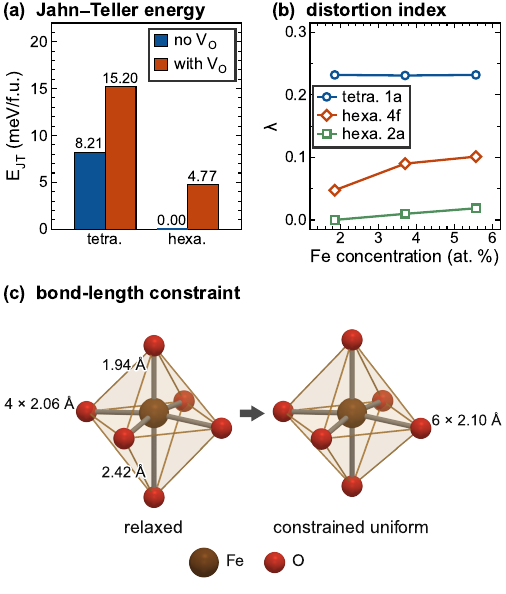}
    \caption{(a) Jahn--Teller (JT) stabilization energy $E_{\mathrm{JT}}$ at 1.85\% Fe, without and with an oxygen vacancy in the first coordination shell of Fe. (b) Distortion index $\lambda$ of the FeO$_6$ octahedron against Fe concentration, for the tetragonal $1a$ site and for the two hexagonal sites, corner-sharing $2a$ and face-sharing $4f$. (c) The tetragonal FeO$_6$ octahedron at 1.85\% Fe, relaxed and with all six bonds constrained to their mean.}
    \label{fig:jt_analysis}
\end{figure}

\label{ln:ejt}To estimate the JT contribution separately from other relaxations, we performed constrained relaxations: (I) from the fully relaxed 1.85\% Fe-substituted cells, take the average Fe--O bond length $\bar d_{\mathrm{Fe\textrm{-}O}}$; (II) build trial structures with all six Fe--O bonds set to $\bar d_{\mathrm{Fe\textrm{-}O}}$; (III) relax again with Fe and its first O shell fixed [Fig.~\ref{fig:jt_analysis}(c)]. The JT stabilization energy is $E_{\mathrm{JT}} = E_{\mathrm{uniform}} - E_{\mathrm{relaxed}}$ per formula unit, that is, the energy gained when the Fe environment relaxes from a uniform octahedron. With six equal bonds Fe is at the centre of its octahedron, so $E_{\mathrm{JT}}$ includes the off-centre displacement of Fe.

\label{ln:ejtval}At 1.85\% Fe, $E_{\mathrm{JT}} = 8.21$~meV/f.u.\ in the tetragonal phase and vanishes in the hexagonal one [Fig.~\ref{fig:jt_analysis}(a)]. In the hexagonal cell Fe can occupy the corner-sharing $2a$ [Ti(1)] site or the face-sharing $4f$ [Ti(2)] site, both observed experimentally~\cite{bottcher2018incorporation}. The value above is for the $2a$ site, which is lower in energy here. With an oxygen vacancy in the first shell Fe becomes fivefold coordinated, and $E_{\mathrm{JT}}$ rises to 15.20 and 4.77~meV/f.u., respectively. \label{ln:vosite}In the tetragonal cell the vacancy is on the apical oxygen nearest Fe, the lowest-energy configuration here and the one also reported in Ref.~\cite{noguchi2020successive}. In the hexagonal cell we take the lowest-energy configuration among those consistent with the Fe--$V_{\mathrm{O}}$ direction fixed by EPR~\cite{bottcher2018incorporation}. In both cells Fe is on a corner-sharing site.

\label{ln:lambda}For fully relaxed structures, the JT distortion index
\begin{equation}
\lambda = \frac{d_{\max}-d_{\min}}{\langle d\rangle}
\end{equation}
with $d_{\max}$ and $d_{\min}$ the extreme Fe--O bond lengths and $\langle d\rangle$ their mean~\cite{rodriguez1993recent,guo2025ultra}, is large in the tetragonal phase and near zero in the hexagonal one [Fig.~\ref{fig:jt_analysis}(b)]. On the tetragonal site $\lambda \approx 0.23$ at all three concentrations (0.232, 0.231 and 0.232 at 1.85, 3.70 and 5.56\% Fe). On the hexagonal $2a$ site the octahedron stays regular (0.000, 0.010 and 0.019), while on the $4f$ site $\lambda$ grows with Fe content, to 0.090 at 3.70\% and 0.101 at 5.56\%.

\label{ln:isym}To verify that the regular octahedron on the $2a$ site is not imposed by the symmetry of the cell, we started from a distorted octahedron ($\lambda = 0.044$, the bond lengths of Ref.~\cite{adeagbo2019theoretical}) and relaxed it with all symmetry operations switched off. The six bonds return to equal length ($\lambda = 0.0009$) and the energy does not drop (Sec.~S5 of the Supplemental Material).

\label{ln:accom}In the tetragonal lattice the octahedron around Fe distorts, whereas in the hexagonal lattice, whose face-sharing connectivity is more compliant, it stays regular~\cite{lufaso2004jahn,adeagbo2019theoretical}. Fe fits the hexagonal framework better, and the JT distortion therefore favours the tetragonal-to-hexagonal transformation.

\subsubsection{Local tolerance factor}
\label{sec:tolerance}

There has been historically conflicting literature regarding the defect chemistry of Fe in BaTiO$_3$, in particular the valence state of Fe. Calculations and spectroscopic characterizations suggest that Fe dopants in BaTiO$_3$ predominantly adopt the high-spin Fe$^{3+}$ oxidation state~\cite{kanagawa2024first,padchasri2021crystal, rajan2017impact,kolodiazhnyi2003analysis,bhide1972mossbauer, ihrig1978phase}, contrasting with the intuitive Fe$^ {4+}$ state. Recent hybrid functional DFT calculations reveal that Fe substitutional defects (Fe$_{\mathrm{Ti}}$) maintain a high-spin configuration ($S=5/2$) across a wide range of Fermi levels in the rhombohedral limit~\cite{kanagawa2024first}. This computational finding is strongly corroborated by experimental characterization. X-ray absorption near-edge structure (XANES) spectroscopy consistently identifies the oxidation state as $+3$ ($>90\%$ purity) in conventionally sintered ceramics~\cite{padchasri2021crystal, rajan2017impact}. Furthermore, Electron Paramagnetic Resonance (EPR) studies unequivocally detect the characteristic $g \approx 2$ signal of high-spin Fe$^{3+}$ centers~\cite{kolodiazhnyi2003analysis}, and M\"ossbauer spectroscopy confirms the isomer shifts typical of high-spin coordination~\cite{bhide1972mossbauer, ihrig1978phase}. 

In addition, it is reported that reduced species (Fe$^{2+}$) can be artificially induced under strongly reducing conditions~\cite{klein1986photorefractive,mazur1997optical}. In addition, in oxygen-deficient regimes, microscopic evidence indicates that oxygen vacancies can locally reduce Fe$^{3+}$ to Fe$^{2+}$, with vacancies clustering around the reduced centers~\cite{chakraborty2013microscopic}. Mixed Fe$^{2+}$/Fe$^{3+}$ valence states have also been reported in reduced samples~\cite{luo2018giant}. Furthermore, both high-spin and low-spin Fe$^{2+}$ are regarded as EPR-silent, rendering EPR evidence insufficient to exclude meaningful Fe$^{2+}$ contributions in realistic synthesis conditions.
Consequently, Fe incorporation, whether predominantly as high-spin Fe$^{3+}$ or with non-negligible Fe$^{2+}$ contributions, invariably reduces the local Goldschmidt tolerance factor relative to pristine BaTiO$_3$. \label{ln:tolradii}We evaluate the local tolerance factor ($t$) using Shannon's ionic radii~\cite{shannon1976revised}, with Ba$^{2+}$ in 12-fold and the $B$-site ions and O$^{2-}$ in 6-fold coordination (Ba$^{2+}$: 1.61~\AA, O$^{2-}$: 1.40~\AA). For dominant high-spin Fe$^{3+}$ ($r = 0.645$~\AA) versus host Ti$^{4+}$ (0.605~\AA),
\begin{equation}
    t_{\mathrm{local}} = \frac{r_{\mathrm{Ba}} + r_{\mathrm{O}}}{\sqrt{2}(r_{\mathrm{Fe}} + r_{\mathrm{O}})} \approx 1.041,
\end{equation}
already a noticeable decrease from $t_{\mathrm{BTO}} \approx 1.062$. For high-spin Fe$^{2+}$ ($r = 0.78$~\AA), $t \approx 0.976$; even for low-spin Fe$^{2+}$ ($r = 0.61$~\AA, though unlikely in oxide environments), $t \approx 1.059$. Any mixture of these valence states therefore yields an effective tolerance factor closer to unity than in undoped BaTiO$_3$. According to the Goldschmidt criterion~\cite{goldschmidt1926gesetze}, $t < 1$ is accommodated by tilting the octahedra, whereas $t > 1$ favors hexagonal stacking with face-sharing octahedra, with more hexagonal layers as $t - 1$ grows~\cite{goodenough2004electronic}. BaTiO$_3$, with $t = 1.06$, adopts the 6H polytype above about 1460~$^{\circ}$C~\cite{ren1994effect}. A larger $B$-site cation brings $t$ towards unity and works against hexagonal stacking~\cite{goodenough2004electronic, langhammer2000crystal}. \label{ln:tolcrit}The criterion was established for corner-sharing perovskites and applies only qualitatively to the hexagonal 6H polytype, which is built from face-sharing Ti$_2$O$_9$ units. By this criterion, the mixed Fe$^{2+}$/Fe$^{3+}$ valence state does not favor hexagonal phases. Experimentally, the hexagonal phase of transition-metal-doped BaTiO$_3$ has been attributed to oxygen vacancies, to Ti$^{3+}$ or Jahn--Teller-active ions, and to metal--metal bonding across the shared faces rather than to the ionic size~\cite{glaister1960investigation, dickinson1961compounds, langhammer2000crystal, bottcher2008phase, keith2004synthesis}. A size argument has been made for Mn$^{4+}$, which is smaller than Ti$^{4+}$ and raises $t$~\cite{ren1994effect}. \label{ln:pal}The tolerance factors of the co-doped samples discussed above~\cite{pal2020origin} are calibrated from the nominal Fe$^{4+}$ content, and Fe$^{4+}$ (0.585~\AA) is smaller than Ti$^{4+}$. Fe$^{3+}$ and Fe$^{2+}$ are larger than Ti$^{4+}$ and lower $t$. \label{ln:tolend}The observed hexagonal stabilization thus occurs in direct contradiction to tolerance-factor predictions.

\label{ln:ctrlmech}For the control hosts the distortion index, $E_{\mathrm{JT}}$ and the local tolerance factor are given and analyzed in Sec.~S9 of the Supplemental Material.

\subsection{Electronic structure}\label{sec:electronic}
\label{ln:esopen}The JT and tolerance-factor arguments above depend on the number of Fe $3d$ electrons, which is set by the way the Fe acceptor is compensated. We examine both in the tetragonal cell without and with an apical $V_{\mathrm{O}}$. Without the vacancy the acceptor leaves a hole in the O $2p$ states, mostly on the six O around Fe, which carry a moment of 0.61~$\mu_B$. With the vacancy this moment falls to 0.17~$\mu_B$ and the hole is filled. The Fe $3d$ shell gains 0.23~$e$ and the Ti across the vacancy 0.18~$e$ (Sec.~S6 of the Supplemental Material). The charge added to Fe goes almost entirely into the spin-down $d_{z^{2}}$ orbital.

\begin{figure*}[htbp]
\includegraphics[width=0.9\textwidth]{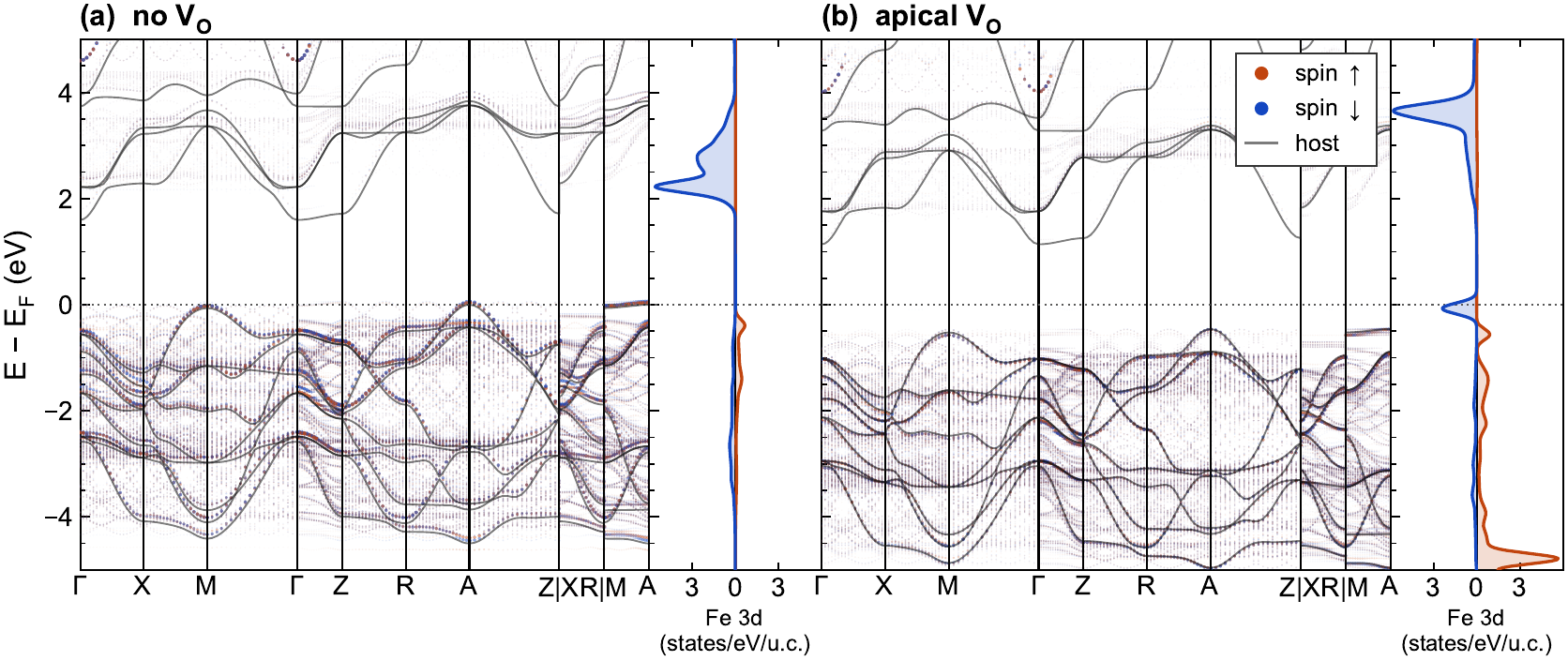}
\caption{Unfolded band structure of tetragonal BaTiO$_3$ with 1.85\% Fe, (a) without and (b) with an apical oxygen vacancy in the first coordination shell of Fe, each with the total Fe $3d$ PDOS in the adjoining side panel. Red and blue markers show spin up and spin down, with opacity giving the spectral weight on the primitive-cell Brillouin zone. Grey lines are the pure BaTiO$_3$ host in the primitive cell, on the same $k$ path and aligned at the VBM.}
\label{fig:bands}
\end{figure*}

\label{ln:bands}Figure~\ref{fig:bands} shows the unfolded band structure of the cell with 1.85\% Fe, without and with the vacancy, each with the Fe $3d$ PDOS in the adjoining side panel. In both cells the unfolded weight follows the bands of the pristine host (grey lines), so 1.85\% Fe remains a dilute perturbation of the tetragonal lattice with or without a vacancy. The weight is smeared around the host lines. States with less than 5\% primitive-cell character account for 20.5\% of the weight without the vacancy and 27.2\% with it. The smearing looks stronger without the vacancy because the spin-up and spin-down bands, red and blue, are split in Fig.~\ref{fig:bands}(a) and coincide in Fig.~\ref{fig:bands}(b).

\label{ln:ef}Without the vacancy $E_{\mathrm{F}}$ lies at the valence-band maximum (VBM), at $-0.001$~eV [Fig.~\ref{fig:bands}(a)], so the hole of the Fe acceptor stays in the host valence band. The Fe $3d$ weight below $E_{\mathrm{F}}$ is spin-up and peaks at $-0.4$~eV. With the vacancy the VBM lies $0.466$~eV below $E_{\mathrm{F}}$ [Fig.~\ref{fig:bands}(b)], and a spin-down Fe $3d$ peak lies in the gap at $-0.08$~eV, just below $E_{\mathrm{F}}$. This peak has no visible counterpart in the band panel; the Fe $3d$ fat band of Sec.~S6 of the Supplemental Material shows the corresponding state. The compensating charge is therefore on Fe and in a single spin channel.

\begin{figure}[htbp]
\includegraphics[width=\columnwidth]{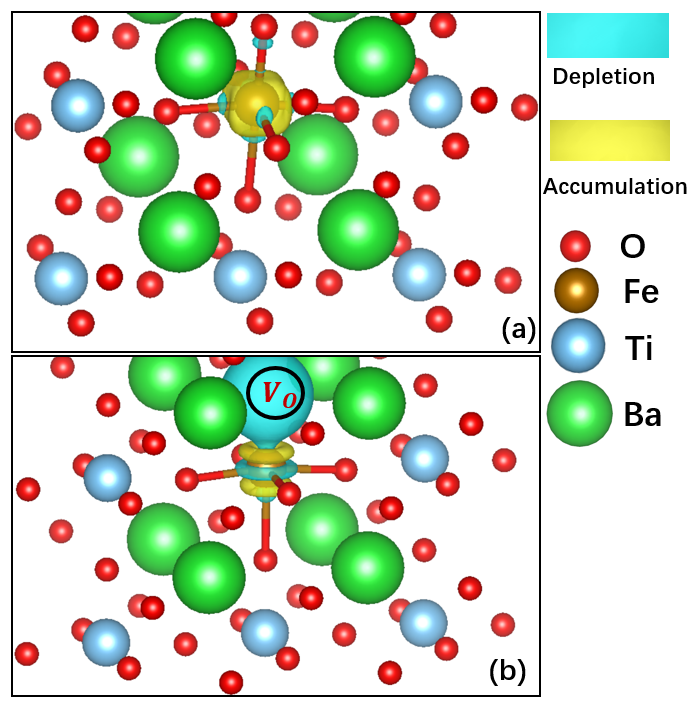}
\caption{Charge-density difference (CDD) isosurfaces ($\pm 0.02~e/\text{\AA}^{3}$) under frozen-ion conditions. (a) Effect of Fe substitution: $\Delta\rho = \rho(\text{Fe}_\text{Ti}) - \rho(\text{pristine})$. (b) Effect of the oxygen vacancy: $\Delta\rho = \rho(\text{Fe}_\text{Ti}{+}V_\text{O}) - \rho(\text{Fe}_\text{Ti})$. Yellow and cyan isosurfaces indicate electron accumulation and depletion, respectively. Atom colors: Ba (green), Ti (blue), Fe (brown), O (red). The vacancy site is circled in (b).}
\label{fig:cdd}
\end{figure}

\label{ln:cddtext}Figure~\ref{fig:cdd} shows the charge-density difference, computed with frozen ions. Fe substitution alone produces weak accumulation on Fe and depletion on the neighbouring O ligands, consistent with Fe--O hybridization~\cite{islam2019effect} [Fig.~\ref{fig:cdd}(a)]. Adding the vacancy produces an accumulation lobe on Fe along the Fe--$V_{\mathrm{O}}$ axis with the shape of $d_{z^{2}}$, and a depletion region at the vacancy site [Fig.~\ref{fig:cdd}(b)]. The compensation electrons stay localized near the vacancy~\cite{kotiuga2019carrier}.

The Fe $3d$ occupation matrices are given below in the basis $(d_{xy}, d_{yz}, d_{z^{2}}, d_{xz}, d_{x^{2}-y^{2}})$, first under frozen-ion conditions and then for the relaxed structures. For the substituted cell the spin-up occupation reads
\begin{equation*}
\begin{pmatrix}
0.966 & 0 & 0 & 0 & 0 \\
0 & 0.970 & 0 & 0 & 0 \\
0 & 0 & 0.919 & 0 & 0 \\
0 & 0 & 0 & 0.970 & 0 \\
0 & 0 & 0 & 0 & 0.983
\end{pmatrix},
\end{equation*}
and the spin-down occupation
\begin{equation*}
\begin{pmatrix}
0.101 & 0 & 0 & 0 & 0 \\
0 & 0.127 & 0 & 0 & 0 \\
0 & 0 & 0.275 & 0 & 0 \\
0 & 0 & 0 & 0.127 & 0 \\
0 & 0 & 0 & 0 & 0.253
\end{pmatrix}.
\end{equation*}
With the vacancy these become
\begin{gather*}
\begin{pmatrix}
0.963 & 0 & 0 & 0 & 0 \\
0 & 0.955 & 0 & 0 & 0 \\
0 & 0 & 0.939 & 0 & 0 \\
0 & 0 & 0 & 0.955 & 0 \\
0 & 0 & 0 & 0 & 1.004
\end{pmatrix},\\
\begin{pmatrix}
0.075 & 0 & 0 & 0 & 0 \\
0 & 0.052 & 0 & 0 & 0 \\
0 & 0 & 0.586 & 0 & 0 \\
0 & 0 & 0 & 0.052 & 0 \\
0 & 0 & 0 & 0 & 0.208
\end{pmatrix}.
\end{gather*}
The total occupation rises from $N_{d} = 5.69$ to $5.79$. The gain of $0.10\,e^{-}$ is almost all in the spin-down channel, where $d_{z^{2}}$ takes $+0.312\,e^{-}$ and no other orbital more than $0.08\,e^{-}$.

\label{ln:relaxocc}Relaxing the ions does not change this. For the substituted cell the spin-up and spin-down occupations are
\begin{gather*}
\begin{pmatrix}
0.975 & 0 & 0 & 0 & 0 \\
0 & 0.977 & 0 & 0 & 0 \\
0 & 0 & 1.024 & 0 & 0 \\
0 & 0 & 0 & 0.977 & 0 \\
0 & 0 & 0 & 0 & 1.034
\end{pmatrix},\\
\begin{pmatrix}
0.090 & 0 & 0 & 0 & 0 \\
0 & 0.112 & 0 & 0 & 0 \\
0 & 0 & 0.258 & 0 & 0 \\
0 & 0 & 0 & 0.112 & 0 \\
0 & 0 & 0 & 0 & 0.251
\end{pmatrix},
\end{gather*}
and with the vacancy
\begin{gather*}
\begin{pmatrix}
0.971 & 0 & 0 & 0 & 0 \\
0 & 0.963 & 0 & 0 & 0 \\
0 & 0 & 0.949 & 0 & 0 \\
0 & 0 & 0 & 0.963 & 0 \\
0 & 0 & 0 & 0 & 1.032
\end{pmatrix},\\
\begin{pmatrix}
0.058 & 0 & 0 & 0 & 0 \\
0 & 0.035 & 0 & 0 & 0 \\
0 & 0 & 0.847 & 0 & 0 \\
0 & 0 & 0 & 0.035 & 0 \\
0 & 0 & 0 & 0 & 0.183
\end{pmatrix}.
\end{gather*}
\begin{figure*}[htbp]
\includegraphics[width=\textwidth]{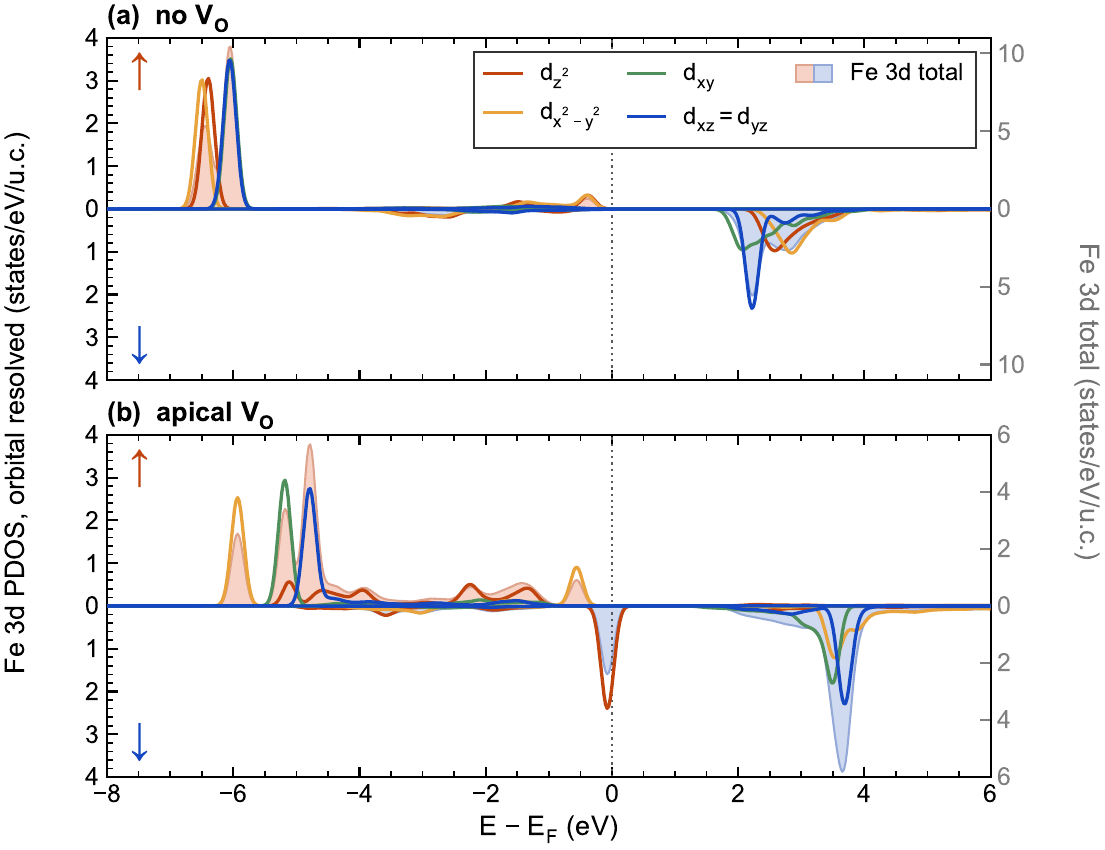}
\caption{Spin-resolved Fe $3d$ projected density of states of tetragonal BaTiO$_3$ with 1.85\% Fe, (a) without and (b) with an apical oxygen vacancy, resolved into the five $3d$ orbitals and broadened to a common width of $0.10$~eV. $d_{xz}$ and $d_{yz}$ are degenerate and drawn once. $E_{\mathrm{F}}$ is at zero.}
\label{fig:pdos}
\end{figure*}

\label{ln:hs3c}In both cells each of the five spin-up orbitals holds about one electron, $4.987$ in total without the vacancy and $4.879$ with it, and the $3d$ moments from the occupation matrices are $4.16$ and $3.72\,\mu_{\mathrm{B}}$. The spin-resolved Fe $3d$ PDOS shows the same [Fig.~\ref{fig:pdos}(a) and \ref{fig:pdos}(b)]. The spin-up shell is thus half filled in both cells.

\label{ln:occ}With the vacancy the spin-down $d_{z^{2}}$ rises from $0.258$ to $0.847$, a gain of $0.59\,e^{-}$ [Fig.~\ref{fig:pdos}(b)], while the other four spin-down orbitals each fall slightly. $d_{z^{2}}$ points at the missing apical oxygen, and its occupied spin-down component is the in-gap level of Fig.~\ref{fig:bands}(b). Spin-summed, $d_{z^{2}}$ rises from $1.282$ to $1.796$ and $N_{d}$ from $5.81$ to $6.04$. The added charge therefore occupies a single orbital. Many-body calculations for Fe in tetragonal BaTiO$_3$ find a mixture of $d^{5}$ and $d^{6}$ configurations, shifted towards $d^{6}$ by a neighbouring vacancy~\cite{li2026redox}. X-ray absorption spectroscopy on oxygen-deficient samples could test this.

\section{Conclusion}\label{sec:conclusion}
We perform systematic first-principles investigations on the tetragonal-to-hexagonal phase transitions in Fe-substituted BaTiO$_3$. Free energy calculations on 270-atom supercells establish a thermodynamic crossover at 2.7--6\% Fe concentration in BaTiO$_3$ on cooling from the sintering temperature, consistent with previous experiments, while neither CaTiO$_3$ nor SrTiO$_3$ exhibits similar behavior. Three possible mechanisms are quantified. It is observed that oxygen vacancies shift the crossover concentration from $2.7$--$6\%$ to $1.1$--$4.7\%$. \label{ln:concl}Moreover, the Jahn-Teller stabilization energy is larger in the tetragonal phase (8.21~meV/f.u.) than in the hexagonal one (0~meV/f.u.), with the distortion index $\lambda$ consistently higher in the tetragonal host across all Fe concentrations, so that Fe fits the hexagonal framework better. \label{ln:concltol}However, local tolerance-factor analysis demonstrates that Fe incorporation does not favor the occurrence of the hexagonal phase.

Orbital-resolved analysis of the Fe~$3d$ occupation matrix reveals an orbital-selective charge redistribution caused by the $V_\text{O}$ formation. The net gain of $0.23\,e$ on the Fe atoms is in the spin-down channel, where the number of electrons in the $d_{z^2}$ orbital is increased by 0.59 by depleting the occupation of the other spin-down orbitals. Spectroscopy on oxygen-deficient samples could test this change in the Fe $3d$ occupation.

\begin{acknowledgments}
The authors wish to acknowledge the financial support of the Deutsche Forschungsgemeinschaft - Project-ID 463184206 - SFB 1548. The authors also gratefully acknowledge the computing time provided to them on the high-performance computer Lichtenberg at the NHR Centers NHR4CES at TU Darmstadt. (Project ID: p0026451) This is funded by the Federal Ministry of Education and Research, and the state governments participating on the basis of the resolutions of the GWK for national high performance computing at universities (www.nhr-verein.de/unsere-partner)
\end{acknowledgments}

\section*{Data availability}
\label{ln:data}All structures, input files and total energies generated in this work are
openly available in Ref.~\cite{li2026dataset}.

\bibliography{FeBTOtet2hex}%

\begin{thebibliography}{84}%
\makeatletter
\providecommand \@ifxundefined [1]{%
 \@ifx{#1\undefined}
}%
\providecommand \@ifnum [1]{%
 \ifnum #1\expandafter \@firstoftwo
 \else \expandafter \@secondoftwo
 \fi
}%
\providecommand \@ifx [1]{%
 \ifx #1\expandafter \@firstoftwo
 \else \expandafter \@secondoftwo
 \fi
}%
\providecommand \natexlab [1]{#1}%
\providecommand \enquote  [1]{``#1''}%
\providecommand \bibnamefont  [1]{#1}%
\providecommand \bibfnamefont [1]{#1}%
\providecommand \citenamefont [1]{#1}%
\providecommand \href@noop [0]{\@secondoftwo}%
\providecommand \href [0]{\begingroup \@sanitize@url \@href}%
\providecommand \@href[1]{\@@startlink{#1}\@@href}%
\providecommand \@@href[1]{\endgroup#1\@@endlink}%
\providecommand \@sanitize@url [0]{\catcode `\\12\catcode `\$12\catcode
  `\&12\catcode `\#12\catcode `\^12\catcode `\_12\catcode `\%12\relax}%
\providecommand \@@startlink[1]{}%
\providecommand \@@endlink[0]{}%
\providecommand \url  [0]{\begingroup\@sanitize@url \@url }%
\providecommand \@url [1]{\endgroup\@href {#1}{\urlprefix }}%
\providecommand \urlprefix  [0]{URL }%
\providecommand \Eprint [0]{\href }%
\providecommand \doibase [0]{https://doi.org/}%
\providecommand \selectlanguage [0]{\@gobble}%
\providecommand \bibinfo  [0]{\@secondoftwo}%
\providecommand \bibfield  [0]{\@secondoftwo}%
\providecommand \translation [1]{[#1]}%
\providecommand \BibitemOpen [0]{}%
\providecommand \bibitemStop [0]{}%
\providecommand \bibitemNoStop [0]{.\EOS\space}%
\providecommand \EOS [0]{\spacefactor3000\relax}%
\providecommand \BibitemShut  [1]{\csname bibitem#1\endcsname}%
\let\auto@bib@innerbib\@empty
\bibitem [{\citenamefont {Catalan}\ and\ \citenamefont
  {Scott}(2009)}]{catalan2009physics}%
  \BibitemOpen
  \bibfield  {author} {\bibinfo {author} {\bibfnamefont {G.}~\bibnamefont
  {Catalan}}\ and\ \bibinfo {author} {\bibfnamefont {J.~F.}\ \bibnamefont
  {Scott}},\ }\bibfield  {title} {\bibinfo {title} {Physics and applications of
  bismuth ferrite},\ }\href@noop {} {\bibfield  {journal} {\bibinfo  {journal}
  {Advanced Materials}\ }\textbf {\bibinfo {volume} {21}},\ \bibinfo {pages}
  {2463} (\bibinfo {year} {2009})}\BibitemShut {NoStop}%
\bibitem [{\citenamefont {Spaldin}\ and\ \citenamefont
  {Ramesh}(2019)}]{spaldin2019advances}%
  \BibitemOpen
  \bibfield  {author} {\bibinfo {author} {\bibfnamefont {N.~A.}\ \bibnamefont
  {Spaldin}}\ and\ \bibinfo {author} {\bibfnamefont {R.}~\bibnamefont
  {Ramesh}},\ }\bibfield  {title} {\bibinfo {title} {Advances in
  magnetoelectric multiferroics},\ }\href@noop {} {\bibfield  {journal}
  {\bibinfo  {journal} {Nature Materials}\ }\textbf {\bibinfo {volume} {18}},\
  \bibinfo {pages} {203} (\bibinfo {year} {2019})}\BibitemShut {NoStop}%
\bibitem [{\citenamefont {Lines}\ and\ \citenamefont
  {Glass}(2001)}]{lines2001principles}%
  \BibitemOpen
  \bibfield  {author} {\bibinfo {author} {\bibfnamefont {M.~E.}\ \bibnamefont
  {Lines}}\ and\ \bibinfo {author} {\bibfnamefont {A.~M.}\ \bibnamefont
  {Glass}},\ }\href@noop {} {\emph {\bibinfo {title} {Principles and
  applications of ferroelectrics and related materials}}}\ (\bibinfo
  {publisher} {Oxford University Press},\ \bibinfo {year} {2001})\BibitemShut
  {NoStop}%
\bibitem [{\citenamefont {Rabe}\ \emph {et~al.}(2007)\citenamefont {Rabe},
  \citenamefont {Dawber}, \citenamefont {Lichtensteiger}, \citenamefont {Ahn},\
  and\ \citenamefont {Triscone}}]{rabe2007modern}%
  \BibitemOpen
  \bibfield  {author} {\bibinfo {author} {\bibfnamefont {K.~M.}\ \bibnamefont
  {Rabe}}, \bibinfo {author} {\bibfnamefont {M.}~\bibnamefont {Dawber}},
  \bibinfo {author} {\bibfnamefont {C.}~\bibnamefont {Lichtensteiger}},
  \bibinfo {author} {\bibfnamefont {C.~H.}\ \bibnamefont {Ahn}},\ and\ \bibinfo
  {author} {\bibfnamefont {J.-M.}\ \bibnamefont {Triscone}},\ }\bibfield
  {title} {\bibinfo {title} {Modern physics of ferroelectrics: Essential
  background},\ }in\ \href@noop {} {\emph {\bibinfo {booktitle} {Physics of
  Ferroelectrics: A Modern Perspective}}}\ (\bibinfo  {publisher} {Springer},\
  \bibinfo {year} {2007})\ pp.\ \bibinfo {pages} {1--30}\BibitemShut {NoStop}%
\bibitem [{\citenamefont {Setter}\ \emph {et~al.}(2006)\citenamefont {Setter},
  \citenamefont {Damjanovic}, \citenamefont {Eng}, \citenamefont {Fox},
  \citenamefont {Gevorgian}, \citenamefont {Hong}, \citenamefont {Kingon},
  \citenamefont {Kohlstedt}, \citenamefont {Park}, \citenamefont {Stephenson}
  \emph {et~al.}}]{setter2006ferroelectric}%
  \BibitemOpen
  \bibfield  {author} {\bibinfo {author} {\bibfnamefont {N.}~\bibnamefont
  {Setter}}, \bibinfo {author} {\bibfnamefont {D.}~\bibnamefont {Damjanovic}},
  \bibinfo {author} {\bibfnamefont {L.}~\bibnamefont {Eng}}, \bibinfo {author}
  {\bibfnamefont {G.}~\bibnamefont {Fox}}, \bibinfo {author} {\bibfnamefont
  {S.}~\bibnamefont {Gevorgian}}, \bibinfo {author} {\bibfnamefont
  {S.}~\bibnamefont {Hong}}, \bibinfo {author} {\bibfnamefont {A.}~\bibnamefont
  {Kingon}}, \bibinfo {author} {\bibfnamefont {H.}~\bibnamefont {Kohlstedt}},
  \bibinfo {author} {\bibfnamefont {N.~Y.}\ \bibnamefont {Park}}, \bibinfo
  {author} {\bibfnamefont {G.~B.}\ \bibnamefont {Stephenson}}, \emph {et~al.},\
  }\bibfield  {title} {\bibinfo {title} {Ferroelectric thin films: Review of
  materials, properties, and applications},\ }\href@noop {} {\bibfield
  {journal} {\bibinfo  {journal} {Journal of Applied Physics}\ }\textbf
  {\bibinfo {volume} {100}},\ \bibinfo {pages} {051606} (\bibinfo {year}
  {2006})}\BibitemShut {NoStop}%
\bibitem [{\citenamefont {Kirby}\ and\ \citenamefont
  {Wechsler}(1991)}]{kirby1991phase}%
  \BibitemOpen
  \bibfield  {author} {\bibinfo {author} {\bibfnamefont {K.~W.}\ \bibnamefont
  {Kirby}}\ and\ \bibinfo {author} {\bibfnamefont {B.~A.}\ \bibnamefont
  {Wechsler}},\ }\bibfield  {title} {\bibinfo {title} {Phase relations in the
  barium titanate—titanium oxide system},\ }\href@noop {} {\bibfield
  {journal} {\bibinfo  {journal} {Journal of the American Ceramic Society}\
  }\textbf {\bibinfo {volume} {74}},\ \bibinfo {pages} {1841} (\bibinfo {year}
  {1991})}\BibitemShut {NoStop}%
\bibitem [{\citenamefont {Liu}\ and\ \citenamefont {Ren}(2009)}]{liu2009large}%
  \BibitemOpen
  \bibfield  {author} {\bibinfo {author} {\bibfnamefont {W.}~\bibnamefont
  {Liu}}\ and\ \bibinfo {author} {\bibfnamefont {X.}~\bibnamefont {Ren}},\
  }\bibfield  {title} {\bibinfo {title} {Large piezoelectric effect in
  {Pb}-free ceramics},\ }\href@noop {} {\bibfield  {journal} {\bibinfo
  {journal} {Physical Review Letters}\ }\textbf {\bibinfo {volume} {103}},\
  \bibinfo {pages} {257602} (\bibinfo {year} {2009})}\BibitemShut {NoStop}%
\bibitem [{\citenamefont {Abel}\ \emph {et~al.}(2019)\citenamefont {Abel},
  \citenamefont {Eltes}, \citenamefont {Ortmann}, \citenamefont {Messner},
  \citenamefont {Castera}, \citenamefont {Wagner}, \citenamefont {Urbonas},
  \citenamefont {Rosa}, \citenamefont {Gutierrez}, \citenamefont {Tulli} \emph
  {et~al.}}]{abel2019large}%
  \BibitemOpen
  \bibfield  {author} {\bibinfo {author} {\bibfnamefont {S.}~\bibnamefont
  {Abel}}, \bibinfo {author} {\bibfnamefont {F.}~\bibnamefont {Eltes}},
  \bibinfo {author} {\bibfnamefont {J.~E.}\ \bibnamefont {Ortmann}}, \bibinfo
  {author} {\bibfnamefont {A.}~\bibnamefont {Messner}}, \bibinfo {author}
  {\bibfnamefont {P.}~\bibnamefont {Castera}}, \bibinfo {author} {\bibfnamefont
  {T.}~\bibnamefont {Wagner}}, \bibinfo {author} {\bibfnamefont
  {D.}~\bibnamefont {Urbonas}}, \bibinfo {author} {\bibfnamefont
  {A.}~\bibnamefont {Rosa}}, \bibinfo {author} {\bibfnamefont {A.~M.}\
  \bibnamefont {Gutierrez}}, \bibinfo {author} {\bibfnamefont {D.}~\bibnamefont
  {Tulli}}, \emph {et~al.},\ }\bibfield  {title} {\bibinfo {title} {Large
  {Pockels} effect in micro-and nanostructured barium titanate integrated on
  silicon},\ }\href@noop {} {\bibfield  {journal} {\bibinfo  {journal} {Nature
  Materials}\ }\textbf {\bibinfo {volume} {18}},\ \bibinfo {pages} {42}
  (\bibinfo {year} {2019})}\BibitemShut {NoStop}%
\bibitem [{\citenamefont {Lee}\ \emph {et~al.}(2010)\citenamefont {Lee},
  \citenamefont {Fang}, \citenamefont {Vlahos}, \citenamefont {Ke},
  \citenamefont {Jung}, \citenamefont {Kourkoutis}, \citenamefont {Kim},
  \citenamefont {Ryan}, \citenamefont {Heeg}, \citenamefont {Roeckerath} \emph
  {et~al.}}]{lee2010strong}%
  \BibitemOpen
  \bibfield  {author} {\bibinfo {author} {\bibfnamefont {J.~H.}\ \bibnamefont
  {Lee}}, \bibinfo {author} {\bibfnamefont {L.}~\bibnamefont {Fang}}, \bibinfo
  {author} {\bibfnamefont {E.}~\bibnamefont {Vlahos}}, \bibinfo {author}
  {\bibfnamefont {X.}~\bibnamefont {Ke}}, \bibinfo {author} {\bibfnamefont
  {Y.~W.}\ \bibnamefont {Jung}}, \bibinfo {author} {\bibfnamefont {L.~F.}\
  \bibnamefont {Kourkoutis}}, \bibinfo {author} {\bibfnamefont {J.-W.}\
  \bibnamefont {Kim}}, \bibinfo {author} {\bibfnamefont {P.~J.}\ \bibnamefont
  {Ryan}}, \bibinfo {author} {\bibfnamefont {T.}~\bibnamefont {Heeg}}, \bibinfo
  {author} {\bibfnamefont {M.}~\bibnamefont {Roeckerath}}, \emph {et~al.},\
  }\bibfield  {title} {\bibinfo {title} {A strong ferroelectric ferromagnet
  created by means of spin--lattice coupling},\ }\href@noop {} {\bibfield
  {journal} {\bibinfo  {journal} {Nature}\ }\textbf {\bibinfo {volume} {466}},\
  \bibinfo {pages} {954} (\bibinfo {year} {2010})}\BibitemShut {NoStop}%
\bibitem [{\citenamefont {Bousquet}\ \emph {et~al.}(2008)\citenamefont
  {Bousquet}, \citenamefont {Dawber}, \citenamefont {Stucki}, \citenamefont
  {Lichtensteiger}, \citenamefont {Hermet}, \citenamefont {Gariglio},
  \citenamefont {Triscone},\ and\ \citenamefont
  {Ghosez}}]{bousquet2008improper}%
  \BibitemOpen
  \bibfield  {author} {\bibinfo {author} {\bibfnamefont {E.}~\bibnamefont
  {Bousquet}}, \bibinfo {author} {\bibfnamefont {M.}~\bibnamefont {Dawber}},
  \bibinfo {author} {\bibfnamefont {N.}~\bibnamefont {Stucki}}, \bibinfo
  {author} {\bibfnamefont {C.}~\bibnamefont {Lichtensteiger}}, \bibinfo
  {author} {\bibfnamefont {P.}~\bibnamefont {Hermet}}, \bibinfo {author}
  {\bibfnamefont {S.}~\bibnamefont {Gariglio}}, \bibinfo {author}
  {\bibfnamefont {J.-M.}\ \bibnamefont {Triscone}},\ and\ \bibinfo {author}
  {\bibfnamefont {P.}~\bibnamefont {Ghosez}},\ }\bibfield  {title} {\bibinfo
  {title} {Improper ferroelectricity in perovskite oxide artificial
  superlattices},\ }\href@noop {} {\bibfield  {journal} {\bibinfo  {journal}
  {Nature}\ }\textbf {\bibinfo {volume} {452}},\ \bibinfo {pages} {732}
  (\bibinfo {year} {2008})}\BibitemShut {NoStop}%
\bibitem [{\citenamefont {Choi}\ \emph {et~al.}(2004)\citenamefont {Choi},
  \citenamefont {Biegalski}, \citenamefont {Li}, \citenamefont {Sharan},
  \citenamefont {Schubert}, \citenamefont {Uecker}, \citenamefont {Reiche},
  \citenamefont {Chen}, \citenamefont {Pan}, \citenamefont {Gopalan} \emph
  {et~al.}}]{choi2004enhancement}%
  \BibitemOpen
  \bibfield  {author} {\bibinfo {author} {\bibfnamefont {K.~J.}\ \bibnamefont
  {Choi}}, \bibinfo {author} {\bibfnamefont {M.}~\bibnamefont {Biegalski}},
  \bibinfo {author} {\bibfnamefont {Y.~L.}\ \bibnamefont {Li}}, \bibinfo
  {author} {\bibfnamefont {A.}~\bibnamefont {Sharan}}, \bibinfo {author}
  {\bibfnamefont {J.}~\bibnamefont {Schubert}}, \bibinfo {author}
  {\bibfnamefont {R.}~\bibnamefont {Uecker}}, \bibinfo {author} {\bibfnamefont
  {P.}~\bibnamefont {Reiche}}, \bibinfo {author} {\bibfnamefont {Y.~B.}\
  \bibnamefont {Chen}}, \bibinfo {author} {\bibfnamefont {X.~Q.}\ \bibnamefont
  {Pan}}, \bibinfo {author} {\bibfnamefont {V.}~\bibnamefont {Gopalan}}, \emph
  {et~al.},\ }\bibfield  {title} {\bibinfo {title} {Enhancement of
  ferroelectricity in strained {BaTiO$_3$} thin films},\ }\href@noop {}
  {\bibfield  {journal} {\bibinfo  {journal} {Science}\ }\textbf {\bibinfo
  {volume} {306}},\ \bibinfo {pages} {1005} (\bibinfo {year}
  {2004})}\BibitemShut {NoStop}%
\bibitem [{\citenamefont {Nova}\ \emph {et~al.}(2017)\citenamefont {Nova},
  \citenamefont {Cartella}, \citenamefont {Cantaluppi}, \citenamefont
  {F{\"o}rst}, \citenamefont {Bossini}, \citenamefont {Mikhaylovskiy},
  \citenamefont {Kimel}, \citenamefont {Merlin},\ and\ \citenamefont
  {Cavalleri}}]{nova2017effective}%
  \BibitemOpen
  \bibfield  {author} {\bibinfo {author} {\bibfnamefont {T.~F.}\ \bibnamefont
  {Nova}}, \bibinfo {author} {\bibfnamefont {A.}~\bibnamefont {Cartella}},
  \bibinfo {author} {\bibfnamefont {A.}~\bibnamefont {Cantaluppi}}, \bibinfo
  {author} {\bibfnamefont {M.}~\bibnamefont {F{\"o}rst}}, \bibinfo {author}
  {\bibfnamefont {D.}~\bibnamefont {Bossini}}, \bibinfo {author} {\bibfnamefont
  {R.~V.}\ \bibnamefont {Mikhaylovskiy}}, \bibinfo {author} {\bibfnamefont
  {A.~V.}\ \bibnamefont {Kimel}}, \bibinfo {author} {\bibfnamefont
  {R.}~\bibnamefont {Merlin}},\ and\ \bibinfo {author} {\bibfnamefont
  {A.}~\bibnamefont {Cavalleri}},\ }\bibfield  {title} {\bibinfo {title} {An
  effective magnetic field from optically driven phonons},\ }\href@noop {}
  {\bibfield  {journal} {\bibinfo  {journal} {Nature Physics}\ }\textbf
  {\bibinfo {volume} {13}},\ \bibinfo {pages} {132} (\bibinfo {year}
  {2017})}\BibitemShut {NoStop}%
\bibitem [{\citenamefont {Li}\ \emph {et~al.}(2019)\citenamefont {Li},
  \citenamefont {Qiu}, \citenamefont {Zhang}, \citenamefont {Baldini},
  \citenamefont {Lu}, \citenamefont {Rappe},\ and\ \citenamefont
  {Nelson}}]{li2019terahertz}%
  \BibitemOpen
  \bibfield  {author} {\bibinfo {author} {\bibfnamefont {X.}~\bibnamefont
  {Li}}, \bibinfo {author} {\bibfnamefont {T.}~\bibnamefont {Qiu}}, \bibinfo
  {author} {\bibfnamefont {J.}~\bibnamefont {Zhang}}, \bibinfo {author}
  {\bibfnamefont {E.}~\bibnamefont {Baldini}}, \bibinfo {author} {\bibfnamefont
  {J.}~\bibnamefont {Lu}}, \bibinfo {author} {\bibfnamefont {A.~M.}\
  \bibnamefont {Rappe}},\ and\ \bibinfo {author} {\bibfnamefont {K.~A.}\
  \bibnamefont {Nelson}},\ }\bibfield  {title} {\bibinfo {title} {Terahertz
  field--induced ferroelectricity in quantum paraelectric {SrTiO$_3$}},\
  }\href@noop {} {\bibfield  {journal} {\bibinfo  {journal} {Science}\ }\textbf
  {\bibinfo {volume} {364}},\ \bibinfo {pages} {1079} (\bibinfo {year}
  {2019})}\BibitemShut {NoStop}%
\bibitem [{\citenamefont {Morrison}\ \emph {et~al.}(1999)\citenamefont
  {Morrison}, \citenamefont {Sinclair},\ and\ \citenamefont
  {West}}]{morrison1999electrical}%
  \BibitemOpen
  \bibfield  {author} {\bibinfo {author} {\bibfnamefont {F.~D.}\ \bibnamefont
  {Morrison}}, \bibinfo {author} {\bibfnamefont {D.~C.}\ \bibnamefont
  {Sinclair}},\ and\ \bibinfo {author} {\bibfnamefont {A.~R.}\ \bibnamefont
  {West}},\ }\bibfield  {title} {\bibinfo {title} {Electrical and structural
  characteristics of lanthanum-doped barium titanate ceramics},\ }\href@noop {}
  {\bibfield  {journal} {\bibinfo  {journal} {Journal of Applied Physics}\
  }\textbf {\bibinfo {volume} {86}},\ \bibinfo {pages} {6355} (\bibinfo {year}
  {1999})}\BibitemShut {NoStop}%
\bibitem [{\citenamefont {Choi}\ \emph {et~al.}(2012)\citenamefont {Choi},
  \citenamefont {Chisholm}, \citenamefont {Singh}, \citenamefont {Choi},
  \citenamefont {Jellison},\ and\ \citenamefont {Lee}}]{choi2012wide}%
  \BibitemOpen
  \bibfield  {author} {\bibinfo {author} {\bibfnamefont {W.~S.}\ \bibnamefont
  {Choi}}, \bibinfo {author} {\bibfnamefont {M.~F.}\ \bibnamefont {Chisholm}},
  \bibinfo {author} {\bibfnamefont {D.~J.}\ \bibnamefont {Singh}}, \bibinfo
  {author} {\bibfnamefont {T.}~\bibnamefont {Choi}}, \bibinfo {author}
  {\bibfnamefont {G.~E.}\ \bibnamefont {Jellison}, \bibfnamefont {Jr.}},\ and\
  \bibinfo {author} {\bibfnamefont {H.~N.}\ \bibnamefont {Lee}},\ }\bibfield
  {title} {\bibinfo {title} {Wide bandgap tunability in complex transition
  metal oxides by site-specific substitution},\ }\href@noop {} {\bibfield
  {journal} {\bibinfo  {journal} {Nature Communications}\ }\textbf {\bibinfo
  {volume} {3}},\ \bibinfo {pages} {689} (\bibinfo {year} {2012})}\BibitemShut
  {NoStop}%
\bibitem [{\citenamefont {Reaney}\ and\ \citenamefont
  {Iddles}(2006)}]{reaney2006microwave}%
  \BibitemOpen
  \bibfield  {author} {\bibinfo {author} {\bibfnamefont {I.~M.}\ \bibnamefont
  {Reaney}}\ and\ \bibinfo {author} {\bibfnamefont {D.}~\bibnamefont
  {Iddles}},\ }\bibfield  {title} {\bibinfo {title} {Microwave dielectric
  ceramics for resonators and filters in mobile phone networks},\ }\href@noop
  {} {\bibfield  {journal} {\bibinfo  {journal} {Journal of the American
  Ceramic Society}\ }\textbf {\bibinfo {volume} {89}},\ \bibinfo {pages} {2063}
  (\bibinfo {year} {2006})}\BibitemShut {NoStop}%
\bibitem [{\citenamefont {Smyth}(2000)}]{smyth2000defect}%
  \BibitemOpen
  \bibfield  {author} {\bibinfo {author} {\bibfnamefont {D.~M.}\ \bibnamefont
  {Smyth}},\ }\href@noop {} {\emph {\bibinfo {title} {The defect chemistry of
  metal oxides}}}\ (\bibinfo  {publisher} {Oxford University Press},\ \bibinfo
  {year} {2000})\BibitemShut {NoStop}%
\bibitem [{\citenamefont {Alexandru}\ \emph {et~al.}(2004)\citenamefont
  {Alexandru}, \citenamefont {Berbecaru}, \citenamefont {Ioachim},
  \citenamefont {Toacsen}, \citenamefont {Banciu}, \citenamefont {Nedelcu},\
  and\ \citenamefont {Ghetu}}]{alexandru2004oxides}%
  \BibitemOpen
  \bibfield  {author} {\bibinfo {author} {\bibfnamefont {H.~V.}\ \bibnamefont
  {Alexandru}}, \bibinfo {author} {\bibfnamefont {C.}~\bibnamefont
  {Berbecaru}}, \bibinfo {author} {\bibfnamefont {A.}~\bibnamefont {Ioachim}},
  \bibinfo {author} {\bibfnamefont {M.~I.}\ \bibnamefont {Toacsen}}, \bibinfo
  {author} {\bibfnamefont {M.~G.}\ \bibnamefont {Banciu}}, \bibinfo {author}
  {\bibfnamefont {L.}~\bibnamefont {Nedelcu}},\ and\ \bibinfo {author}
  {\bibfnamefont {D.}~\bibnamefont {Ghetu}},\ }\bibfield  {title} {\bibinfo
  {title} {Oxides ferroelectric {(Ba,Sr)TiO$_3$} for microwave devices},\
  }\href@noop {} {\bibfield  {journal} {\bibinfo  {journal} {Materials Science
  and Engineering: B}\ }\textbf {\bibinfo {volume} {109}},\ \bibinfo {pages}
  {152} (\bibinfo {year} {2004})}\BibitemShut {NoStop}%
\bibitem [{\citenamefont {Klein}\ \emph {et~al.}(2023)\citenamefont {Klein},
  \citenamefont {Albe}, \citenamefont {Bein}, \citenamefont {Clemens},
  \citenamefont {Creutz}, \citenamefont {Erhart}, \citenamefont {Frericks},
  \citenamefont {Ghorbani}, \citenamefont {Hofmann}, \citenamefont {Huang}
  \emph {et~al.}}]{klein2023fermi}%
  \BibitemOpen
  \bibfield  {author} {\bibinfo {author} {\bibfnamefont {A.}~\bibnamefont
  {Klein}}, \bibinfo {author} {\bibfnamefont {K.}~\bibnamefont {Albe}},
  \bibinfo {author} {\bibfnamefont {N.}~\bibnamefont {Bein}}, \bibinfo {author}
  {\bibfnamefont {O.}~\bibnamefont {Clemens}}, \bibinfo {author} {\bibfnamefont
  {K.~A.}\ \bibnamefont {Creutz}}, \bibinfo {author} {\bibfnamefont
  {P.}~\bibnamefont {Erhart}}, \bibinfo {author} {\bibfnamefont
  {M.}~\bibnamefont {Frericks}}, \bibinfo {author} {\bibfnamefont
  {E.}~\bibnamefont {Ghorbani}}, \bibinfo {author} {\bibfnamefont {J.~P.}\
  \bibnamefont {Hofmann}}, \bibinfo {author} {\bibfnamefont {B.}~\bibnamefont
  {Huang}}, \emph {et~al.},\ }\bibfield  {title} {\bibinfo {title} {The {Fermi}
  energy as common parameter to describe charge compensation mechanisms: A path
  to {Fermi} level engineering of oxide electroceramics},\ }\href@noop {}
  {\bibfield  {journal} {\bibinfo  {journal} {Journal of Electroceramics}\
  }\textbf {\bibinfo {volume} {51}},\ \bibinfo {pages} {147} (\bibinfo {year}
  {2023})}\BibitemShut {NoStop}%
\bibitem [{\citenamefont {Keith}\ \emph {et~al.}(2004)\citenamefont {Keith},
  \citenamefont {Rampling}, \citenamefont {Sarma}, \citenamefont {Alford},\
  and\ \citenamefont {Sinclair}}]{keith2004synthesis}%
  \BibitemOpen
  \bibfield  {author} {\bibinfo {author} {\bibfnamefont {G.~M.}\ \bibnamefont
  {Keith}}, \bibinfo {author} {\bibfnamefont {M.~J.}\ \bibnamefont {Rampling}},
  \bibinfo {author} {\bibfnamefont {K.}~\bibnamefont {Sarma}}, \bibinfo
  {author} {\bibfnamefont {N.~M.}\ \bibnamefont {Alford}},\ and\ \bibinfo
  {author} {\bibfnamefont {D.~C.}\ \bibnamefont {Sinclair}},\ }\bibfield
  {title} {\bibinfo {title} {Synthesis and characterisation of doped
  {6H-BaTiO$_3$} ceramics},\ }\href@noop {} {\bibfield  {journal} {\bibinfo
  {journal} {Journal of the European Ceramic Society}\ }\textbf {\bibinfo
  {volume} {24}},\ \bibinfo {pages} {1721} (\bibinfo {year}
  {2004})}\BibitemShut {NoStop}%
\bibitem [{\citenamefont {Xu}\ \emph {et~al.}(2009)\citenamefont {Xu},
  \citenamefont {Yin}, \citenamefont {Lin}, \citenamefont {Xia}, \citenamefont
  {Wan}, \citenamefont {Yin}, \citenamefont {Bai}, \citenamefont {Du},\ and\
  \citenamefont {Liu}}]{xu2009room}%
  \BibitemOpen
  \bibfield  {author} {\bibinfo {author} {\bibfnamefont {B.}~\bibnamefont
  {Xu}}, \bibinfo {author} {\bibfnamefont {K.~B.}\ \bibnamefont {Yin}},
  \bibinfo {author} {\bibfnamefont {J.}~\bibnamefont {Lin}}, \bibinfo {author}
  {\bibfnamefont {Y.~D.}\ \bibnamefont {Xia}}, \bibinfo {author} {\bibfnamefont
  {X.~G.}\ \bibnamefont {Wan}}, \bibinfo {author} {\bibfnamefont
  {J.}~\bibnamefont {Yin}}, \bibinfo {author} {\bibfnamefont {X.~J.}\
  \bibnamefont {Bai}}, \bibinfo {author} {\bibfnamefont {J.}~\bibnamefont
  {Du}},\ and\ \bibinfo {author} {\bibfnamefont {Z.~G.}\ \bibnamefont {Liu}},\
  }\bibfield  {title} {\bibinfo {title} {Room-temperature ferromagnetism and
  ferroelectricity in {Fe}-doped {BaTiO$_3$}},\ }\href@noop {} {\bibfield
  {journal} {\bibinfo  {journal} {Physical Review B}\ }\textbf {\bibinfo
  {volume} {79}},\ \bibinfo {pages} {134109} (\bibinfo {year}
  {2009})}\BibitemShut {NoStop}%
\bibitem [{\citenamefont {Mangalam}\ \emph {et~al.}(2009)\citenamefont
  {Mangalam}, \citenamefont {Ray}, \citenamefont {Waghmare}, \citenamefont
  {Sundaresan},\ and\ \citenamefont {Rao}}]{mangalam2009multiferroic}%
  \BibitemOpen
  \bibfield  {author} {\bibinfo {author} {\bibfnamefont {R.~V.~K.}\
  \bibnamefont {Mangalam}}, \bibinfo {author} {\bibfnamefont {N.}~\bibnamefont
  {Ray}}, \bibinfo {author} {\bibfnamefont {U.~V.}\ \bibnamefont {Waghmare}},
  \bibinfo {author} {\bibfnamefont {A.}~\bibnamefont {Sundaresan}},\ and\
  \bibinfo {author} {\bibfnamefont {C.~N.~R.}\ \bibnamefont {Rao}},\ }\bibfield
   {title} {\bibinfo {title} {Multiferroic properties of nanocrystalline
  {BaTiO$_3$}},\ }\href@noop {} {\bibfield  {journal} {\bibinfo  {journal}
  {Solid State Communications}\ }\textbf {\bibinfo {volume} {149}},\ \bibinfo
  {pages} {1} (\bibinfo {year} {2009})}\BibitemShut {NoStop}%
\bibitem [{\citenamefont {Ray}\ \emph {et~al.}(2008)\citenamefont {Ray},
  \citenamefont {Mahadevan}, \citenamefont {Mandal}, \citenamefont
  {Krishnakumar}, \citenamefont {Kuroda}, \citenamefont {Sasaki}, \citenamefont
  {Taniyama},\ and\ \citenamefont {Itoh}}]{ray2008high}%
  \BibitemOpen
  \bibfield  {author} {\bibinfo {author} {\bibfnamefont {S.}~\bibnamefont
  {Ray}}, \bibinfo {author} {\bibfnamefont {P.}~\bibnamefont {Mahadevan}},
  \bibinfo {author} {\bibfnamefont {S.}~\bibnamefont {Mandal}}, \bibinfo
  {author} {\bibfnamefont {S.~R.}\ \bibnamefont {Krishnakumar}}, \bibinfo
  {author} {\bibfnamefont {C.~S.}\ \bibnamefont {Kuroda}}, \bibinfo {author}
  {\bibfnamefont {T.}~\bibnamefont {Sasaki}}, \bibinfo {author} {\bibfnamefont
  {T.}~\bibnamefont {Taniyama}},\ and\ \bibinfo {author} {\bibfnamefont
  {M.}~\bibnamefont {Itoh}},\ }\bibfield  {title} {\bibinfo {title} {High
  temperature ferromagnetism in single crystalline dilute {Fe}-doped
  {BaTiO$_3$}},\ }\href@noop {} {\bibfield  {journal} {\bibinfo  {journal}
  {Physical Review B}\ }\textbf {\bibinfo {volume} {77}},\ \bibinfo {pages}
  {104416} (\bibinfo {year} {2008})}\BibitemShut {NoStop}%
\bibitem [{\citenamefont {Chakraborty}\ \emph {et~al.}(2011)\citenamefont
  {Chakraborty}, \citenamefont {Ray},\ and\ \citenamefont
  {Itoh}}]{chakraborty2011defect}%
  \BibitemOpen
  \bibfield  {author} {\bibinfo {author} {\bibfnamefont {T.}~\bibnamefont
  {Chakraborty}}, \bibinfo {author} {\bibfnamefont {S.}~\bibnamefont {Ray}},\
  and\ \bibinfo {author} {\bibfnamefont {M.}~\bibnamefont {Itoh}},\ }\bibfield
  {title} {\bibinfo {title} {Defect-induced magnetism: Test of dilute magnetism
  in {Fe}-doped hexagonal {BaTiO$_3$} single crystals},\ }\href@noop {}
  {\bibfield  {journal} {\bibinfo  {journal} {Physical Review B}\ }\textbf
  {\bibinfo {volume} {83}},\ \bibinfo {pages} {144407} (\bibinfo {year}
  {2011})}\BibitemShut {NoStop}%
\bibitem [{\citenamefont {Nguyen}\ \emph {et~al.}(2011)\citenamefont {Nguyen},
  \citenamefont {Dang}, \citenamefont {Chuang}, \citenamefont {Thanh},
  \citenamefont {Hu}, \citenamefont {Chen}, \citenamefont {Lam}, \citenamefont
  {Lee},\ and\ \citenamefont {Hong}}]{nguyen2011tetragonal}%
  \BibitemOpen
  \bibfield  {author} {\bibinfo {author} {\bibfnamefont {H.~M.}\ \bibnamefont
  {Nguyen}}, \bibinfo {author} {\bibfnamefont {N.~V.}\ \bibnamefont {Dang}},
  \bibinfo {author} {\bibfnamefont {P.-Y.}\ \bibnamefont {Chuang}}, \bibinfo
  {author} {\bibfnamefont {T.~D.}\ \bibnamefont {Thanh}}, \bibinfo {author}
  {\bibfnamefont {C.-W.}\ \bibnamefont {Hu}}, \bibinfo {author} {\bibfnamefont
  {T.-Y.}\ \bibnamefont {Chen}}, \bibinfo {author} {\bibfnamefont {V.~D.}\
  \bibnamefont {Lam}}, \bibinfo {author} {\bibfnamefont {C.-H.}\ \bibnamefont
  {Lee}},\ and\ \bibinfo {author} {\bibfnamefont {L.~V.}\ \bibnamefont
  {Hong}},\ }\bibfield  {title} {\bibinfo {title} {Tetragonal and hexagonal
  polymorphs of {BaTi$_{1-x}$Fe$_x$O$_{3-\delta}$} multiferroics using x-ray
  and {Raman} analyses},\ }\href@noop {} {\bibfield  {journal} {\bibinfo
  {journal} {Applied Physics Letters}\ }\textbf {\bibinfo {volume} {99}},\
  \bibinfo {pages} {202501} (\bibinfo {year} {2011})}\BibitemShut {NoStop}%
\bibitem [{\citenamefont {Tho}\ \emph {et~al.}(2024)\citenamefont {Tho},
  \citenamefont {Karpinsky}, \citenamefont {Husain}, \citenamefont {Khien},
  \citenamefont {Silibin}, \citenamefont {Thanh}, \citenamefont {Lee},
  \citenamefont {Nhi}, \citenamefont {Lam}, \citenamefont {Manh} \emph
  {et~al.}}]{tho2024crystal}%
  \BibitemOpen
  \bibfield  {author} {\bibinfo {author} {\bibfnamefont {P.~T.}\ \bibnamefont
  {Tho}}, \bibinfo {author} {\bibfnamefont {D.~V.}\ \bibnamefont {Karpinsky}},
  \bibinfo {author} {\bibfnamefont {S.}~\bibnamefont {Husain}}, \bibinfo
  {author} {\bibfnamefont {N.~V.}\ \bibnamefont {Khien}}, \bibinfo {author}
  {\bibfnamefont {M.~V.}\ \bibnamefont {Silibin}}, \bibinfo {author}
  {\bibfnamefont {T.~D.}\ \bibnamefont {Thanh}}, \bibinfo {author}
  {\bibfnamefont {B.~W.}\ \bibnamefont {Lee}}, \bibinfo {author} {\bibfnamefont
  {B.~D.}\ \bibnamefont {Nhi}}, \bibinfo {author} {\bibfnamefont {D.~S.}\
  \bibnamefont {Lam}}, \bibinfo {author} {\bibfnamefont {D.~H.}\ \bibnamefont
  {Manh}}, \emph {et~al.},\ }\bibfield  {title} {\bibinfo {title} {Crystal
  structure and magnetic properties of {Fe} doped {BaTiO$_3$} at morphotropic
  phase boundary},\ }\href@noop {} {\bibfield  {journal} {\bibinfo  {journal}
  {Ceramics International}\ }\textbf {\bibinfo {volume} {50}},\ \bibinfo
  {pages} {50337} (\bibinfo {year} {2024})}\BibitemShut {NoStop}%
\bibitem [{\citenamefont {Langhammer}\ \emph {et~al.}(2000)\citenamefont
  {Langhammer}, \citenamefont {M{\"u}ller}, \citenamefont {Felgner},\ and\
  \citenamefont {Abicht}}]{langhammer2000crystal}%
  \BibitemOpen
  \bibfield  {author} {\bibinfo {author} {\bibfnamefont {H.~T.}\ \bibnamefont
  {Langhammer}}, \bibinfo {author} {\bibfnamefont {T.}~\bibnamefont
  {M{\"u}ller}}, \bibinfo {author} {\bibfnamefont {K.-H.}\ \bibnamefont
  {Felgner}},\ and\ \bibinfo {author} {\bibfnamefont {H.-P.}\ \bibnamefont
  {Abicht}},\ }\bibfield  {title} {\bibinfo {title} {Crystal structure and
  related properties of manganese-doped barium titanate ceramics},\ }\href@noop
  {} {\bibfield  {journal} {\bibinfo  {journal} {Journal of the American
  Ceramic Society}\ }\textbf {\bibinfo {volume} {83}},\ \bibinfo {pages} {605}
  (\bibinfo {year} {2000})}\BibitemShut {NoStop}%
\bibitem [{\citenamefont {Zorko}\ \emph {et~al.}(2015)\citenamefont {Zorko},
  \citenamefont {Pregelj}, \citenamefont {Gomil{\v{s}}ek}, \citenamefont
  {Jagli{\v{c}}i{\'c}}, \citenamefont {Paji{\'c}}, \citenamefont {Telling},
  \citenamefont {Ar{\v{c}}on}, \citenamefont {Mikulska},\ and\ \citenamefont
  {Valant}}]{zorko2015strain}%
  \BibitemOpen
  \bibfield  {author} {\bibinfo {author} {\bibfnamefont {A.}~\bibnamefont
  {Zorko}}, \bibinfo {author} {\bibfnamefont {M.}~\bibnamefont {Pregelj}},
  \bibinfo {author} {\bibfnamefont {M.}~\bibnamefont {Gomil{\v{s}}ek}},
  \bibinfo {author} {\bibfnamefont {Z.}~\bibnamefont {Jagli{\v{c}}i{\'c}}},
  \bibinfo {author} {\bibfnamefont {D.}~\bibnamefont {Paji{\'c}}}, \bibinfo
  {author} {\bibfnamefont {M.}~\bibnamefont {Telling}}, \bibinfo {author}
  {\bibfnamefont {I.}~\bibnamefont {Ar{\v{c}}on}}, \bibinfo {author}
  {\bibfnamefont {I.}~\bibnamefont {Mikulska}},\ and\ \bibinfo {author}
  {\bibfnamefont {M.}~\bibnamefont {Valant}},\ }\bibfield  {title} {\bibinfo
  {title} {Strain-induced extrinsic high-temperature ferromagnetism in the
  {Fe}-doped hexagonal barium titanate},\ }\href@noop {} {\bibfield  {journal}
  {\bibinfo  {journal} {Scientific Reports}\ }\textbf {\bibinfo {volume} {5}},\
  \bibinfo {pages} {7703} (\bibinfo {year} {2015})}\BibitemShut {NoStop}%
\bibitem [{\citenamefont {Becerro}\ \emph {et~al.}(2002)\citenamefont
  {Becerro}, \citenamefont {Redfern}, \citenamefont {Carpenter}, \citenamefont
  {Knight},\ and\ \citenamefont {Seifert}}]{becerro2002displacive}%
  \BibitemOpen
  \bibfield  {author} {\bibinfo {author} {\bibfnamefont {A.~I.}\ \bibnamefont
  {Becerro}}, \bibinfo {author} {\bibfnamefont {S.~A.~T.}\ \bibnamefont
  {Redfern}}, \bibinfo {author} {\bibfnamefont {M.~A.}\ \bibnamefont
  {Carpenter}}, \bibinfo {author} {\bibfnamefont {K.~S.}\ \bibnamefont
  {Knight}},\ and\ \bibinfo {author} {\bibfnamefont {F.}~\bibnamefont
  {Seifert}},\ }\bibfield  {title} {\bibinfo {title} {Displacive phase
  transitions in and strain analysis of {Fe}-doped {CaTiO$_3$} perovskites at
  high temperatures by neutron diffraction},\ }\href@noop {} {\bibfield
  {journal} {\bibinfo  {journal} {Journal of Solid State Chemistry}\ }\textbf
  {\bibinfo {volume} {167}},\ \bibinfo {pages} {459} (\bibinfo {year}
  {2002})}\BibitemShut {NoStop}%
\bibitem [{\citenamefont {Shafique}\ \emph {et~al.}(2021)\citenamefont
  {Shafique}, \citenamefont {Aldaghfag}, \citenamefont {Kashif}, \citenamefont
  {Zahid}, \citenamefont {Yaseen}, \citenamefont {Iqbal}, \citenamefont
  {Misbah},\ and\ \citenamefont {Neffati}}]{shafique2021magnetic}%
  \BibitemOpen
  \bibfield  {author} {\bibinfo {author} {\bibfnamefont {H.}~\bibnamefont
  {Shafique}}, \bibinfo {author} {\bibfnamefont {S.~A.}\ \bibnamefont
  {Aldaghfag}}, \bibinfo {author} {\bibfnamefont {M.}~\bibnamefont {Kashif}},
  \bibinfo {author} {\bibfnamefont {M.}~\bibnamefont {Zahid}}, \bibinfo
  {author} {\bibfnamefont {M.}~\bibnamefont {Yaseen}}, \bibinfo {author}
  {\bibfnamefont {J.}~\bibnamefont {Iqbal}}, \bibinfo {author} {\bibfnamefont
  {A.}~\bibnamefont {Misbah}},\ and\ \bibinfo {author} {\bibfnamefont
  {R.}~\bibnamefont {Neffati}},\ }\bibfield  {title} {\bibinfo {title}
  {Magnetic and optical characteristics of {Fe} doped {SrTiO$_3$} perovskite
  compound: a first principle study},\ }\href@noop {} {\bibfield  {journal}
  {\bibinfo  {journal} {Chalcogenide Letters}\ }\textbf {\bibinfo {volume}
  {18}},\ \bibinfo {pages} {589} (\bibinfo {year} {2021})}\BibitemShut
  {NoStop}%
\bibitem [{\citenamefont {Pal}\ \emph {et~al.}(2020)\citenamefont {Pal},
  \citenamefont {Rudrapal}, \citenamefont {Mahana}, \citenamefont {Yadav},
  \citenamefont {Paramanik}, \citenamefont {Mishra}, \citenamefont {Singh},
  \citenamefont {Sheet}, \citenamefont {Topwal}, \citenamefont {Chaudhuri}
  \emph {et~al.}}]{pal2020origin}%
  \BibitemOpen
  \bibfield  {author} {\bibinfo {author} {\bibfnamefont {P.}~\bibnamefont
  {Pal}}, \bibinfo {author} {\bibfnamefont {K.}~\bibnamefont {Rudrapal}},
  \bibinfo {author} {\bibfnamefont {S.}~\bibnamefont {Mahana}}, \bibinfo
  {author} {\bibfnamefont {S.}~\bibnamefont {Yadav}}, \bibinfo {author}
  {\bibfnamefont {T.}~\bibnamefont {Paramanik}}, \bibinfo {author}
  {\bibfnamefont {S.}~\bibnamefont {Mishra}}, \bibinfo {author} {\bibfnamefont
  {K.}~\bibnamefont {Singh}}, \bibinfo {author} {\bibfnamefont
  {G.}~\bibnamefont {Sheet}}, \bibinfo {author} {\bibfnamefont
  {D.}~\bibnamefont {Topwal}}, \bibinfo {author} {\bibfnamefont {A.~R.}\
  \bibnamefont {Chaudhuri}}, \emph {et~al.},\ }\bibfield  {title} {\bibinfo
  {title} {Origin and tuning of room-temperature multiferroicity in {Fe}-doped
  {BaTiO$_3$}},\ }\href@noop {} {\bibfield  {journal} {\bibinfo  {journal}
  {Physical Review B}\ }\textbf {\bibinfo {volume} {101}},\ \bibinfo {pages}
  {064409} (\bibinfo {year} {2020})}\BibitemShut {NoStop}%
\bibitem [{\citenamefont {Sambrano}\ \emph {et~al.}(2005)\citenamefont
  {Sambrano}, \citenamefont {Orhan}, \citenamefont {Gurgel}, \citenamefont
  {Campos}, \citenamefont {G{\'o}es}, \citenamefont {Paiva-Santos},
  \citenamefont {Varela},\ and\ \citenamefont
  {Longo}}]{sambrano2005theoretical}%
  \BibitemOpen
  \bibfield  {author} {\bibinfo {author} {\bibfnamefont {J.~R.}\ \bibnamefont
  {Sambrano}}, \bibinfo {author} {\bibfnamefont {E.}~\bibnamefont {Orhan}},
  \bibinfo {author} {\bibfnamefont {M.~F.~C.}\ \bibnamefont {Gurgel}}, \bibinfo
  {author} {\bibfnamefont {A.~B.}\ \bibnamefont {Campos}}, \bibinfo {author}
  {\bibfnamefont {M.~S.}\ \bibnamefont {G{\'o}es}}, \bibinfo {author}
  {\bibfnamefont {C.~O.}\ \bibnamefont {Paiva-Santos}}, \bibinfo {author}
  {\bibfnamefont {J.~A.}\ \bibnamefont {Varela}},\ and\ \bibinfo {author}
  {\bibfnamefont {E.}~\bibnamefont {Longo}},\ }\bibfield  {title} {\bibinfo
  {title} {Theoretical analysis of the structural deformation in {Mn}-doped
  {BaTiO$_3$}},\ }\href@noop {} {\bibfield  {journal} {\bibinfo  {journal}
  {Chemical Physics Letters}\ }\textbf {\bibinfo {volume} {402}},\ \bibinfo
  {pages} {491} (\bibinfo {year} {2005})}\BibitemShut {NoStop}%
\bibitem [{\citenamefont {Yin}\ and\ \citenamefont {Qu}(2014)}]{yin2014ab}%
  \BibitemOpen
  \bibfield  {author} {\bibinfo {author} {\bibfnamefont {B.}~\bibnamefont
  {Yin}}\ and\ \bibinfo {author} {\bibfnamefont {S.}~\bibnamefont {Qu}},\
  }\bibfield  {title} {\bibinfo {title} {Ab initio study of the effects of
  interfacial structure on the ferroelectric, magnetic, and magnetoelectric
  coupling properties of {Fe/BaTiO$_3$} multiferroic tunnel junctions},\
  }\href@noop {} {\bibfield  {journal} {\bibinfo  {journal} {Physical Review
  B}\ }\textbf {\bibinfo {volume} {89}},\ \bibinfo {pages} {014106} (\bibinfo
  {year} {2014})}\BibitemShut {NoStop}%
\bibitem [{\citenamefont {Yang}\ \emph {et~al.}(2017)\citenamefont {Yang},
  \citenamefont {Lin}, \citenamefont {Yang}, \citenamefont {Liao},
  \citenamefont {Chen},\ and\ \citenamefont {Wang}}]{yang2017first}%
  \BibitemOpen
  \bibfield  {author} {\bibinfo {author} {\bibfnamefont {F.}~\bibnamefont
  {Yang}}, \bibinfo {author} {\bibfnamefont {S.}~\bibnamefont {Lin}}, \bibinfo
  {author} {\bibfnamefont {L.}~\bibnamefont {Yang}}, \bibinfo {author}
  {\bibfnamefont {J.}~\bibnamefont {Liao}}, \bibinfo {author} {\bibfnamefont
  {Y.}~\bibnamefont {Chen}},\ and\ \bibinfo {author} {\bibfnamefont {C.-Z.}\
  \bibnamefont {Wang}},\ }\bibfield  {title} {\bibinfo {title}
  {First-principles investigation of metal-doped cubic {BaTiO$_3$}},\
  }\href@noop {} {\bibfield  {journal} {\bibinfo  {journal} {Materials Research
  Bulletin}\ }\textbf {\bibinfo {volume} {96}},\ \bibinfo {pages} {372}
  (\bibinfo {year} {2017})}\BibitemShut {NoStop}%
\bibitem [{\citenamefont {Adeagbo}\ \emph {et~al.}(2019)\citenamefont
  {Adeagbo}, \citenamefont {Ben~Hamed}, \citenamefont {Nayak}, \citenamefont
  {B{\"o}ttcher}, \citenamefont {Langhammer},\ and\ \citenamefont
  {Hergert}}]{adeagbo2019theoretical}%
  \BibitemOpen
  \bibfield  {author} {\bibinfo {author} {\bibfnamefont {W.~A.}\ \bibnamefont
  {Adeagbo}}, \bibinfo {author} {\bibfnamefont {H.}~\bibnamefont {Ben~Hamed}},
  \bibinfo {author} {\bibfnamefont {S.~K.}\ \bibnamefont {Nayak}}, \bibinfo
  {author} {\bibfnamefont {R.}~\bibnamefont {B{\"o}ttcher}}, \bibinfo {author}
  {\bibfnamefont {H.~T.}\ \bibnamefont {Langhammer}},\ and\ \bibinfo {author}
  {\bibfnamefont {W.}~\bibnamefont {Hergert}},\ }\bibfield  {title} {\bibinfo
  {title} {Theoretical investigation of iron incorporation in hexagonal barium
  titanate},\ }\href@noop {} {\bibfield  {journal} {\bibinfo  {journal}
  {Physical Review B}\ }\textbf {\bibinfo {volume} {100}},\ \bibinfo {pages}
  {184108} (\bibinfo {year} {2019})}\BibitemShut {NoStop}%
\bibitem [{\citenamefont {Li}\ \emph {et~al.}(2021)\citenamefont {Li},
  \citenamefont {He}, \citenamefont {Xiong}, \citenamefont {Zou}, \citenamefont
  {Liu}, \citenamefont {Chen}, \citenamefont {Zhang}, \citenamefont {Yan},
  \citenamefont {Xiao}, \citenamefont {Tang} \emph
  {et~al.}}]{li2021coexistence}%
  \BibitemOpen
  \bibfield  {author} {\bibinfo {author} {\bibfnamefont {G.}~\bibnamefont
  {Li}}, \bibinfo {author} {\bibfnamefont {C.}~\bibnamefont {He}}, \bibinfo
  {author} {\bibfnamefont {Y.}~\bibnamefont {Xiong}}, \bibinfo {author}
  {\bibfnamefont {Z.}~\bibnamefont {Zou}}, \bibinfo {author} {\bibfnamefont
  {Y.}~\bibnamefont {Liu}}, \bibinfo {author} {\bibfnamefont {Q.}~\bibnamefont
  {Chen}}, \bibinfo {author} {\bibfnamefont {W.}~\bibnamefont {Zhang}},
  \bibinfo {author} {\bibfnamefont {S.}~\bibnamefont {Yan}}, \bibinfo {author}
  {\bibfnamefont {Y.}~\bibnamefont {Xiao}}, \bibinfo {author} {\bibfnamefont
  {M.}~\bibnamefont {Tang}}, \emph {et~al.},\ }\bibfield  {title} {\bibinfo
  {title} {Coexistence of ferroelectricity and metallicity in {M}-doped
  {BaTiO$_3$} ({M} = {Al}, {V}, {Cr}, {Fe}, {Ni}, and {Nb}): First-principles
  study},\ }\href@noop {} {\bibfield  {journal} {\bibinfo  {journal} {Materials
  Today Communications}\ }\textbf {\bibinfo {volume} {27}},\ \bibinfo {pages}
  {102394} (\bibinfo {year} {2021})}\BibitemShut {NoStop}%
\bibitem [{\citenamefont {Zunger}(2018)}]{zunger2018inverse}%
  \BibitemOpen
  \bibfield  {author} {\bibinfo {author} {\bibfnamefont {A.}~\bibnamefont
  {Zunger}},\ }\bibfield  {title} {\bibinfo {title} {Inverse design in search
  of materials with target functionalities},\ }\href@noop {} {\bibfield
  {journal} {\bibinfo  {journal} {Nature Reviews Chemistry}\ }\textbf {\bibinfo
  {volume} {2}},\ \bibinfo {pages} {0121} (\bibinfo {year} {2018})}\BibitemShut
  {NoStop}%
\bibitem [{\citenamefont {Curtarolo}\ \emph {et~al.}(2013)\citenamefont
  {Curtarolo}, \citenamefont {Hart}, \citenamefont {Nardelli}, \citenamefont
  {Mingo}, \citenamefont {Sanvito},\ and\ \citenamefont
  {Levy}}]{curtarolo2013high}%
  \BibitemOpen
  \bibfield  {author} {\bibinfo {author} {\bibfnamefont {S.}~\bibnamefont
  {Curtarolo}}, \bibinfo {author} {\bibfnamefont {G.~L.~W.}\ \bibnamefont
  {Hart}}, \bibinfo {author} {\bibfnamefont {M.~B.}\ \bibnamefont {Nardelli}},
  \bibinfo {author} {\bibfnamefont {N.}~\bibnamefont {Mingo}}, \bibinfo
  {author} {\bibfnamefont {S.}~\bibnamefont {Sanvito}},\ and\ \bibinfo {author}
  {\bibfnamefont {O.}~\bibnamefont {Levy}},\ }\bibfield  {title} {\bibinfo
  {title} {The high-throughput highway to computational materials design},\
  }\href@noop {} {\bibfield  {journal} {\bibinfo  {journal} {Nature Materials}\
  }\textbf {\bibinfo {volume} {12}},\ \bibinfo {pages} {191} (\bibinfo {year}
  {2013})}\BibitemShut {NoStop}%
\bibitem [{\citenamefont {Bartel}\ \emph {et~al.}(2019)\citenamefont {Bartel},
  \citenamefont {Sutton}, \citenamefont {Goldsmith}, \citenamefont {Ouyang},
  \citenamefont {Musgrave}, \citenamefont {Ghiringhelli},\ and\ \citenamefont
  {Scheffler}}]{bartel2019new}%
  \BibitemOpen
  \bibfield  {author} {\bibinfo {author} {\bibfnamefont {C.~J.}\ \bibnamefont
  {Bartel}}, \bibinfo {author} {\bibfnamefont {C.}~\bibnamefont {Sutton}},
  \bibinfo {author} {\bibfnamefont {B.~R.}\ \bibnamefont {Goldsmith}}, \bibinfo
  {author} {\bibfnamefont {R.}~\bibnamefont {Ouyang}}, \bibinfo {author}
  {\bibfnamefont {C.~B.}\ \bibnamefont {Musgrave}}, \bibinfo {author}
  {\bibfnamefont {L.~M.}\ \bibnamefont {Ghiringhelli}},\ and\ \bibinfo {author}
  {\bibfnamefont {M.}~\bibnamefont {Scheffler}},\ }\bibfield  {title} {\bibinfo
  {title} {New tolerance factor to predict the stability of perovskite oxides
  and halides},\ }\href@noop {} {\bibfield  {journal} {\bibinfo  {journal}
  {Science Advances}\ }\textbf {\bibinfo {volume} {5}},\ \bibinfo {pages}
  {eaav0693} (\bibinfo {year} {2019})}\BibitemShut {NoStop}%
\bibitem [{\citenamefont {Goldschmidt}(1926)}]{goldschmidt1926gesetze}%
  \BibitemOpen
  \bibfield  {author} {\bibinfo {author} {\bibfnamefont {V.~M.}\ \bibnamefont
  {Goldschmidt}},\ }\bibfield  {title} {\bibinfo {title} {{Die Gesetze der
  Krystallochemie}},\ }\href@noop {} {\bibfield  {journal} {\bibinfo  {journal}
  {Naturwissenschaften}\ }\textbf {\bibinfo {volume} {14}},\ \bibinfo {pages}
  {477} (\bibinfo {year} {1926})}\BibitemShut {NoStop}%
\bibitem [{\citenamefont {Kresse}\ and\ \citenamefont
  {Furthm\"uller}(1996)}]{PhysRevB.54.11169}%
  \BibitemOpen
  \bibfield  {author} {\bibinfo {author} {\bibfnamefont {G.}~\bibnamefont
  {Kresse}}\ and\ \bibinfo {author} {\bibfnamefont {J.}~\bibnamefont
  {Furthm\"uller}},\ }\bibfield  {title} {\bibinfo {title} {Efficient iterative
  schemes for ab initio total-energy calculations using a plane-wave basis
  set},\ }\href {https://doi.org/10.1103/PhysRevB.54.11169} {\bibfield
  {journal} {\bibinfo  {journal} {Physical Review B}\ }\textbf {\bibinfo
  {volume} {54}},\ \bibinfo {pages} {11169} (\bibinfo {year}
  {1996})}\BibitemShut {NoStop}%
\bibitem [{\citenamefont {Bl{\"o}chl}(1994)}]{blochl1994projector}%
  \BibitemOpen
  \bibfield  {author} {\bibinfo {author} {\bibfnamefont {P.~E.}\ \bibnamefont
  {Bl{\"o}chl}},\ }\bibfield  {title} {\bibinfo {title} {Projector
  augmented-wave method},\ }\href@noop {} {\bibfield  {journal} {\bibinfo
  {journal} {Physical Review B}\ }\textbf {\bibinfo {volume} {50}},\ \bibinfo
  {pages} {17953} (\bibinfo {year} {1994})}\BibitemShut {NoStop}%
\bibitem [{\citenamefont {Perdew}\ \emph {et~al.}(1996)\citenamefont {Perdew},
  \citenamefont {Burke},\ and\ \citenamefont
  {Ernzerhof}}]{perdew1996generalized}%
  \BibitemOpen
  \bibfield  {author} {\bibinfo {author} {\bibfnamefont {J.~P.}\ \bibnamefont
  {Perdew}}, \bibinfo {author} {\bibfnamefont {K.}~\bibnamefont {Burke}},\ and\
  \bibinfo {author} {\bibfnamefont {M.}~\bibnamefont {Ernzerhof}},\ }\bibfield
  {title} {\bibinfo {title} {Generalized gradient approximation made simple},\
  }\href@noop {} {\bibfield  {journal} {\bibinfo  {journal} {Physical Review
  Letters}\ }\textbf {\bibinfo {volume} {77}},\ \bibinfo {pages} {3865}
  (\bibinfo {year} {1996})}\BibitemShut {NoStop}%
\bibitem [{\citenamefont {Morgan}(2017)}]{Morgan_JOSS2017b}%
  \BibitemOpen
  \bibfield  {author} {\bibinfo {author} {\bibfnamefont {B.~J.}\ \bibnamefont
  {Morgan}},\ }\bibfield  {title} {\bibinfo {title} {bsym: A basic symmetry
  module},\ }\href {https://doi.org/10.21105/joss.00370} {\bibfield  {journal}
  {\bibinfo  {journal} {The Journal of Open Source Software}\ }\textbf
  {\bibinfo {volume} {2}},\ \bibinfo {pages} {370} (\bibinfo {year}
  {2017})}\BibitemShut {NoStop}%
\bibitem [{\citenamefont {Nayak}\ \emph {et~al.}(2015)\citenamefont {Nayak},
  \citenamefont {Langhammer}, \citenamefont {Adeagbo}, \citenamefont {Hergert},
  \citenamefont {M{\"u}ller},\ and\ \citenamefont
  {B{\"o}ttcher}}]{nayak2015chromium}%
  \BibitemOpen
  \bibfield  {author} {\bibinfo {author} {\bibfnamefont {S.~K.}\ \bibnamefont
  {Nayak}}, \bibinfo {author} {\bibfnamefont {H.~T.}\ \bibnamefont
  {Langhammer}}, \bibinfo {author} {\bibfnamefont {W.~A.}\ \bibnamefont
  {Adeagbo}}, \bibinfo {author} {\bibfnamefont {W.}~\bibnamefont {Hergert}},
  \bibinfo {author} {\bibfnamefont {T.}~\bibnamefont {M{\"u}ller}},\ and\
  \bibinfo {author} {\bibfnamefont {R.}~\bibnamefont {B{\"o}ttcher}},\
  }\bibfield  {title} {\bibinfo {title} {Chromium point defects in hexagonal
  {BaTiO$_3$}: A comparative study of first-principles calculations and
  experiments},\ }\href@noop {} {\bibfield  {journal} {\bibinfo  {journal}
  {Physical Review B}\ }\textbf {\bibinfo {volume} {91}},\ \bibinfo {pages}
  {155105} (\bibinfo {year} {2015})}\BibitemShut {NoStop}%
\bibitem [{\citenamefont {B{\"o}ttcher}\ \emph {et~al.}(2005)\citenamefont
  {B{\"o}ttcher}, \citenamefont {Erdem}, \citenamefont {Langhammer},
  \citenamefont {M{\"u}ller},\ and\ \citenamefont
  {Abicht}}]{bottcher2005incorporation}%
  \BibitemOpen
  \bibfield  {author} {\bibinfo {author} {\bibfnamefont {R.}~\bibnamefont
  {B{\"o}ttcher}}, \bibinfo {author} {\bibfnamefont {E.}~\bibnamefont {Erdem}},
  \bibinfo {author} {\bibfnamefont {H.~T.}\ \bibnamefont {Langhammer}},
  \bibinfo {author} {\bibfnamefont {T.}~\bibnamefont {M{\"u}ller}},\ and\
  \bibinfo {author} {\bibfnamefont {H.-P.}\ \bibnamefont {Abicht}},\ }\bibfield
   {title} {\bibinfo {title} {Incorporation of chromium into hexagonal barium
  titanate: an electron paramagnetic resonance study},\ }\href@noop {}
  {\bibfield  {journal} {\bibinfo  {journal} {Journal of Physics: Condensed
  Matter}\ }\textbf {\bibinfo {volume} {17}},\ \bibinfo {pages} {2763}
  (\bibinfo {year} {2005})}\BibitemShut {NoStop}%
\bibitem [{\citenamefont {Ren}(2004)}]{ren2004large}%
  \BibitemOpen
  \bibfield  {author} {\bibinfo {author} {\bibfnamefont {X.}~\bibnamefont
  {Ren}},\ }\bibfield  {title} {\bibinfo {title} {Large electric-field-induced
  strain in ferroelectric crystals by point-defect-mediated reversible domain
  switching},\ }\href@noop {} {\bibfield  {journal} {\bibinfo  {journal}
  {Nature Materials}\ }\textbf {\bibinfo {volume} {3}},\ \bibinfo {pages} {91}
  (\bibinfo {year} {2004})}\BibitemShut {NoStop}%
\bibitem [{\citenamefont {Noguchi}\ \emph {et~al.}(2020)\citenamefont
  {Noguchi}, \citenamefont {Taniguchi}, \citenamefont {Inoue},\ and\
  \citenamefont {Miyayama}}]{noguchi2020successive}%
  \BibitemOpen
  \bibfield  {author} {\bibinfo {author} {\bibfnamefont {Y.}~\bibnamefont
  {Noguchi}}, \bibinfo {author} {\bibfnamefont {Y.}~\bibnamefont {Taniguchi}},
  \bibinfo {author} {\bibfnamefont {R.}~\bibnamefont {Inoue}},\ and\ \bibinfo
  {author} {\bibfnamefont {M.}~\bibnamefont {Miyayama}},\ }\bibfield  {title}
  {\bibinfo {title} {Successive redox-mediated visible-light
  ferrophotovoltaics},\ }\href@noop {} {\bibfield  {journal} {\bibinfo
  {journal} {Nature Communications}\ }\textbf {\bibinfo {volume} {11}},\
  \bibinfo {pages} {966} (\bibinfo {year} {2020})}\BibitemShut {NoStop}%
\bibitem [{\citenamefont {B{\"o}ttcher}\ \emph {et~al.}(2018)\citenamefont
  {B{\"o}ttcher}, \citenamefont {Langhammer}, \citenamefont {K{\"u}cker},
  \citenamefont {Eisenschmidt},\ and\ \citenamefont
  {Ebbinghaus}}]{bottcher2018incorporation}%
  \BibitemOpen
  \bibfield  {author} {\bibinfo {author} {\bibfnamefont {R.}~\bibnamefont
  {B{\"o}ttcher}}, \bibinfo {author} {\bibfnamefont {H.~T.}\ \bibnamefont
  {Langhammer}}, \bibinfo {author} {\bibfnamefont {S.}~\bibnamefont
  {K{\"u}cker}}, \bibinfo {author} {\bibfnamefont {C.}~\bibnamefont
  {Eisenschmidt}},\ and\ \bibinfo {author} {\bibfnamefont {S.~G.}\ \bibnamefont
  {Ebbinghaus}},\ }\bibfield  {title} {\bibinfo {title} {On the incorporation
  of iron into hexagonal barium titanate: {I. Electron} paramagnetic resonance
  ({EPR}) study},\ }\href@noop {} {\bibfield  {journal} {\bibinfo  {journal}
  {Journal of Physics: Condensed Matter}\ }\textbf {\bibinfo {volume} {30}},\
  \bibinfo {pages} {425701} (\bibinfo {year} {2018})}\BibitemShut {NoStop}%
\bibitem [{\citenamefont {Langhammer}\ \emph {et~al.}(2020)\citenamefont
  {Langhammer}, \citenamefont {Walther}, \citenamefont {B{\"o}ttcher},\ and\
  \citenamefont {Ebbinghaus}}]{langhammer2020incorporation}%
  \BibitemOpen
  \bibfield  {author} {\bibinfo {author} {\bibfnamefont {H.~T.}\ \bibnamefont
  {Langhammer}}, \bibinfo {author} {\bibfnamefont {T.}~\bibnamefont {Walther}},
  \bibinfo {author} {\bibfnamefont {R.}~\bibnamefont {B{\"o}ttcher}},\ and\
  \bibinfo {author} {\bibfnamefont {S.~G.}\ \bibnamefont {Ebbinghaus}},\
  }\bibfield  {title} {\bibinfo {title} {On the incorporation of iron into
  hexagonal barium titanate: {II. Magnetic} moment, electron paramagnetic
  resonance ({EPR}) and optical transmission},\ }\href@noop {} {\bibfield
  {journal} {\bibinfo  {journal} {Journal of Physics: Condensed Matter}\
  }\textbf {\bibinfo {volume} {32}},\ \bibinfo {pages} {385702} (\bibinfo
  {year} {2020})}\BibitemShut {NoStop}%
\bibitem [{\citenamefont {Ehsan}\ \emph {et~al.}(2021)\citenamefont {Ehsan},
  \citenamefont {Arrigoni}, \citenamefont {Madsen}, \citenamefont {Blaha},\
  and\ \citenamefont {Tr{\"o}ster}}]{ehsan2021first}%
  \BibitemOpen
  \bibfield  {author} {\bibinfo {author} {\bibfnamefont {S.}~\bibnamefont
  {Ehsan}}, \bibinfo {author} {\bibfnamefont {M.}~\bibnamefont {Arrigoni}},
  \bibinfo {author} {\bibfnamefont {G.~K.~H.}\ \bibnamefont {Madsen}}, \bibinfo
  {author} {\bibfnamefont {P.}~\bibnamefont {Blaha}},\ and\ \bibinfo {author}
  {\bibfnamefont {A.}~\bibnamefont {Tr{\"o}ster}},\ }\bibfield  {title}
  {\bibinfo {title} {First-principles self-consistent phonon approach to the
  study of the vibrational properties and structural phase transition of
  {BaTiO$_3$}},\ }\href@noop {} {\bibfield  {journal} {\bibinfo  {journal}
  {Physical Review B}\ }\textbf {\bibinfo {volume} {103}},\ \bibinfo {pages}
  {094108} (\bibinfo {year} {2021})}\BibitemShut {NoStop}%
\bibitem [{\citenamefont {Sahasrabuddhe}\ \emph {et~al.}(2026)\citenamefont
  {Sahasrabuddhe}, \citenamefont {Zheng}, \citenamefont {Lin}, \citenamefont
  {Kaplan}, \citenamefont {Mathis}, \citenamefont {Huck}, \citenamefont
  {Petretto}, \citenamefont {George}, \citenamefont {Marzari}, \citenamefont
  {Persson}, \citenamefont {Hautier},\ and\ \citenamefont
  {Jain}}]{sahasrabuddhe2026phonon}%
  \BibitemOpen
  \bibfield  {author} {\bibinfo {author} {\bibfnamefont {H.}~\bibnamefont
  {Sahasrabuddhe}}, \bibinfo {author} {\bibfnamefont {J.}~\bibnamefont
  {Zheng}}, \bibinfo {author} {\bibfnamefont {C.}~\bibnamefont {Lin}}, \bibinfo
  {author} {\bibfnamefont {A.}~\bibnamefont {Kaplan}}, \bibinfo {author}
  {\bibfnamefont {T.}~\bibnamefont {Mathis}}, \bibinfo {author} {\bibfnamefont
  {P.}~\bibnamefont {Huck}}, \bibinfo {author} {\bibfnamefont {G.}~\bibnamefont
  {Petretto}}, \bibinfo {author} {\bibfnamefont {J.}~\bibnamefont {George}},
  \bibinfo {author} {\bibfnamefont {N.}~\bibnamefont {Marzari}}, \bibinfo
  {author} {\bibfnamefont {K.~A.}\ \bibnamefont {Persson}}, \bibinfo {author}
  {\bibfnamefont {G.}~\bibnamefont {Hautier}},\ and\ \bibinfo {author}
  {\bibfnamefont {A.}~\bibnamefont {Jain}},\ }\href
  {https://doi.org/10.26434/chemrxiv.15004632/v1} {\bibinfo {title} {A
  high-throughput ab initio database of harmonic phonon properties for
  inorganic crystals}} (\bibinfo {year} {2026}),\ \bibinfo {note} {{ChemRxiv},
  \href{https://doi.org/10.26434/chemrxiv.15004632/v1}{doi:10.26434/chemrxiv.15004632/v1}}\BibitemShut
  {NoStop}%
\bibitem [{\citenamefont {Jain}\ \emph {et~al.}(2013)\citenamefont {Jain},
  \citenamefont {Ong}, \citenamefont {Hautier}, \citenamefont {Chen},
  \citenamefont {Richards}, \citenamefont {Dacek}, \citenamefont {Cholia},
  \citenamefont {Gunter}, \citenamefont {Skinner}, \citenamefont {Ceder},\ and\
  \citenamefont {Persson}}]{jain2013commentary}%
  \BibitemOpen
  \bibfield  {author} {\bibinfo {author} {\bibfnamefont {A.}~\bibnamefont
  {Jain}}, \bibinfo {author} {\bibfnamefont {S.~P.}\ \bibnamefont {Ong}},
  \bibinfo {author} {\bibfnamefont {G.}~\bibnamefont {Hautier}}, \bibinfo
  {author} {\bibfnamefont {W.}~\bibnamefont {Chen}}, \bibinfo {author}
  {\bibfnamefont {W.~D.}\ \bibnamefont {Richards}}, \bibinfo {author}
  {\bibfnamefont {S.}~\bibnamefont {Dacek}}, \bibinfo {author} {\bibfnamefont
  {S.}~\bibnamefont {Cholia}}, \bibinfo {author} {\bibfnamefont
  {D.}~\bibnamefont {Gunter}}, \bibinfo {author} {\bibfnamefont
  {D.}~\bibnamefont {Skinner}}, \bibinfo {author} {\bibfnamefont
  {G.}~\bibnamefont {Ceder}},\ and\ \bibinfo {author} {\bibfnamefont {K.~A.}\
  \bibnamefont {Persson}},\ }\bibfield  {title} {\bibinfo {title} {Commentary:
  The {Materials Project}: A materials genome approach to accelerating
  materials innovation},\ }\href@noop {} {\bibfield  {journal} {\bibinfo
  {journal} {APL Materials}\ }\textbf {\bibinfo {volume} {1}},\ \bibinfo
  {pages} {011002} (\bibinfo {year} {2013})}\BibitemShut {NoStop}%
\bibitem [{\citenamefont {Dudarev}\ \emph {et~al.}(1998)\citenamefont
  {Dudarev}, \citenamefont {Botton}, \citenamefont {Savrasov}, \citenamefont
  {Humphreys},\ and\ \citenamefont {Sutton}}]{PhysRevB.57.1505}%
  \BibitemOpen
  \bibfield  {author} {\bibinfo {author} {\bibfnamefont {S.~L.}\ \bibnamefont
  {Dudarev}}, \bibinfo {author} {\bibfnamefont {G.~A.}\ \bibnamefont {Botton}},
  \bibinfo {author} {\bibfnamefont {S.~Y.}\ \bibnamefont {Savrasov}}, \bibinfo
  {author} {\bibfnamefont {C.~J.}\ \bibnamefont {Humphreys}},\ and\ \bibinfo
  {author} {\bibfnamefont {A.~P.}\ \bibnamefont {Sutton}},\ }\bibfield  {title}
  {\bibinfo {title} {Electron-energy-loss spectra and the structural stability
  of nickel oxide: An {LSDA+U} study},\ }\href
  {https://doi.org/10.1103/PhysRevB.57.1505} {\bibfield  {journal} {\bibinfo
  {journal} {Physical Review B}\ }\textbf {\bibinfo {volume} {57}},\ \bibinfo
  {pages} {1505} (\bibinfo {year} {1998})}\BibitemShut {NoStop}%
\bibitem [{\citenamefont {Meng}\ \emph {et~al.}(2016)\citenamefont {Meng},
  \citenamefont {Liu}, \citenamefont {Huo}, \citenamefont {Guo}, \citenamefont
  {Cao}, \citenamefont {Peng}, \citenamefont {Dearden}, \citenamefont {Gonze},
  \citenamefont {Yang}, \citenamefont {Wang} \emph {et~al.}}]{meng2016density}%
  \BibitemOpen
  \bibfield  {author} {\bibinfo {author} {\bibfnamefont {Y.}~\bibnamefont
  {Meng}}, \bibinfo {author} {\bibfnamefont {X.-W.}\ \bibnamefont {Liu}},
  \bibinfo {author} {\bibfnamefont {C.-F.}\ \bibnamefont {Huo}}, \bibinfo
  {author} {\bibfnamefont {W.-P.}\ \bibnamefont {Guo}}, \bibinfo {author}
  {\bibfnamefont {D.-B.}\ \bibnamefont {Cao}}, \bibinfo {author} {\bibfnamefont
  {Q.}~\bibnamefont {Peng}}, \bibinfo {author} {\bibfnamefont {A.}~\bibnamefont
  {Dearden}}, \bibinfo {author} {\bibfnamefont {X.}~\bibnamefont {Gonze}},
  \bibinfo {author} {\bibfnamefont {Y.}~\bibnamefont {Yang}}, \bibinfo {author}
  {\bibfnamefont {J.}~\bibnamefont {Wang}}, \emph {et~al.},\ }\bibfield
  {title} {\bibinfo {title} {When density functional approximations meet iron
  oxides},\ }\href@noop {} {\bibfield  {journal} {\bibinfo  {journal} {Journal
  of Chemical Theory and Computation}\ }\textbf {\bibinfo {volume} {12}},\
  \bibinfo {pages} {5132} (\bibinfo {year} {2016})}\BibitemShut {NoStop}%
\bibitem [{\citenamefont {Grau-Crespo}\ \emph {et~al.}(2006)\citenamefont
  {Grau-Crespo}, \citenamefont {Cor{\`a}}, \citenamefont {Sokol}, \citenamefont
  {de~Leeuw},\ and\ \citenamefont {Catlow}}]{grau2006electronic}%
  \BibitemOpen
  \bibfield  {author} {\bibinfo {author} {\bibfnamefont {R.}~\bibnamefont
  {Grau-Crespo}}, \bibinfo {author} {\bibfnamefont {F.}~\bibnamefont
  {Cor{\`a}}}, \bibinfo {author} {\bibfnamefont {A.~A.}\ \bibnamefont {Sokol}},
  \bibinfo {author} {\bibfnamefont {N.~H.}\ \bibnamefont {de~Leeuw}},\ and\
  \bibinfo {author} {\bibfnamefont {C.~R.~A.}\ \bibnamefont {Catlow}},\
  }\bibfield  {title} {\bibinfo {title} {Electronic structure and magnetic
  coupling in {FeSbO$_4$}: A {DFT} study using hybrid functionals and {GGA+U}
  methods},\ }\href@noop {} {\bibfield  {journal} {\bibinfo  {journal}
  {Physical Review B}\ }\textbf {\bibinfo {volume} {73}},\ \bibinfo {pages}
  {035116} (\bibinfo {year} {2006})}\BibitemShut {NoStop}%
\bibitem [{\citenamefont {Padchasri}\ \emph {et~al.}(2021)\citenamefont
  {Padchasri}, \citenamefont {Triamnak}, \citenamefont {Sareein}, \citenamefont
  {Jutimoosik}, \citenamefont {Tongsaeng}, \citenamefont {Bootchanont},
  \citenamefont {Kidkhunthod}, \citenamefont {Rujirawat}, \citenamefont
  {Manyum},\ and\ \citenamefont {Yimnirun}}]{padchasri2021crystal}%
  \BibitemOpen
  \bibfield  {author} {\bibinfo {author} {\bibfnamefont {J.}~\bibnamefont
  {Padchasri}}, \bibinfo {author} {\bibfnamefont {N.}~\bibnamefont {Triamnak}},
  \bibinfo {author} {\bibfnamefont {T.}~\bibnamefont {Sareein}}, \bibinfo
  {author} {\bibfnamefont {J.}~\bibnamefont {Jutimoosik}}, \bibinfo {author}
  {\bibfnamefont {S.}~\bibnamefont {Tongsaeng}}, \bibinfo {author}
  {\bibfnamefont {A.}~\bibnamefont {Bootchanont}}, \bibinfo {author}
  {\bibfnamefont {P.}~\bibnamefont {Kidkhunthod}}, \bibinfo {author}
  {\bibfnamefont {S.}~\bibnamefont {Rujirawat}}, \bibinfo {author}
  {\bibfnamefont {P.}~\bibnamefont {Manyum}},\ and\ \bibinfo {author}
  {\bibfnamefont {R.}~\bibnamefont {Yimnirun}},\ }\bibfield  {title} {\bibinfo
  {title} {Crystal structure and {XANES} study of {Fe}-substituted barium
  titanate ceramics prepared by conventional solid-state technique},\
  }\href@noop {} {\bibfield  {journal} {\bibinfo  {journal} {Radiation Physics
  and Chemistry}\ }\textbf {\bibinfo {volume} {188}},\ \bibinfo {pages}
  {109657} (\bibinfo {year} {2021})}\BibitemShut {NoStop}%
\bibitem [{\citenamefont {Rajan}\ \emph {et~al.}(2017)\citenamefont {Rajan},
  \citenamefont {Gazzali},\ and\ \citenamefont
  {Chandrasekaran}}]{rajan2017impact}%
  \BibitemOpen
  \bibfield  {author} {\bibinfo {author} {\bibfnamefont {S.}~\bibnamefont
  {Rajan}}, \bibinfo {author} {\bibfnamefont {P.~M.~M.}\ \bibnamefont
  {Gazzali}},\ and\ \bibinfo {author} {\bibfnamefont {G.}~\bibnamefont
  {Chandrasekaran}},\ }\bibfield  {title} {\bibinfo {title} {Impact of {Fe} on
  structural modification and room temperature magnetic ordering in
  {BaTiO$_3$}},\ }\href@noop {} {\bibfield  {journal} {\bibinfo  {journal}
  {Spectrochimica Acta Part A: Molecular and Biomolecular Spectroscopy}\
  }\textbf {\bibinfo {volume} {171}},\ \bibinfo {pages} {80} (\bibinfo {year}
  {2017})}\BibitemShut {NoStop}%
\bibitem [{\citenamefont {Kolodiazhnyi}\ and\ \citenamefont
  {Petric}(2003)}]{kolodiazhnyi2003analysis}%
  \BibitemOpen
  \bibfield  {author} {\bibinfo {author} {\bibfnamefont {T.}~\bibnamefont
  {Kolodiazhnyi}}\ and\ \bibinfo {author} {\bibfnamefont {A.}~\bibnamefont
  {Petric}},\ }\bibfield  {title} {\bibinfo {title} {Analysis of point defects
  in polycrystalline {BaTiO$_3$} by electron paramagnetic resonance},\
  }\href@noop {} {\bibfield  {journal} {\bibinfo  {journal} {Journal of Physics
  and Chemistry of Solids}\ }\textbf {\bibinfo {volume} {64}},\ \bibinfo
  {pages} {953} (\bibinfo {year} {2003})}\BibitemShut {NoStop}%
\bibitem [{\citenamefont {Shannon}(1976)}]{shannon1976revised}%
  \BibitemOpen
  \bibfield  {author} {\bibinfo {author} {\bibfnamefont {R.~D.}\ \bibnamefont
  {Shannon}},\ }\bibfield  {title} {\bibinfo {title} {Revised effective ionic
  radii and systematic studies of interatomic distances in halides and
  chalcogenides},\ }\href@noop {} {\bibfield  {journal} {\bibinfo  {journal}
  {Acta Crystallographica Section A}\ }\textbf {\bibinfo {volume} {32}},\
  \bibinfo {pages} {751} (\bibinfo {year} {1976})}\BibitemShut {NoStop}%
\bibitem [{\citenamefont {Ku}\ \emph {et~al.}(2010)\citenamefont {Ku},
  \citenamefont {Berlijn},\ and\ \citenamefont {Lee}}]{ku2010unfolding}%
  \BibitemOpen
  \bibfield  {author} {\bibinfo {author} {\bibfnamefont {W.}~\bibnamefont
  {Ku}}, \bibinfo {author} {\bibfnamefont {T.}~\bibnamefont {Berlijn}},\ and\
  \bibinfo {author} {\bibfnamefont {C.-C.}\ \bibnamefont {Lee}},\ }\bibfield
  {title} {\bibinfo {title} {Unfolding first-principles band structures},\
  }\href@noop {} {\bibfield  {journal} {\bibinfo  {journal} {Physical Review
  Letters}\ }\textbf {\bibinfo {volume} {104}},\ \bibinfo {pages} {216401}
  (\bibinfo {year} {2010})}\BibitemShut {NoStop}%
\bibitem [{\citenamefont {Herath}\ \emph {et~al.}(2020)\citenamefont {Herath},
  \citenamefont {Tavadze}, \citenamefont {He}, \citenamefont {Bousquet},
  \citenamefont {Singh}, \citenamefont {Mu{\~n}oz},\ and\ \citenamefont
  {Romero}}]{herath2020pyprocar}%
  \BibitemOpen
  \bibfield  {author} {\bibinfo {author} {\bibfnamefont {U.}~\bibnamefont
  {Herath}}, \bibinfo {author} {\bibfnamefont {P.}~\bibnamefont {Tavadze}},
  \bibinfo {author} {\bibfnamefont {X.}~\bibnamefont {He}}, \bibinfo {author}
  {\bibfnamefont {E.}~\bibnamefont {Bousquet}}, \bibinfo {author}
  {\bibfnamefont {S.}~\bibnamefont {Singh}}, \bibinfo {author} {\bibfnamefont
  {F.}~\bibnamefont {Mu{\~n}oz}},\ and\ \bibinfo {author} {\bibfnamefont
  {A.~H.}\ \bibnamefont {Romero}},\ }\bibfield  {title} {\bibinfo {title}
  {{PyProcar}: A {Python} library for electronic structure
  pre/post-processing},\ }\href@noop {} {\bibfield  {journal} {\bibinfo
  {journal} {Computer Physics Communications}\ }\textbf {\bibinfo {volume}
  {251}},\ \bibinfo {pages} {107080} (\bibinfo {year} {2020})}\BibitemShut
  {NoStop}%
\bibitem [{\citenamefont {Fritsch}\ and\ \citenamefont
  {Carlson}(1980)}]{fritsch1980monotone}%
  \BibitemOpen
  \bibfield  {author} {\bibinfo {author} {\bibfnamefont {F.~N.}\ \bibnamefont
  {Fritsch}}\ and\ \bibinfo {author} {\bibfnamefont {R.~E.}\ \bibnamefont
  {Carlson}},\ }\bibfield  {title} {\bibinfo {title} {Monotone piecewise cubic
  interpolation},\ }\href {https://doi.org/10.1137/0717021} {\bibfield
  {journal} {\bibinfo  {journal} {SIAM Journal on Numerical Analysis}\ }\textbf
  {\bibinfo {volume} {17}},\ \bibinfo {pages} {238} (\bibinfo {year}
  {1980})}\BibitemShut {NoStop}%
\bibitem [{\citenamefont {Bersuker}(2006)}]{bersuker2006jahn}%
  \BibitemOpen
  \bibfield  {author} {\bibinfo {author} {\bibfnamefont {I.~B.}\ \bibnamefont
  {Bersuker}},\ }\href@noop {} {\emph {\bibinfo {title} {The {Jahn--Teller}
  Effect}}}\ (\bibinfo  {publisher} {Cambridge University Press},\ \bibinfo
  {year} {2006})\BibitemShut {NoStop}%
\bibitem [{\citenamefont {Bersuker}(2021)}]{bersuker2020jahn}%
  \BibitemOpen
  \bibfield  {author} {\bibinfo {author} {\bibfnamefont {I.~B.}\ \bibnamefont
  {Bersuker}},\ }\bibfield  {title} {\bibinfo {title} {{Jahn--Teller} and
  pseudo-{Jahn--Teller} effects: from particular features to general tools in
  exploring molecular and solid state properties},\ }\href@noop {} {\bibfield
  {journal} {\bibinfo  {journal} {Chemical Reviews}\ }\textbf {\bibinfo
  {volume} {121}},\ \bibinfo {pages} {1463} (\bibinfo {year}
  {2021})}\BibitemShut {NoStop}%
\bibitem [{\citenamefont
  {Rodr{\'\i}guez-Carvajal}(1993)}]{rodriguez1993recent}%
  \BibitemOpen
  \bibfield  {author} {\bibinfo {author} {\bibfnamefont {J.}~\bibnamefont
  {Rodr{\'\i}guez-Carvajal}},\ }\bibfield  {title} {\bibinfo {title} {Recent
  advances in magnetic structure determination by neutron powder diffraction},\
  }\href@noop {} {\bibfield  {journal} {\bibinfo  {journal} {Physica B:
  Condensed Matter}\ }\textbf {\bibinfo {volume} {192}},\ \bibinfo {pages} {55}
  (\bibinfo {year} {1993})}\BibitemShut {NoStop}%
\bibitem [{\citenamefont {Guo}\ \emph {et~al.}(2025)\citenamefont {Guo},
  \citenamefont {Wang}, \citenamefont {Niu}, \citenamefont {Ruan},
  \citenamefont {Yang}, \citenamefont {Wang}, \citenamefont {Wang},
  \citenamefont {Wang}, \citenamefont {Jiang}, \citenamefont {Li} \emph
  {et~al.}}]{guo2025ultra}%
  \BibitemOpen
  \bibfield  {author} {\bibinfo {author} {\bibfnamefont {M.}~\bibnamefont
  {Guo}}, \bibinfo {author} {\bibfnamefont {C.}~\bibnamefont {Wang}}, \bibinfo
  {author} {\bibfnamefont {Y.}~\bibnamefont {Niu}}, \bibinfo {author}
  {\bibfnamefont {M.}~\bibnamefont {Ruan}}, \bibinfo {author} {\bibfnamefont
  {D.}~\bibnamefont {Yang}}, \bibinfo {author} {\bibfnamefont {F.}~\bibnamefont
  {Wang}}, \bibinfo {author} {\bibfnamefont {H.}~\bibnamefont {Wang}}, \bibinfo
  {author} {\bibfnamefont {N.}~\bibnamefont {Wang}}, \bibinfo {author}
  {\bibfnamefont {Y.}~\bibnamefont {Jiang}}, \bibinfo {author} {\bibfnamefont
  {T.}~\bibnamefont {Li}}, \emph {et~al.},\ }\bibfield  {title} {\bibinfo
  {title} {Ultra-fast charging high-voltage spinel
  {LiNi$_{0.5}$Mn$_{1.5}$O$_4$} batteries enabled by {Mn--O} bond regulating
  strategy to defeat {Jahn--Teller} distortion},\ }\href@noop {} {\bibfield
  {journal} {\bibinfo  {journal} {Advanced Energy Materials}\ }\textbf
  {\bibinfo {volume} {15}},\ \bibinfo {pages} {2502226} (\bibinfo {year}
  {2025})}\BibitemShut {NoStop}%
\bibitem [{\citenamefont {Lufaso}\ and\ \citenamefont
  {Woodward}(2004)}]{lufaso2004jahn}%
  \BibitemOpen
  \bibfield  {author} {\bibinfo {author} {\bibfnamefont {M.~W.}\ \bibnamefont
  {Lufaso}}\ and\ \bibinfo {author} {\bibfnamefont {P.~M.}\ \bibnamefont
  {Woodward}},\ }\bibfield  {title} {\bibinfo {title} {{Jahn--Teller}
  distortions, cation ordering and octahedral tilting in perovskites},\
  }\href@noop {} {\bibfield  {journal} {\bibinfo  {journal} {Acta
  Crystallographica Section B}\ }\textbf {\bibinfo {volume} {60}},\ \bibinfo
  {pages} {10} (\bibinfo {year} {2004})}\BibitemShut {NoStop}%
\bibitem [{\citenamefont {Kanagawa}\ \emph {et~al.}(2024)\citenamefont
  {Kanagawa}, \citenamefont {Hirai},\ and\ \citenamefont
  {Hirose}}]{kanagawa2024first}%
  \BibitemOpen
  \bibfield  {author} {\bibinfo {author} {\bibfnamefont {T.}~\bibnamefont
  {Kanagawa}}, \bibinfo {author} {\bibfnamefont {D.}~\bibnamefont {Hirai}},\
  and\ \bibinfo {author} {\bibfnamefont {S.}~\bibnamefont {Hirose}},\
  }\bibfield  {title} {\bibinfo {title} {First-principles calculations on
  charge states and solubility of impurity defects in {BaTiO$_3$}},\
  }\href@noop {} {\bibfield  {journal} {\bibinfo  {journal} {Journal of Applied
  Physics}\ }\textbf {\bibinfo {volume} {136}},\ \bibinfo {pages} {034103}
  (\bibinfo {year} {2024})}\BibitemShut {NoStop}%
\bibitem [{\citenamefont {Bhide}\ and\ \citenamefont
  {Hegde}(1972)}]{bhide1972mossbauer}%
  \BibitemOpen
  \bibfield  {author} {\bibinfo {author} {\bibfnamefont {V.~G.}\ \bibnamefont
  {Bhide}}\ and\ \bibinfo {author} {\bibfnamefont {M.~S.}\ \bibnamefont
  {Hegde}},\ }\bibfield  {title} {\bibinfo {title} {{M{\"o}ssbauer} effect for
  {$^{57}$Fe} in ferroelectric lead titanate},\ }\href@noop {} {\bibfield
  {journal} {\bibinfo  {journal} {Physical Review B}\ }\textbf {\bibinfo
  {volume} {5}},\ \bibinfo {pages} {3488} (\bibinfo {year} {1972})}\BibitemShut
  {NoStop}%
\bibitem [{\citenamefont {Ihrig}(1978)}]{ihrig1978phase}%
  \BibitemOpen
  \bibfield  {author} {\bibinfo {author} {\bibfnamefont {H.}~\bibnamefont
  {Ihrig}},\ }\bibfield  {title} {\bibinfo {title} {The phase stability of
  {BaTiO$_3$} as a function of doped 3d elements: an experimental study},\
  }\href@noop {} {\bibfield  {journal} {\bibinfo  {journal} {Journal of Physics
  C: Solid State Physics}\ }\textbf {\bibinfo {volume} {11}},\ \bibinfo {pages}
  {819} (\bibinfo {year} {1978})}\BibitemShut {NoStop}%
\bibitem [{\citenamefont {Klein}\ and\ \citenamefont
  {Schwartz}(1986)}]{klein1986photorefractive}%
  \BibitemOpen
  \bibfield  {author} {\bibinfo {author} {\bibfnamefont {M.~B.}\ \bibnamefont
  {Klein}}\ and\ \bibinfo {author} {\bibfnamefont {R.~N.}\ \bibnamefont
  {Schwartz}},\ }\bibfield  {title} {\bibinfo {title} {Photorefractive effect
  in {BaTiO$_3$}: microscopic origins},\ }\href@noop {} {\bibfield  {journal}
  {\bibinfo  {journal} {Journal of the Optical Society of America B}\ }\textbf
  {\bibinfo {volume} {3}},\ \bibinfo {pages} {293} (\bibinfo {year}
  {1986})}\BibitemShut {NoStop}%
\bibitem [{\citenamefont {Mazur}\ \emph {et~al.}(1997)\citenamefont {Mazur},
  \citenamefont {Schirmer},\ and\ \citenamefont
  {Mendricks}}]{mazur1997optical}%
  \BibitemOpen
  \bibfield  {author} {\bibinfo {author} {\bibfnamefont {A.}~\bibnamefont
  {Mazur}}, \bibinfo {author} {\bibfnamefont {O.~F.}\ \bibnamefont
  {Schirmer}},\ and\ \bibinfo {author} {\bibfnamefont {S.}~\bibnamefont
  {Mendricks}},\ }\bibfield  {title} {\bibinfo {title} {Optical absorption and
  light-induced charge transport of {Fe$^{2+}$} in {BaTiO$_3$}},\ }\href@noop
  {} {\bibfield  {journal} {\bibinfo  {journal} {Applied Physics Letters}\
  }\textbf {\bibinfo {volume} {70}},\ \bibinfo {pages} {2395} (\bibinfo {year}
  {1997})}\BibitemShut {NoStop}%
\bibitem [{\citenamefont {Chakraborty}\ \emph {et~al.}(2013)\citenamefont
  {Chakraborty}, \citenamefont {Meneghini}, \citenamefont {Aquilanti},\ and\
  \citenamefont {Ray}}]{chakraborty2013microscopic}%
  \BibitemOpen
  \bibfield  {author} {\bibinfo {author} {\bibfnamefont {T.}~\bibnamefont
  {Chakraborty}}, \bibinfo {author} {\bibfnamefont {C.}~\bibnamefont
  {Meneghini}}, \bibinfo {author} {\bibfnamefont {G.}~\bibnamefont
  {Aquilanti}},\ and\ \bibinfo {author} {\bibfnamefont {S.}~\bibnamefont
  {Ray}},\ }\bibfield  {title} {\bibinfo {title} {Microscopic distribution of
  metal dopants and anion vacancies in {Fe}-doped {BaTiO$_{3-\delta}$} single
  crystals},\ }\href@noop {} {\bibfield  {journal} {\bibinfo  {journal}
  {Journal of Physics: Condensed Matter}\ }\textbf {\bibinfo {volume} {25}},\
  \bibinfo {pages} {236002} (\bibinfo {year} {2013})}\BibitemShut {NoStop}%
\bibitem [{\citenamefont {Luo}\ \emph {et~al.}(2018)\citenamefont {Luo},
  \citenamefont {Wang}, \citenamefont {Tian}, \citenamefont {Song},
  \citenamefont {Zhao}, \citenamefont {Cai}, \citenamefont {Feng},\ and\
  \citenamefont {Li}}]{luo2018giant}%
  \BibitemOpen
  \bibfield  {author} {\bibinfo {author} {\bibfnamefont {B.}~\bibnamefont
  {Luo}}, \bibinfo {author} {\bibfnamefont {X.}~\bibnamefont {Wang}}, \bibinfo
  {author} {\bibfnamefont {E.}~\bibnamefont {Tian}}, \bibinfo {author}
  {\bibfnamefont {H.}~\bibnamefont {Song}}, \bibinfo {author} {\bibfnamefont
  {Q.}~\bibnamefont {Zhao}}, \bibinfo {author} {\bibfnamefont {Z.}~\bibnamefont
  {Cai}}, \bibinfo {author} {\bibfnamefont {W.}~\bibnamefont {Feng}},\ and\
  \bibinfo {author} {\bibfnamefont {L.}~\bibnamefont {Li}},\ }\bibfield
  {title} {\bibinfo {title} {Giant permittivity and low dielectric loss of {Fe}
  doped {BaTiO$_3$} ceramics: Experimental and first-principles calculations},\
  }\href@noop {} {\bibfield  {journal} {\bibinfo  {journal} {Journal of the
  European Ceramic Society}\ }\textbf {\bibinfo {volume} {38}},\ \bibinfo
  {pages} {1562} (\bibinfo {year} {2018})}\BibitemShut {NoStop}%
\bibitem [{\citenamefont {Goodenough}(2004)}]{goodenough2004electronic}%
  \BibitemOpen
  \bibfield  {author} {\bibinfo {author} {\bibfnamefont {J.~B.}\ \bibnamefont
  {Goodenough}},\ }\bibfield  {title} {\bibinfo {title} {Electronic and ionic
  transport properties and other physical aspects of perovskites},\ }\href
  {https://doi.org/10.1088/0034-4885/67/11/R01} {\bibfield  {journal} {\bibinfo
   {journal} {Reports on Progress in Physics}\ }\textbf {\bibinfo {volume}
  {67}},\ \bibinfo {pages} {1915} (\bibinfo {year} {2004})}\BibitemShut
  {NoStop}%
\bibitem [{\citenamefont {Ren}\ \emph {et~al.}(1994)\citenamefont {Ren},
  \citenamefont {Ishida},\ and\ \citenamefont {Mineta}}]{ren1994effect}%
  \BibitemOpen
  \bibfield  {author} {\bibinfo {author} {\bibfnamefont {F.}~\bibnamefont
  {Ren}}, \bibinfo {author} {\bibfnamefont {S.}~\bibnamefont {Ishida}},\ and\
  \bibinfo {author} {\bibfnamefont {S.}~\bibnamefont {Mineta}},\ }\bibfield
  {title} {\bibinfo {title} {Effect of manganese addition on phase stability of
  hexagonal {BaTiO$_3$}},\ }\href {https://doi.org/10.2109/jcersj.102.105}
  {\bibfield  {journal} {\bibinfo  {journal} {Journal of the Ceramic Society of
  Japan}\ }\textbf {\bibinfo {volume} {102}},\ \bibinfo {pages} {105} (\bibinfo
  {year} {1994})}\BibitemShut {NoStop}%
\bibitem [{\citenamefont {Glaister}\ and\ \citenamefont
  {Kay}(1960)}]{glaister1960investigation}%
  \BibitemOpen
  \bibfield  {author} {\bibinfo {author} {\bibfnamefont {R.~M.}\ \bibnamefont
  {Glaister}}\ and\ \bibinfo {author} {\bibfnamefont {H.~F.}\ \bibnamefont
  {Kay}},\ }\bibfield  {title} {\bibinfo {title} {An investigation of the
  cubic-hexagonal transition in barium titanate},\ }\href
  {https://doi.org/10.1088/0370-1328/76/5/317} {\bibfield  {journal} {\bibinfo
  {journal} {Proceedings of the Physical Society}\ }\textbf {\bibinfo {volume}
  {76}},\ \bibinfo {pages} {763} (\bibinfo {year} {1960})}\BibitemShut
  {NoStop}%
\bibitem [{\citenamefont {Dickson}\ \emph {et~al.}(1961)\citenamefont
  {Dickson}, \citenamefont {Katz},\ and\ \citenamefont
  {Ward}}]{dickinson1961compounds}%
  \BibitemOpen
  \bibfield  {author} {\bibinfo {author} {\bibfnamefont {J.~G.}\ \bibnamefont
  {Dickson}}, \bibinfo {author} {\bibfnamefont {L.}~\bibnamefont {Katz}},\ and\
  \bibinfo {author} {\bibfnamefont {R.}~\bibnamefont {Ward}},\ }\bibfield
  {title} {\bibinfo {title} {Compounds with the hexagonal barium titanate
  structure},\ }\href {https://doi.org/10.1021/ja01475a012} {\bibfield
  {journal} {\bibinfo  {journal} {Journal of the American Chemical Society}\
  }\textbf {\bibinfo {volume} {83}},\ \bibinfo {pages} {3026} (\bibinfo {year}
  {1961})}\BibitemShut {NoStop}%
\bibitem [{\citenamefont {B{\"o}ttcher}\ \emph {et~al.}(2008)\citenamefont
  {B{\"o}ttcher}, \citenamefont {Langhammer}, \citenamefont {M{\"u}ller},\ and\
  \citenamefont {Abicht}}]{bottcher2008phase}%
  \BibitemOpen
  \bibfield  {author} {\bibinfo {author} {\bibfnamefont {R.}~\bibnamefont
  {B{\"o}ttcher}}, \bibinfo {author} {\bibfnamefont {H.~T.}\ \bibnamefont
  {Langhammer}}, \bibinfo {author} {\bibfnamefont {T.}~\bibnamefont
  {M{\"u}ller}},\ and\ \bibinfo {author} {\bibfnamefont {H.-P.}\ \bibnamefont
  {Abicht}},\ }\bibfield  {title} {\bibinfo {title} {{3C}--{6H} phase
  transition in {BaTiO$_3$} induced by {Fe} ions: an electron paramagnetic
  resonance study},\ }\href {https://doi.org/10.1088/0953-8984/20/50/505209}
  {\bibfield  {journal} {\bibinfo  {journal} {Journal of Physics: Condensed
  Matter}\ }\textbf {\bibinfo {volume} {20}},\ \bibinfo {pages} {505209}
  (\bibinfo {year} {2008})}\BibitemShut {NoStop}%
\bibitem [{\citenamefont {Islam}\ \emph {et~al.}(2019)\citenamefont {Islam},
  \citenamefont {Momin},\ and\ \citenamefont {Nesa}}]{islam2019effect}%
  \BibitemOpen
  \bibfield  {author} {\bibinfo {author} {\bibfnamefont {M.~A.}\ \bibnamefont
  {Islam}}, \bibinfo {author} {\bibfnamefont {M.~A.}\ \bibnamefont {Momin}},\
  and\ \bibinfo {author} {\bibfnamefont {M.}~\bibnamefont {Nesa}},\ }\bibfield
  {title} {\bibinfo {title} {Effect of {Fe} doping on the structural, optical
  and electronic properties of {BaTiO$_3$}: {DFT} based calculation},\
  }\href@noop {} {\bibfield  {journal} {\bibinfo  {journal} {Chinese Journal of
  Physics}\ }\textbf {\bibinfo {volume} {60}},\ \bibinfo {pages} {731}
  (\bibinfo {year} {2019})}\BibitemShut {NoStop}%
\bibitem [{\citenamefont {Kotiuga}\ \emph {et~al.}(2019)\citenamefont
  {Kotiuga}, \citenamefont {Zhang}, \citenamefont {Li}, \citenamefont
  {Rodolakis}, \citenamefont {Zhou}, \citenamefont {Sutarto}, \citenamefont
  {He}, \citenamefont {Wang}, \citenamefont {Sun}, \citenamefont {Wang} \emph
  {et~al.}}]{kotiuga2019carrier}%
  \BibitemOpen
  \bibfield  {author} {\bibinfo {author} {\bibfnamefont {M.}~\bibnamefont
  {Kotiuga}}, \bibinfo {author} {\bibfnamefont {Z.}~\bibnamefont {Zhang}},
  \bibinfo {author} {\bibfnamefont {J.}~\bibnamefont {Li}}, \bibinfo {author}
  {\bibfnamefont {F.}~\bibnamefont {Rodolakis}}, \bibinfo {author}
  {\bibfnamefont {H.}~\bibnamefont {Zhou}}, \bibinfo {author} {\bibfnamefont
  {R.}~\bibnamefont {Sutarto}}, \bibinfo {author} {\bibfnamefont
  {F.}~\bibnamefont {He}}, \bibinfo {author} {\bibfnamefont {Q.}~\bibnamefont
  {Wang}}, \bibinfo {author} {\bibfnamefont {Y.}~\bibnamefont {Sun}}, \bibinfo
  {author} {\bibfnamefont {Y.}~\bibnamefont {Wang}}, \emph {et~al.},\
  }\bibfield  {title} {\bibinfo {title} {Carrier localization in perovskite
  nickelates from oxygen vacancies},\ }\href@noop {} {\bibfield  {journal}
  {\bibinfo  {journal} {Proceedings of the National Academy of Sciences}\
  }\textbf {\bibinfo {volume} {116}},\ \bibinfo {pages} {21992} (\bibinfo
  {year} {2019})}\BibitemShut {NoStop}%
\bibitem [{\citenamefont {Li}\ \emph {et~al.}(2026{\natexlab{a}})\citenamefont
  {Li}, \citenamefont {Zerdoumi}, \citenamefont {Wang}, \citenamefont {Xie},\
  and\ \citenamefont {Zhang}}]{li2026redox}%
  \BibitemOpen
  \bibfield  {author} {\bibinfo {author} {\bibfnamefont {Z.}~\bibnamefont
  {Li}}, \bibinfo {author} {\bibfnamefont {H.}~\bibnamefont {Zerdoumi}},
  \bibinfo {author} {\bibfnamefont {H.}~\bibnamefont {Wang}}, \bibinfo {author}
  {\bibfnamefont {R.}~\bibnamefont {Xie}},\ and\ \bibinfo {author}
  {\bibfnamefont {H.}~\bibnamefont {Zhang}},\ }\bibfield  {title} {\bibinfo
  {title} {Redox behaviour of {Fe} impurities in {BaTiO}$_3$ based on many-body
  calculations},\ }\href@noop {} {\bibfield  {journal} {\bibinfo  {journal}
  {arXiv preprint arXiv:2605.24747}\ } (\bibinfo {year}
  {2026}{\natexlab{a}})}\BibitemShut {NoStop}%
\bibitem [{\citenamefont {Li}\ \emph {et~al.}(2026{\natexlab{b}})\citenamefont
  {Li}, \citenamefont {Xie},\ and\ \citenamefont {Zhang}}]{li2026dataset}%
  \BibitemOpen
  \bibfield  {author} {\bibinfo {author} {\bibfnamefont {Z.}~\bibnamefont
  {Li}}, \bibinfo {author} {\bibfnamefont {R.}~\bibnamefont {Xie}},\ and\
  \bibinfo {author} {\bibfnamefont {H.}~\bibnamefont {Zhang}},\ }\href
  {https://doi.org/10.5281/zenodo.22228797} {\bibinfo {title} {{Origin of the
  tetragonal-to-hexagonal phase transitions in Fe-doped BaTiO$_3$: structures,
  inputs and total energies}}} (\bibinfo {year} {2026}{\natexlab{b}}),\
  \bibinfo {note} {{Zenodo},
  \href{https://doi.org/10.5281/zenodo.22228797}{doi:10.5281/zenodo.22228797}}\BibitemShut
  {NoStop}%
\end{thebibliography}%


\begin{thebibliography}{10}

\bibitem{ren2004large}
Xiaobing Ren.
\newblock Large electric-field-induced strain in ferroelectric crystals by
  point-defect-mediated reversible domain switching.
\newblock {\em Nature Materials}, 3(2):91--94, 2004.

\bibitem{mestric2005iron}
H.~Me{\v s}tri{\'c}, R.-A. Eichel, T.~Kloss, K.-P. Dinse, So. Laubach, St.
  Laubach, P.~C. Schmidt, K.~A. Sch{\"o}nau, M.~Knapp, and H.~Ehrenberg.
\newblock Iron-oxygen vacancy defect centers in {PbTiO$_3$}: {Newman}
  superposition model analysis and density functional calculations.
\newblock {\em Physical Review B}, 71(13):134109, 2005.

\bibitem{erhart2013formation}
Paul Erhart, Petra Tr{\"a}skelin, and Karsten Albe.
\newblock Formation and switching of defect dipoles in acceptor-doped lead
  titanate: A kinetic model based on first-principles calculations.
\newblock {\em Physical Review B}, 88(2):024107, 2013.

\bibitem{noguchi2020successive}
Yuji Noguchi, Yuki Taniguchi, Ryotaro Inoue, and Masaru Miyayama.
\newblock Successive redox-mediated visible-light ferrophotovoltaics.
\newblock {\em Nature Communications}, 11(1):966, 2020.

\bibitem{adeagbo2019theoretical}
Waheed~A Adeagbo, Hichem Ben~Hamed, Sanjeev~K Nayak, Rolf B{\"o}ttcher, Hans~T
  Langhammer, and Wolfram Hergert.
\newblock Theoretical investigation of iron incorporation in hexagonal barium
  titanate.
\newblock {\em Physical Review B}, 100(18):184108, 2019.

\bibitem{bottcher2018incorporation}
R~B{\"o}ttcher, H.~T. Langhammer, S~K{\"u}cker, C~Eisenschmidt, and S.~G.
  Ebbinghaus.
\newblock On the incorporation of iron into hexagonal barium titanate: {I.
  Electron} paramagnetic resonance ({EPR}) study.
\newblock {\em Journal of Physics: Condensed Matter}, 30(42):425701, 2018.

\bibitem{langhammer2020incorporation}
H~T Langhammer, T~Walther, R~B{\"o}ttcher, and S~G Ebbinghaus.
\newblock On the incorporation of iron into hexagonal barium titanate: {II.
  Magnetic} moment, electron paramagnetic resonance ({EPR}) and optical
  transmission.
\newblock {\em Journal of Physics: Condensed Matter}, 32(38):385702, 2020.

\bibitem{dawson2013phase}
J.~A. Dawson, C.~L. Freeman, J.~H. Harding, and D.~C. Sinclair.
\newblock Phase stabilisation of hexagonal barium titanate doped with
  transition metals: A computational study.
\newblock {\em Journal of Solid State Chemistry}, 200:310--316, 2013.

\bibitem{grey1998structure}
Ian~E. Grey, Christina Li, Lachlan M.~D. Cranswick, Robert~S. Roth, and
  Terrell~A. Vanderah.
\newblock Structure analysis of the {6H}--{Ba(Ti, Fe$^{3+}$,
  Fe$^{4+}$)O$_{3-\delta}$} solid solution.
\newblock {\em Journal of Solid State Chemistry}, 135(2):312--321, 1998.

\bibitem{grey1998accurate}
Ian~E. Grey, Lachlan M.~D. Cranswick, and Christina Li.
\newblock Accurate site occupancies for light atoms from powder {X}-ray data?
  {Oxygen}/vacancy ordering in {6H-BaFe$_{0.67}$Ti$_{0.33}$O$_{3-\delta}$}
  ($\delta$ = 0.08 and 0.32).
\newblock {\em Journal of Applied Crystallography}, 31(5):692--699, 1998.

\bibitem{chakraborty2013microscopic}
Tanushree Chakraborty, Carlo Meneghini, Giuliana Aquilanti, and Sugata Ray.
\newblock Microscopic distribution of metal dopants and anion vacancies in
  {Fe}-doped {BaTiO$_{3-\delta}$} single crystals.
\newblock {\em Journal of Physics: Condensed Matter}, 25(23):236002, 2013.

\bibitem{valant2013cation}
Matjaz Valant, Iztok Ar{\v{c}}on, Iuliia Mikulska, and Darja Lisjak.
\newblock Cation order--disorder transition in {Fe}-doped {6H-BaTiO$_3$} for
  dilute room-temperature ferromagnetism.
\newblock {\em Chemistry of Materials}, 25(17):3544--3550, 2013.

\bibitem{bottcher2005incorporation}
R~B{\"o}ttcher, E~Erdem, H~T Langhammer, T~M{\"u}ller, and H-P Abicht.
\newblock Incorporation of chromium into hexagonal barium titanate: an electron
  paramagnetic resonance study.
\newblock {\em Journal of Physics: Condensed Matter}, 17(17):2763, 2005.

\bibitem{nayak2015chromium}
Sanjeev~K. Nayak, Hans~T. Langhammer, Waheed~A. Adeagbo, Wolfram Hergert,
  Thomas M{\"u}ller, and Rolf B{\"o}ttcher.
\newblock Chromium point defects in hexagonal {BaTiO$_3$}: A comparative study
  of first-principles calculations and experiments.
\newblock {\em Physical Review B}, 91(15):155105, 2015.

\bibitem{bottcher2008phase}
R.~B{\"o}ttcher, H.~T. Langhammer, T.~M{\"u}ller, and H.-P. Abicht.
\newblock {3C}--{6H} phase transition in {BaTiO$_3$} induced by {Fe} ions: an
  electron paramagnetic resonance study.
\newblock {\em Journal of Physics: Condensed Matter}, 20(50):505209, 2008.

\bibitem{becerro2000transition}
A.~I. Becerro, F.~Seifert, R.~J. Angel, S.~R{\'\i}os, and C.~McCammon.
\newblock Displacive phase transitions and spontaneous strains in oxygen
  deficient {CaFe$_x$Ti$_{1-x}$O$_{3-x/2}$} perovskites ($0 \le x \le 0.40$).
\newblock {\em Journal of Physics: Condensed Matter}, 12(15):3661--3670, 2000.

\bibitem{vracar2007jahn}
M.~Vra{\v{c}}ar, A.~Kuzmin, R.~Merkle, J.~Purans, E.~A. Kotomin, J.~Maier, and
  O.~Mathon.
\newblock {Jahn--Teller} distortion around {Fe$^{4+}$} in
  {Sr(Fe$_x$Ti$_{1-x}$)O$_{3-\delta}$} from x-ray absorption spectroscopy,
  x-ray diffraction, and vibrational spectroscopy.
\newblock {\em Physical Review B}, 76(17):174107, 2007.

\bibitem{mccammon2000short}
C.~A. McCammon, A.~I. Becerro, F.~Langenhorst, R.~J. Angel, S.~Marion, and
  F.~Seifert.
\newblock Short-range ordering of oxygen vacancies in
  {CaFe$_x$Ti$_{1-x}$O$_{3-x/2}$} perovskites ($0 < x < 0.4$).
\newblock {\em Journal of Physics: Condensed Matter}, 12(13):2969--2984, 2000.

\bibitem{waerenborgh2001fe4}
J.~C. Waerenborgh, F.~M. Figueiredo, J.~R. Frade, M.~T. Colomer, and J.~R.
  Jurado.
\newblock {Fe$^{4+}$} content and ordering of anion vacancies in partially
  reduced {AFe$_x$Ti$_{1-x}$O$_{3-y}$} ({A} = {Ca}, {Sr}; $x \le 0.6$)
  perovskites. {An $^{57}$Fe M{\"o}ssbauer} spectroscopy study.
\newblock {\em Journal of Physics: Condensed Matter}, 13(35):8171--8187, 2001.

\bibitem{shannon1976revised}
Robert~D Shannon.
\newblock Revised effective ionic radii and systematic studies of interatomic
  distances in halides and chalcogenides.
\newblock {\em Acta Crystallographica Section A}, 32(5):751--767, 1976.

\end{thebibliography}

\end{document}


\begin{center}
    \Large \textbf{Supplementary Material for:\\ Origin of the tetragonal-to-hexagonal phase transitions in Fe-doped BaTiO$_3$}

    \vspace{0.5cm}

    \large Zhiyuan Li, Ruiwen Xie,  Hongbin Zhang\\

    \date{\today}
\end{center}

\setcounter{section}{0}
\setcounter{figure}{0}
\setcounter{table}{0}
\setcounter{page}{1}

\newcommand{\TODO}[1]{\textcolor{red}{[\textbf{TODO} #1]}}
\newcommand{\VO}{$V_{\mathrm{O}}$}
\newcommand{\mfu}{meV/f.u.}

\section{Configurational averaging}\label{si:averaging}

\begin{figure}[htbp]
\centering
\includegraphics[width=\textwidth]{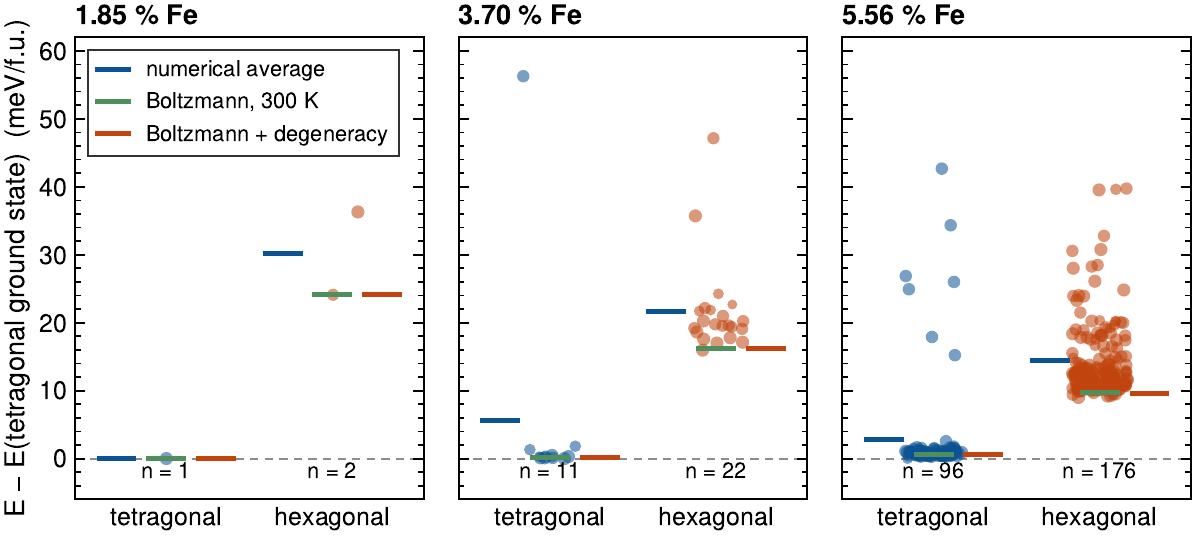}
\caption{All symmetry-inequivalent configurations at 1.85, 3.70 and
5.56\,\% Fe in the tetragonal and hexagonal phases. Marker area scales with the
logarithm of the configurational degeneracy, and points are jittered horizontally.
The three horizontal bars in each column are the three averages.}
\label{fig:si_landscape}
\end{figure}

Besides the degeneracy-weighted average of Eq.~(1) of the main text, two other
averages were evaluated: the simple arithmetic average of all computed
configurations,
\begin{equation}
\langle E \rangle_{\text{simple}} = \frac{1}{N} \sum_{i=1}^{N} E_i ,
\end{equation}
where $N$ is the total number of configurations, and the Boltzmann average
without degeneracies,
\begin{equation}
\langle E \rangle_{\text{Boltzmann}} = \frac{\sum_{i} e^{-E_i/(k_B T)} \cdot E_i}{\sum_{i} e^{-E_i/(k_B T)}} ,
\end{equation}
at the same $T = 300$~K.
Table~\ref{tab:deltaE} gives the three averages for each composition, and Fig.~\ref{fig:si_landscape} shows the distribution of configuration-specific total energies. The arithmetic mean overweights the
high-lying configurations. The two Boltzmann schemes differ little.

\begin{table}[htbp]
\centering
\caption{Averaged total energies of the tetragonal and the hexagonal phase,
$\langle E\rangle_{\mathrm{tet}}$ and $\langle E\rangle_{\mathrm{hex}}$ in eV
per formula unit, and their difference $\Delta E = E_{\mathrm{hex}} -
E_{\mathrm{tet}}$ in \mfu, under the arithmetic mean, the Boltzmann average at
300~K and the degeneracy-weighted Boltzmann average at 300~K. Both supercells
contain 54 formula units (270 atoms, 269 with a vacancy), and the supercell
energy is divided by 54. $N$ is the number of symmetry-inequivalent
configurations (tetragonal / hexagonal).}
\label{tab:deltaE}
\begin{tabular}{@{}llclccc@{}}
\toprule
& Fe & $N$ & scheme & $\langle E\rangle_{\mathrm{tet}}$ & $\langle E\rangle_{\mathrm{hex}}$ & $\Delta E$ \\
\midrule
BaTiO$_3$
& 0\,\%    & 1 / 1    & all three     & $-39.9369$ & $-39.9126$ & 24.28 \\[2pt]
& 1.85\,\% & 1 / 2    & arithmetic    & $-39.7895$ & $-39.7593$ & 30.20 \\
&          &          & Boltzmann     & $-39.7895$ & $-39.7654$ & 24.10 \\
&          &          & deg.-weighted & $-39.7895$ & $-39.7654$ & 24.10 \\[2pt]
& 3.70\,\% & 11 / 22  & arithmetic    & $-39.6350$ & $-39.6189$ & 16.17 \\
&          &          & Boltzmann     & $-39.6404$ & $-39.6243$ & 16.06 \\
&          &          & deg.-weighted & $-39.6404$ & $-39.6243$ & 16.07 \\[2pt]
& 5.56\,\% & 96 / 176 & arithmetic    & $-39.4883$ & $-39.4766$ & 11.70 \\
&          &          & Boltzmann     & $-39.4905$ & $-39.4814$ &  9.09 \\
&          &          & deg.-weighted & $-39.4905$ & $-39.4815$ &  9.04 \\
\midrule
BaTiO$_3$ + \VO
& 1.85\,\% & 3 / 3    & arithmetic    & $-39.6630$ & $-39.6398$ & 23.28 \\
&          &          & Boltzmann     & $-39.6662$ & $-39.6472$ & 19.05 \\
&          &          & deg.-weighted & $-39.6662$ & $-39.6472$ & 19.05 \\[2pt]
& 3.70\,\% & 68 / 104 & arithmetic    & $-39.5358$ & $-39.5166$ & 19.20 \\
&          &          & Boltzmann     & $-39.5421$ & $-39.5304$ & 11.70 \\
&          &          & deg.-weighted & $-39.5421$ & $-39.5304$ & 11.64 \\
\bottomrule
\end{tabular}
\end{table}

\section{Sensitivity to the Hubbard $U$}\label{si:hubbard}

\begin{figure}[htbp]
\centering
\includegraphics[width=\textwidth]{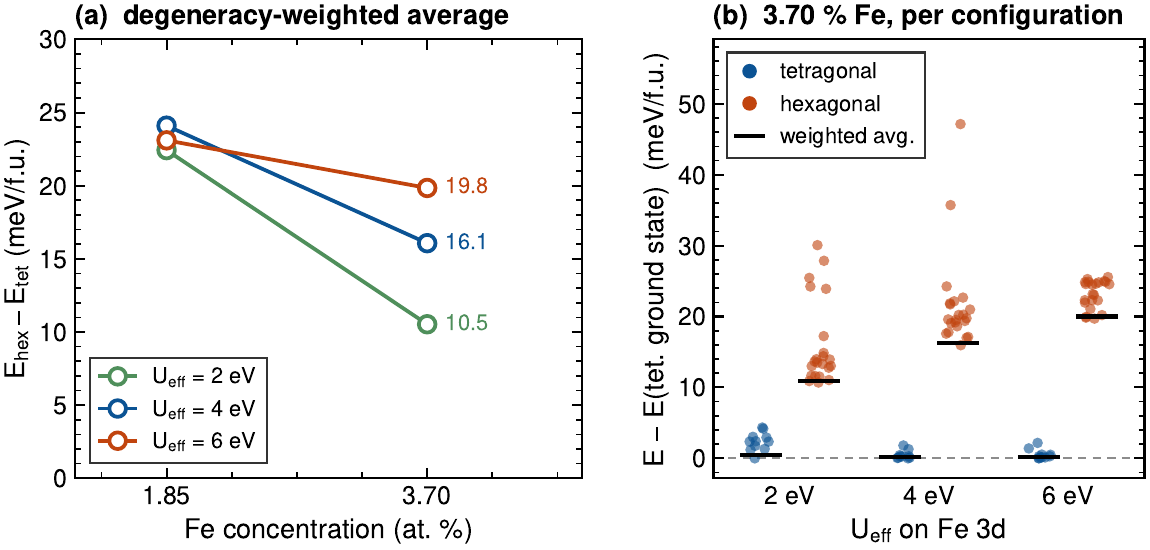}
\caption{(a) $\Delta E$ against Fe content at $U_{\mathrm{eff}} = 2$, 4 and
6~eV. (b) The configuration-resolved energies at 3.70\,\% Fe.}
\label{fig:si_hubbard}
\end{figure}

The ensembles at 1.85 and 3.70\,\% Fe were recomputed at $U_{\mathrm{eff}} = 2$ and 6~eV, each run
seeded from the converged $U = 4$~eV structure of the same configuration. $\Delta E$ falls from 24.3~meV/f.u.\ in the pristine host to 22.4 and 10.5~meV/f.u.\ at 1.85 and 3.70\,\% Fe with $U_{\mathrm{eff}} = 2$~eV, to 24.1 and 16.1~meV/f.u.\ with 4~eV, and to 23.1 and 19.9~meV/f.u.\ with 6~eV (Fig.~\ref{fig:si_hubbard}). $\Delta E$ decreases with the Fe content for all three values of $U$.

\section{Free energy under each averaging scheme}\label{si:freeenergy}

\begin{figure}[htbp]
\centering
\includegraphics[width=\textwidth]{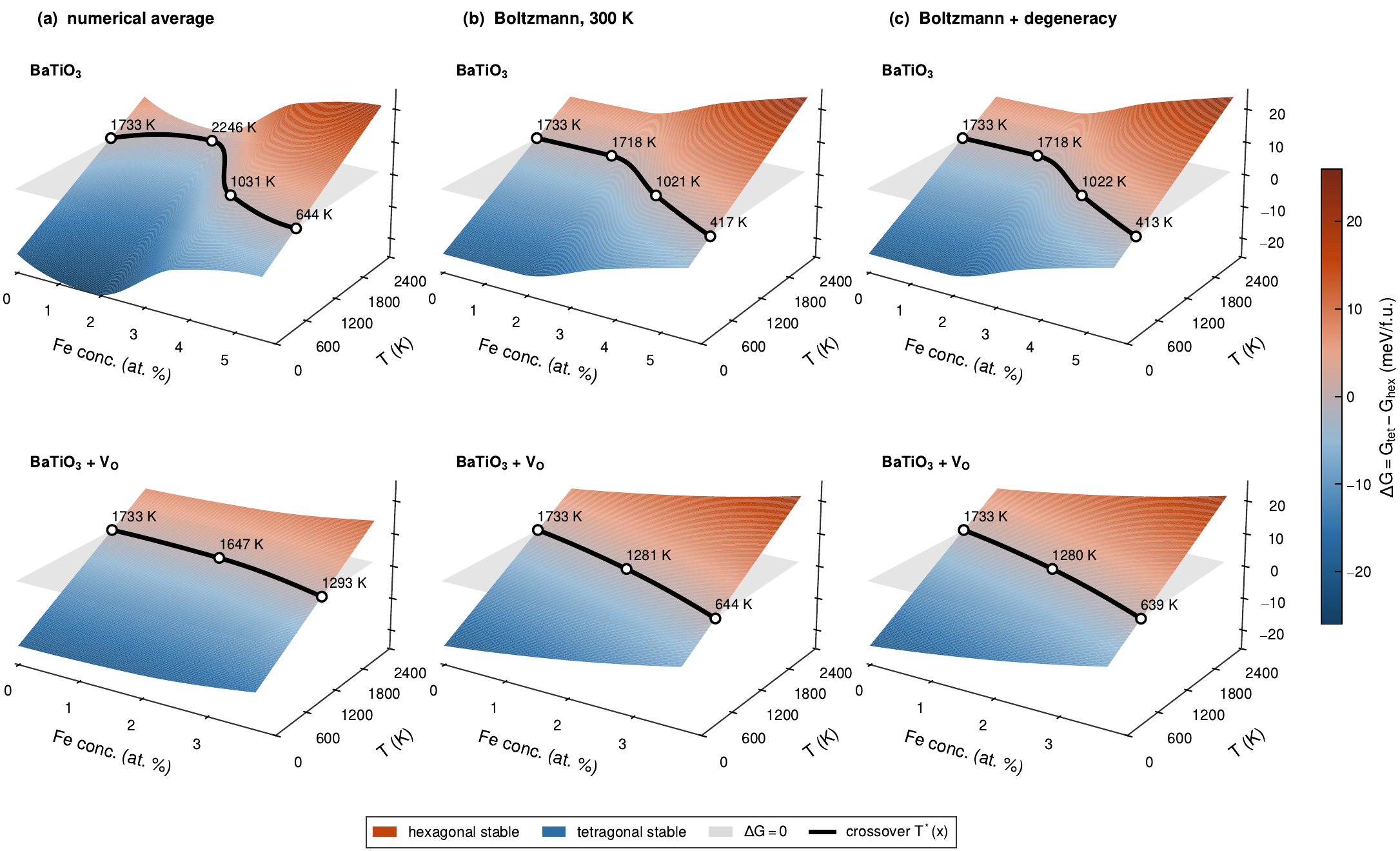}
\caption{$G_{\mathrm{tet}} - G_{\mathrm{hex}}$, that is $-\Delta G(x,T)$ of
Eq.~(2) of the main text, under each averaging scheme, for BaTiO$_3$ and for
BaTiO$_3$ with an oxygen vacancy.}
\label{fig:si_freeenergy}
\end{figure}

Figure~\ref{fig:si_freeenergy} shows the free-energy surface of Fig.~3 of the main text for each averaging scheme, with $T^{*}(x)$ marked. The two Boltzmann schemes agree to within 5~K, and the arithmetic mean differs from them by up to about 650~K. In every panel $T^{*}$ decreases with Fe content from 1.85 to 5.56\,\% Fe (to 3.70\,\% with a vacancy).

\section{Test of the relaxation method and the supercell shape}\label{si:host}

\begin{figure}[htbp]
\centering
\includegraphics[width=\textwidth]{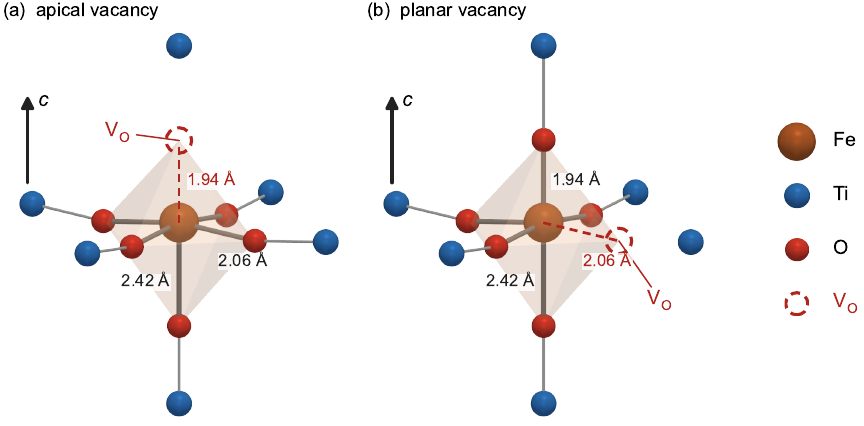}
\caption{The FeO$_6$ octahedron of tetragonal BaTiO$_3$ with
1.85\,\% Fe, relaxed without a vacancy, with the oxygen that is removed drawn
as a dashed circle ($V_{\mathrm{O}}$): (a) the apical vacancy, on the short
Fe--O bond along $c$, and (b) a planar vacancy. The Fe--O bond lengths of the
cell without a vacancy are given; the six nearest Ti are shown.}
\label{fig:si_tilt}
\end{figure}

For an acceptor in tetragonal BaTiO$_3$ and PbTiO$_3$, measurement and calculation agree that
the compensating oxygen vacancy occupies the apical site nearest to the
impurity~\cite{ren2004large, mestric2005iron, erhart2013formation,
noguchi2020successive}. The apical site nearest Fe (apical-near), on the short Fe--O bond along $c$, and a planar site are
drawn in Fig.~\ref{fig:si_tilt}. Our results are consistent with those of Ref.~\cite{noguchi2020successive}. With their relaxation scheme, in which the ions within 6~\AA{} of the defect are relaxed, the apical-near vacancy in the cell of the main text is 238~meV lower than the planar one, and this is the site they report from calculation and experiment. We also compared the two sites in three supercells
and under three relaxation schemes (Table~\ref{tab:ninecell}).

\begin{table}[htbp]
\centering
\caption{$\Delta E = E(\text{apical-near}) - E(\text{planar})$ in meV. Negative
means the apical-near vacancy is the more stable.}
\label{tab:ninecell}
\begin{tabular}{lccc}
\toprule
host freedom & $3\times3\times6$ & $3\times3\times3$
& $3\sqrt{2}\times3\sqrt{2}\times3$ \\
\midrule
ions within 6~\AA{} free~\cite{noguchi2020successive} & $+109$ & $-93$ & $-238$ \\
ions free (ISIF~=~2)            & $+44$  & $-20$ & $+64$  \\
ions and volume free (ISIF~=~8) & $+64$  & $-65$ & $+112$ \\
\bottomrule
\end{tabular}
\end{table}

In the anisotropic $3\times3\times6$ cell, of axial ratio 2.1, the planar site is the lower one under all three relaxation schemes, and in the nearly cubic $3\times3\times3$ cell the apical site. In the $3\sqrt{2}\times3\sqrt{2}\times3$ cell, of axial ratio 0.74 and used in the main text, the apical site is the lower one only when the ions within 6~\AA{} of the
defect are relaxed, as in Ref.~\cite{noguchi2020successive}. The calculations
of Refs.~\cite{mestric2005iron, erhart2013formation} also keep the lattice
parameters fixed. The lowest-energy vacancy site can thus change with the supercell shape
and the relaxation scheme. Considering the experimental and computational studies cited above, and for consistency between the hosts, the calculations of the main text use the scheme
of Ref.~\cite{noguchi2020successive} in all cells.

\section{The FeO$_6$ octahedron on the hexagonal $2a$ site}\label{si:jt}

In the hexagonal cells relaxed here the FeO$_6$ octahedron on the $2a$ site is
regular, $\lambda = 0$, whereas Ref.~\cite{adeagbo2019theoretical} reports a
distorted one for the same site. The relaxation protocols differ, so the two
were compared directly: the six Fe--O bonds were set to the values of that
reference and relaxed again with all symmetry operations switched off. The bonds
return to equal length (Table~\ref{tab:isym}) and the energy rises by 0.45~meV, within the force convergence tolerance.

\begin{table}[htbp]
\centering
\caption{Fe--O bond lengths before and after relaxation with all symmetry
operations switched off, for Fe on the hexagonal $2a$ site.}
\label{tab:isym}
\begin{tabular}{lcccccc c}
\toprule
& \multicolumn{6}{c}{Fe--O (\AA)} & $\lambda$ \\
\midrule
starting & 2.0194 & 2.0394 & 2.0394 & 2.0994 & 2.0994 & 2.1094 & 0.0435 \\
relaxed  & 2.0586 & 2.0589 & 2.0599 & 2.0599 & 2.0600 & 2.0605 & 0.0009 \\
\bottomrule
\end{tabular}
\end{table}

\section{Fe $3d$ fat band and charge distribution}\label{si:fatband}

\begin{figure}[htbp]
\centering
\includegraphics[width=\textwidth]{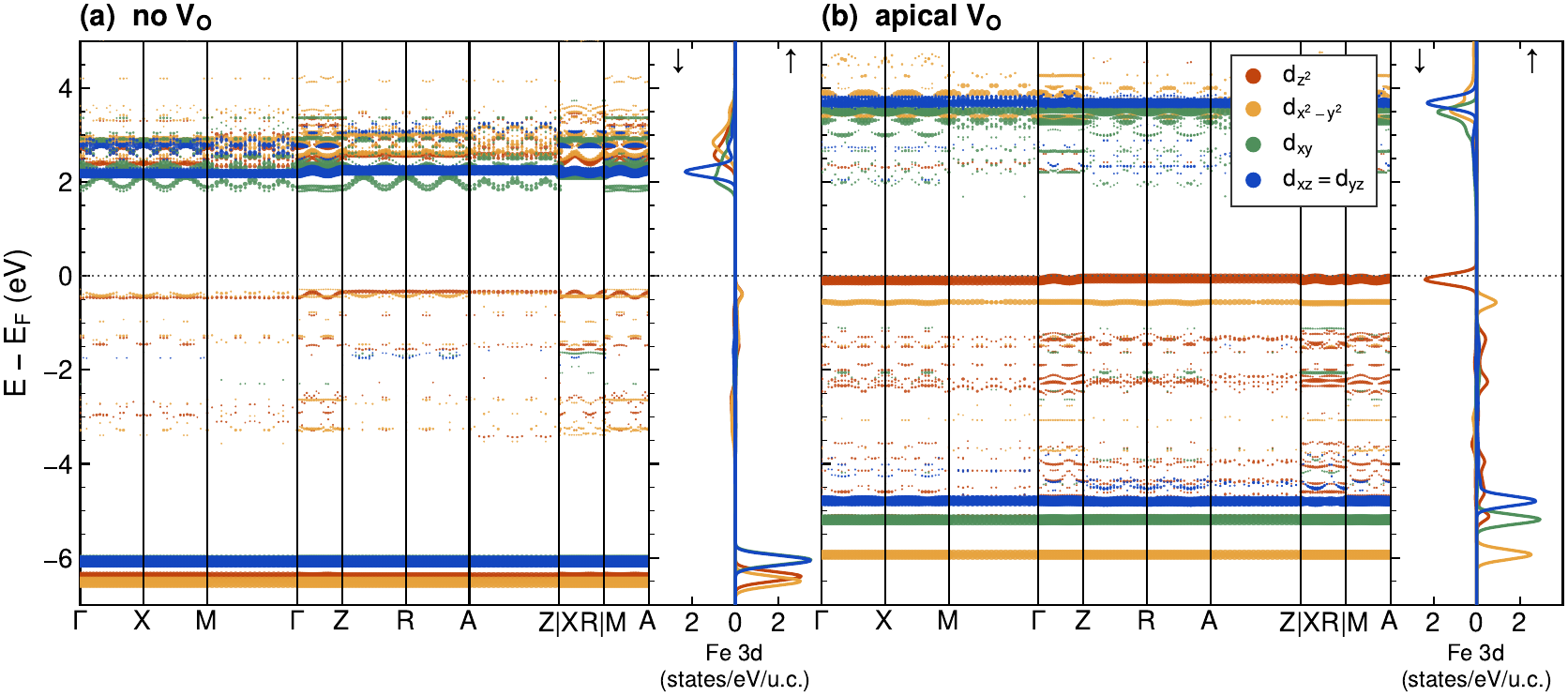}
\caption{Fe $3d$ fat band of tetragonal BaTiO$_3$ with 1.85\,\% Fe and an
apical oxygen vacancy, with the orbital-resolved Fe $3d$ projected density of
states beside it. Marker size is the Fe $3d$ weight. The flat in-gap band just
below $E_{\mathrm{F}}$ is the spin-down $d_{z^2}$ state of main-text
Fig.~5(b).}
\label{fig:si_fatband}
\end{figure}

Table~\ref{tab:charge} gives the charges and moments, integrated in the PAW
spheres, of tetragonal BaTiO$_3$ with 1.85\,\% Fe without and with the apical
vacancy. Without the vacancy the acceptor leaves a hole in the O $2p$ states,
and the O atoms carry a moment of 0.91~$\mu_B$, 0.61~$\mu_B$ of it on the six O
around Fe. With the vacancy the moment on O falls to 0.18~$\mu_B$, the Fe
$3d$ shell gains 0.23~$e$, and the $3d$ shell of the Ti across the vacancy
gains 0.18~$e$. The in-gap state that holds the charge added to Fe is shown in
Fig.~\ref{fig:si_fatband}.

\begin{table}[htbp]
\centering
\caption{Charges and moments, integrated in the PAW spheres, of tetragonal
BaTiO$_3$ with 1.85\,\% Fe without and with the apical vacancy (the cells of
main-text Sec.~III~C).}
\label{tab:charge}
\begin{tabular}{@{}lcc@{}}
\toprule
& no $V_{\mathrm{O}}$ & apical $V_{\mathrm{O}}$ \\
\midrule
Fe $3d$ charge ($e$)                                  & 5.79 & 6.02 \\
Fe moment ($\mu_B$)                                   & 4.21 & 3.75 \\
moment on the O around Fe ($\mu_B$; 6 and 5 O)        & 0.61 & 0.17 \\
moment on all O ($\mu_B$)                             & 0.91 & 0.18 \\
$3d$ charge of the Ti across the vacancy site ($e$)   & 1.91 & 2.09 \\
moment of the cell ($\mu_B$)                          & 5.19 & 3.99 \\
\bottomrule
\end{tabular}
\end{table}

\section{Structures of the SrTiO$_3$ and CaTiO$_3$ hosts}\label{si:cto}

\begin{figure}[htbp]
\centering
\includegraphics[width=\textwidth]{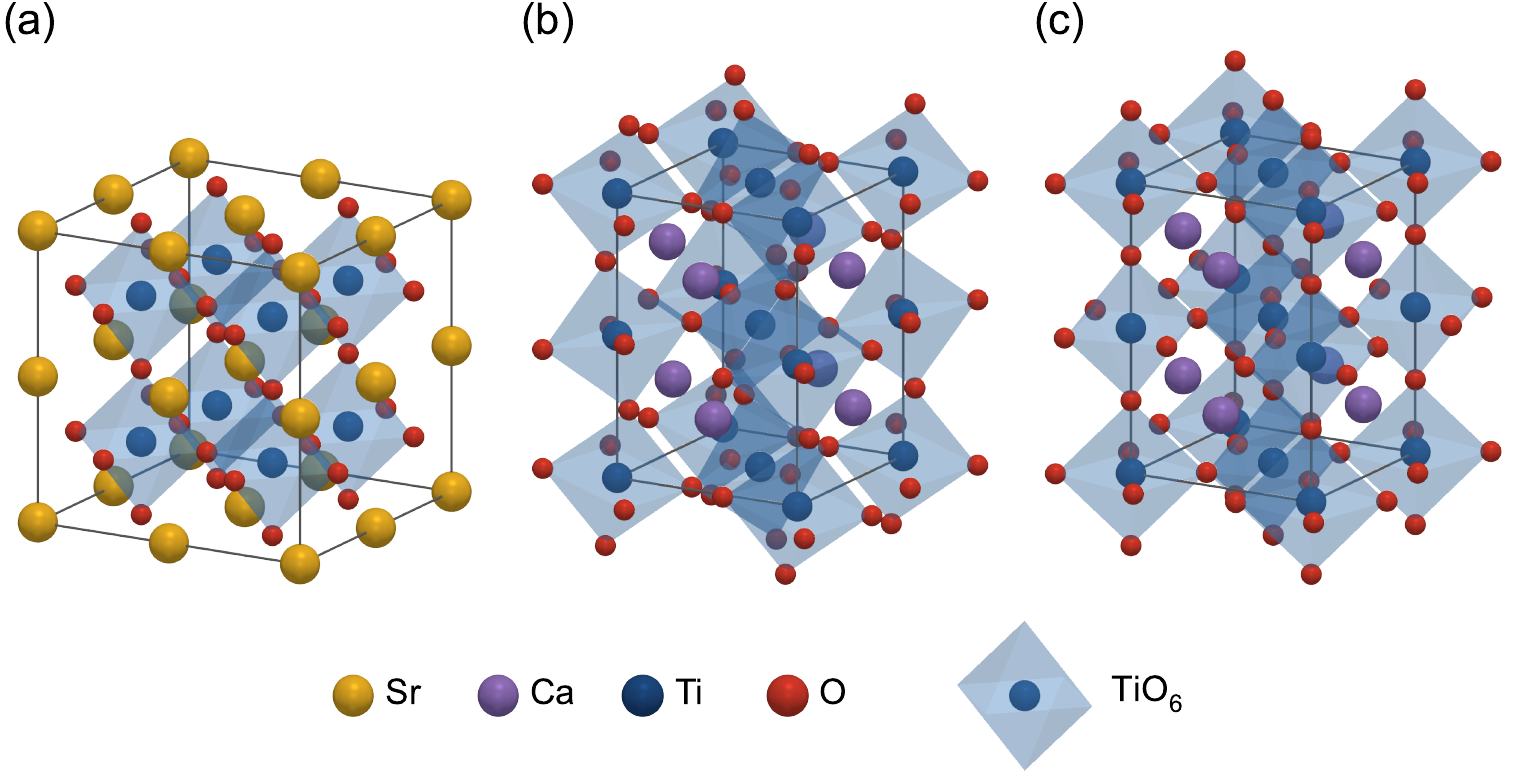}
\caption{Relaxed structures of the control hosts, drawn to the same scale: (a) cubic SrTiO$_3$ ($Pm\bar{3}m$),
$2\times2\times2$ primitive cells, (b) orthorhombic CaTiO$_3$ ($Pbnm$) and
(c) tetragonal CaTiO$_3$ ($I4/mcm$), one conventional cell each. TiO$_6$
octahedra are drawn translucent, with Ti at their centres. In (b) the Ca atoms
lie within 0.23~\AA{} of the cell faces and are drawn on both faces. The hypothetical
hexagonal SrTiO$_3$ has the framework of hexagonal BaTiO$_3$ [Fig.~1(a) of the
main text] with Sr on the A site and is not shown separately.}
\label{fig:si_cto}
\end{figure}

The SrTiO$_3$ and CaTiO$_3$ hosts used as controls in the main text are shown
in Fig.~\ref{fig:si_cto}.

\section{The oxygen vacancy next to Fe in hexagonal BaTiO$_3$}\label{si:hexvo}

The studies of the oxygen vacancy next to Fe in hexagonal BaTiO$_3$ are
listed in Table~\ref{tab:hexvo}. EPR on single crystals finds one associate
with the vacancy in the first shell, with Fe on Ti(2) or Ti(1) and the vacancy
on O(2)~\cite{bottcher2018incorporation}. Q-band EPR then finds only Fe on
Ti(1) with the vacancy on O(2)~\cite{langhammer2020incorporation}. Calculations
favour Fe on Ti(2) with the vacancy on O(1)~\cite{dawson2013phase,
adeagbo2019theoretical}, as do ours. Diffraction at 12.5--67\,\% Fe finds the
vacancies confined to O(1)~\cite{grey1998structure, grey1998accurate}, and
EXAFS finds them next to Fe in Fe$_2$O$_9$
dimers~\cite{chakraborty2013microscopic, valant2013cation}. For Cr,
calculations and EPR place the vacancy outside the first
shell~\cite{bottcher2005incorporation, nayak2015chromium}.

\begin{figure}[htbp]
\centering
\includegraphics[width=\textwidth]{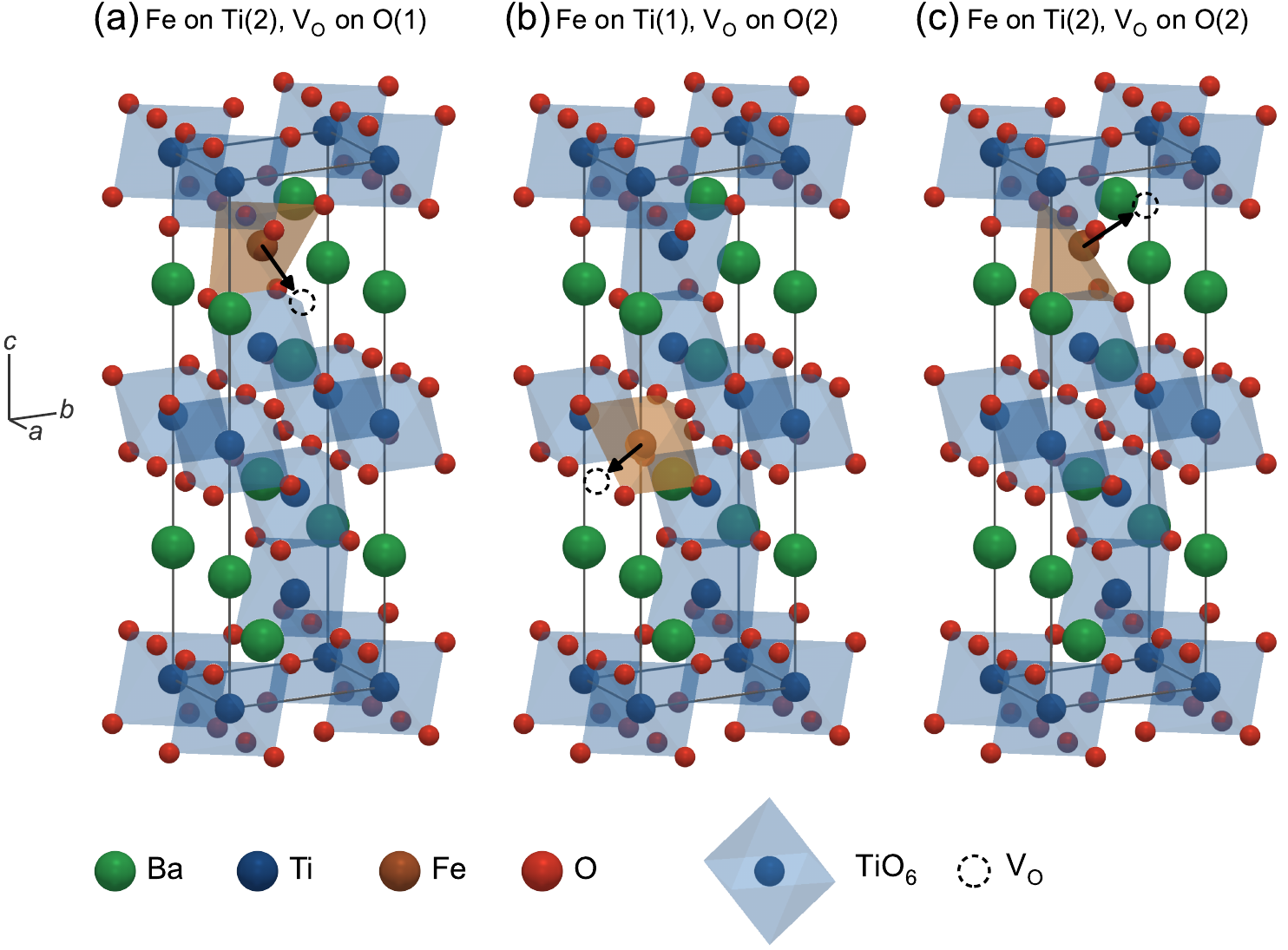}
\caption{The three Fe--$V_{\mathrm{O}}$ pairs in hexagonal BaTiO$_3$ with the
vacancy in the first shell of Fe, drawn in the 6H cell of main-text
Fig.~1(a): (a) Fe on Ti(2), $V_{\mathrm{O}}$ on O(1); (b) Fe on Ti(1),
$V_{\mathrm{O}}$ on O(2); (c) Fe on Ti(2), $V_{\mathrm{O}}$ on O(2). The
FeO$_5$ unit is shaded, the vacancy is drawn as a dashed circle, and the arrow
points from Fe to the vacancy.}
\label{fig:si_dipoles}
\end{figure}

The three Fe--$V_{\mathrm{O}}$ pairs with the vacancy in the first shell of Fe
are drawn in Fig.~\ref{fig:si_dipoles}. The EPR and DFT results for them are
compared in Table~\ref{tab:hexvo_eprdft}. In both calculations pair (a) is
the lowest, pair (b) is 0.5--0.6~eV higher and pair (c) 0.7~eV higher. EPR
observes a pair with the vacancy on O(2) and excludes
pair (a)~\cite{langhammer2020incorporation}. This difference between
calculation and EPR was noted in Ref.~\cite{adeagbo2019theoretical}, where the
configuration that matches the EPR angle was found to be too high in energy.
All three pairs are included in the configurational average of Sec.~III~B~1 of
the main text.

\begin{table}[htbp]
\centering
\caption{Fe site and oxygen vacancy next to Fe in hexagonal 6H-BaTiO$_3$, in
order of publication. For DFT, the lowest-energy first-shell complex. Ti(1) is the corner-sharing $2a$ site and
Ti(2) the face-sharing $4f$ site; O(1) is the oxygen on the shared face and
O(2) the other one (some studies use the opposite labels; they are converted
here).}
\label{tab:hexvo}
\small
\begin{tabular}{@{}l>{\raggedright\arraybackslash}p{0.27\linewidth}>{\raggedright\arraybackslash}p{0.2\linewidth}>{\raggedright\arraybackslash}p{0.2\linewidth}@{}}
\toprule
study & method & Fe site & $V_{\mathrm{O}}$ \\
\midrule
Grey 1998~\cite{grey1998structure, grey1998accurate} & X-ray and neutron diffraction & Ti(1), Ti(2) & O(1) \\
B\"ottcher 2008~\cite{bottcher2008phase} & EPR, ceramics & Ti(1) & not in first shell \\
Dawson 2013~\cite{dawson2013phase} & interatomic potentials & Ti(2), two Fe & O(1) \\
Chakraborty 2013~\cite{chakraborty2013microscopic} & EXAFS, single crystals & Ti(2), Fe$_2$O$_9$ dimers & around Fe$^{2+}$ \\
Valant 2013~\cite{valant2013cation} & EXAFS, ceramics & Ti(2), Fe$_2$O$_9$ dimers & O(1), after annealing at 1500~$^{\circ}$C \\
B\"ottcher 2018~\cite{bottcher2018incorporation} & EPR, ceramics and single crystals & Ti(2) or Ti(1) & O(2) \\
Adeagbo 2019~\cite{adeagbo2019theoretical} & DFT & Ti(2) & O(1) \\
Langhammer 2020~\cite{langhammer2020incorporation} & Q-band EPR, single crystals & Ti(1) & O(2) \\
this work & DFT+$U$ & Ti(2) & O(1) \\
\bottomrule
\end{tabular}
\end{table}

\begin{table}[htbp]
\centering
\caption{EPR and DFT studies of the three first-shell Fe--$V_{\mathrm{O}}$
pairs of Fig.~\ref{fig:si_dipoles} in hexagonal BaTiO$_3$, in order of
publication. Energies are in eV relative to pair (a). Those of
Ref.~\cite{adeagbo2019theoretical} are for the Fe$^{3+}$--$V_{\mathrm{O}}$
complex (Table~II there). Ours are for the cell with 1.85\,\% Fe.}
\label{tab:hexvo_eprdft}
\small
\begin{tabular}{@{}l>{\raggedright\arraybackslash}p{0.18\linewidth}>{\centering\arraybackslash}p{0.165\linewidth}>{\centering\arraybackslash}p{0.165\linewidth}>{\centering\arraybackslash}p{0.165\linewidth}@{}}
\toprule
study & method & (a) Fe on Ti(2), $V_{\mathrm{O}}$ on O(1) & (b) Fe on Ti(1), $V_{\mathrm{O}}$ on O(2) & (c) Fe on Ti(2), $V_{\mathrm{O}}$ on O(2) \\
\midrule
B\"ottcher 2018~\cite{bottcher2018incorporation} & EPR, single crystal, 2\,\% & -- & \multicolumn{2}{c}{one pair observed, (b) or (c)} \\
Adeagbo 2019~\cite{adeagbo2019theoretical} & DFT, 1.85\,\% & 0 & 0.64 & 0.73 \\
Langhammer 2020~\cite{langhammer2020incorporation} & Q-band EPR, single crystal, 5\,\% & excluded & observed & -- \\
this work & DFT+$U$, 1.85\,\% & 0 & 0.51 & 0.70 \\
\bottomrule
\end{tabular}
\end{table}

\section{Fe in the control hosts SrTiO$_3$ and CaTiO$_3$}\label{si:controls}

The distortion index $\lambda$ of the FeO$_6$ octahedron and the Jahn--Teller
energy $E_{\mathrm{JT}}$ in the two control hosts are listed in
Table~\ref{tab:ctrl_jt}. They were obtained in the same way as for BaTiO$_3$
in Sec.~III~B~2 of the main text. In SrTiO$_3$ the FeO$_6$ octahedron is
regular in the cubic phase ($\lambda = 0.001$,
$E_{\mathrm{JT}} = 0.00$~meV/f.u.). In the hypothetical hexagonal phase it is
as regular on the $2a$ site and more distorted on the $4f$ site
($\lambda = 0.027$, $E_{\mathrm{JT}} = 2.73$~meV/f.u.). Neither hexagonal site
is more regular than the cubic one. In BaTiO$_3$, for comparison, $\lambda$
decreases from 0.232 in the tetragonal phase to zero on the hexagonal $2a$
site. In CaTiO$_3$ the octahedron is distorted in both phases, slightly less
in the tetragonal one ($\lambda = 0.064$ compared with 0.077,
$E_{\mathrm{JT}} = 1.88$ compared with 2.34~meV/f.u.). The difference of
0.5~meV/f.u.\ is small compared with the 92~meV/f.u.\ between the two phases
(Sec.~III~A of the main text). Its sign is consistent with the observation
that Fe lowers the temperature of the orthorhombic-to-tetragonal transition of
CaTiO$_3$~\cite{becerro2000transition}.

\begin{table}[htbp]
\centering
\caption{Distortion index $\lambda$ of the FeO$_6$ octahedron and
Jahn--Teller energy $E_{\mathrm{JT}}$ in the control hosts, with the
BaTiO$_3$ values of the main text for comparison. $\lambda$ is the average
over all Fe atoms on the given site in all configurations with one and with
two Fe per supercell (1.85 and 3.70\,\% Fe in SrTiO$_3$ and BaTiO$_3$, 2.08
and 4.17\,\% in CaTiO$_3$). $E_{\mathrm{JT}}$ is for one Fe per supercell.}
\label{tab:ctrl_jt}
\begin{tabular}{@{}llccc@{}}
\toprule
host & phase, Fe site & $\lambda$, one Fe & $\lambda$, two Fe & $E_{\mathrm{JT}}$ (\mfu) \\
\midrule
SrTiO$_3$ & cubic             & 0.001 & 0.006 & 0.00 \\
          & hexagonal, $2a$   & 0.000 & 0.005 & 0.18 \\
          & hexagonal, $4f$   & 0.027 & 0.091 & 2.73 \\[2pt]
CaTiO$_3$ & orthorhombic      & 0.077 & 0.084 & 2.34 \\
          & tetragonal        & 0.064 & 0.067 & 1.88 \\
\midrule
BaTiO$_3$ & tetragonal        & 0.232 & 0.231 & 8.21 \\
          & hexagonal, $2a$   & 0.000 & 0.010 & 0.00 \\
\bottomrule
\end{tabular}
\end{table}

The local tolerance factor is evaluated as in Sec.~III~B~3 of the main text
(Table~\ref{tab:ctrl_tol}). In SrTiO$_3$, Fe on the Ti site is a mixture of
Fe$^{3+}$ and Fe$^{4+}$ whose ratio depends on the oxygen
treatment~\cite{vracar2007jahn}. In CaTiO$_3$ it is Fe$^{3+}$ after treatment
in CO--CO$_2$~\cite{mccammon2000short}, and up to about 20\,\% of it is
Fe$^{4+}$ after annealing in air~\cite{waerenborgh2001fe4}. For both valence
states the local tolerance factor is close to unity in SrTiO$_3$ and below
unity in CaTiO$_3$. In BaTiO$_3$ it is 1.062 for Ti$^{4+}$ and 1.041 for
Fe$^{3+}$.

\begin{table}[htbp]
\centering
\caption{Local Goldschmidt tolerance factor in the control hosts for
Ti$^{4+}$, high-spin Fe$^{3+}$ and Fe$^{4+}$ on the $B$ site, with the Shannon
radii~\cite{shannon1976revised} of the main text and Sr$^{2+}$ 1.44~\AA,
Ca$^{2+}$ 1.34~\AA\ (12-fold), Fe$^{4+}$ 0.585~\AA\ (6-fold).}
\label{tab:ctrl_tol}
\begin{tabular}{@{}lccc@{}}
\toprule
host & Ti$^{4+}$ & Fe$^{3+}$ (HS) & Fe$^{4+}$ \\
\midrule
SrTiO$_3$ & 1.002 & 0.982 & 1.012 \\
CaTiO$_3$ & 0.966 & 0.947 & 0.976 \\
\bottomrule
\end{tabular}
\end{table}

\bibliographystyle{unsrt}
\bibliography{FeBTOtet2hex} %